\pdfoutput=1

\documentclass[12pt]{article}
\usepackage[a4paper,left=1in, right=1in, top=1in,bottom=1in]{geometry}
\usepackage{amsmath,amssymb,amsthm}

\usepackage[authoryear]{natbib}
\usepackage[hidelinks,colorlinks=true,allcolors=magenta]{hyperref}
\usepackage{shortex}
\usepackage{caption}
\usepackage{float}
\usepackage{multirow}
\usepackage{blindtext}
\usepackage{dsfont}
\usepackage[title]{appendix}
\usepackage{graphicx}

\usepackage{setspace}
\usepackage{import}
\usepackage{xifthen}
\usepackage{pdfpages}
\usepackage{transparent}

\newcounter{xxx}
\newcommand{\schedule}{\mcB}
\newcommand{\statespace}{\mcX}

\newcommand{\E}{\mathbb{E}}

\newcommand{\AIS}{\mathrm{AIS}}
\newcommand{\SIR}{\mathrm{SIR}}
\newcommand{\AR}{\mathrm{AR}}

\newcommand{\DAR}{\mathrm{DAR}}
\newcommand{\SMC}{\mathrm{SMC}}

\newcommand{\CESS}{\mathrm{CESS}}

\newcommand{\eff}{\mathrm{eff}}

\newcommand{\pre}{\mathrm{pre}}
\newcommand{\post}{\mathrm{post}}
\newcommand{\KL}{\mathrm{KL}}
\newcommand{\PT}{\mathrm{PT}}

\newcommand{\ELBO}{\mathrm{ELBO}}

\newcommand{\iidsim}{\stackrel{\mathrm{i.i.d.}}\sim}

\newtheorem{theorem}{Theorem}
\newtheorem{corollary}{Corollary}

\newtheorem{assumption}{Assumption}

\usepackage{authblk}
\usepackage{todonotes}

\title{Optimised Annealed Sequential Monte Carlo Samplers}
\author[1]{Saifuddin Syed\footnote{\textit{Corresponding author.} Address: 3182 - 2207 Main Mall, Vancouver, BC, V6T 1Z4, Canada. \\
Email: \texttt{saif.syed@stat.ubc.ca}.}}
\author[1]{Alexandre Bouchard-C\^{o}t\'{e}}
\author[1]{Kevin Chern}
\author[2]{Arnaud Doucet}
\affil[1]{Department of Statistics, University of British Columbia}
\affil[2]{Department of Statistics, University of Oxford}

\date{}

\usepackage{fancyhdr}
\begin{document}

\maketitle
\begin{abstract}
\noindent Annealed Sequential Monte Carlo (ASMC) samplers are special cases of SMC samplers where the sequence of distributions can be embedded in a smooth path of distributions. Using this underlying path and a performance model based on the variance of the normalising constant estimator, we systematically study dense-schedule limits.
From our theory emerges a notion of global barrier, capturing the inherent complexity of normalising constant approximation under our performance model. We then turn the resulting approximations into surrogate objective functions of algorithm performance, using them to guide method development. This leads to novel adaptive methods, Optimised Annealed SMC (OASMC), which address practical difficulties inherent in previous adaptive SMC methods.  
First, our OASMC algorithms are predictable: they produce a sequence of increasingly precise estimates at deterministic, known times.
Second, Optimised Annealed Importance Sampling (OAIS), a special case of OASMC, enables schedule adaptation at a memory cost constant in the number of particles, requiring significantly less communication. Finally, these characteristics make OAIS highly efficient on GPUs. 
We provide an open-source, high-performance GPU implementation of our method and demonstrate up to a hundred-fold speed improvement compared to 
state-of-the-art adaptive AIS methods.  \bigskip


\end{abstract}

\section{Introduction}
\label{sec:intro}

Annealed Importance Sampling (AIS) \citep{neal_annealed_2001} and Sequential Monte Carlo (SMC) samplers \citep{del_moral_sequential_2006} are widely used methods for approximating normalising constants of complex \emph{target} probability distributions $\pi$ (e.g., posterior distribution). These methods take as input an unnormalised density $\gamma(x)$ specifying the target density $\pi(x) = \gamma(x) / Z$, a tractable \emph{reference distribution} $\eta$ (for example, a prior distribution), and a sequence of intermediate distributions $\pi^{(0)}, \pi^{(1)}, \dots, \pi^{(T)}$ interpolating between $\pi^{(0)}=\eta$ and $\pi^{(T)}=\pi$. The output is an unbiased and consistent approximation to the normalising constant $Z$. Here we focus on a special case, \emph{annealed SMC} (ASMC), where $\pi^{(t)}=\pi_{\beta_t}$ arises from a discretisation of a continuum of \emph{annealing distributions} $\pi_\beta$ parameterised by the \emph{annealing parameter} $\beta \in [0, 1]$, satisfying $\pi_\beta=\eta$ at $\beta=0$ and $\pi_\beta=\pi$ at $\beta=1$.

In this work, we first develop a theoretical framework in \cref{sec:variance-finite-schedule} exploiting the assumption that the discrete sequence $(\pi^{(t)})_{t=0}^T$ is embedded in a continuum $(\pi_\beta)_{\beta \in [0, 1]}$. Our framework refines and extends previous work based on the ``perfect transitions'' assumption \citep{neal_annealed_2001,grosse2013annealing,dai2022invitation,Omar2024} and enables a comprehensive analysis of the performance of the normalising constant estimator associated with ASMC and AIS as a function of several key algorithmic tuning parameters: the annealing schedule, the resampling schedule, and the number of particles. 

Our analysis of ASMC in the dense schedule limit $T\to\infty$ (\cref{sec:dense_schedule_limit}) reveals an analogue of the local and global barriers $\lambda(\beta)$ and $\Lambda$ developed in the context of Parallel Tempering (PT) \citep{geyer1991markov,hukushima1996exchange,syed2019non}, a popular annealing-based MCMC algorithm.\footnote{We will use $\lambda$ and $\Lambda$ for the local and global barriers developed in this work, and denote their PT counterparts by $\lambda_\text{PT}$ and $\Lambda_\text{PT}$.} The local barrier encodes the difficulty of an infinitesimal importance sampling problem in the neighbourhood of the annealing parameter $\beta$, and generalises the standard deviation of the energy statistic \citep{gelman1998simulating,grosse2013annealing,rotskoff2017statistical,chopin2023connection,Omar2024,barzegar2024optimal} to a broader class of SMC algorithms. The global barrier encodes the overall difficulty of a problem when approached via AIS or ASMC when $T$ is large. The local and global barriers are key factors in the performance of particle algorithms (see \cref{sec:global_barrier_discussion} and \cite{syed2019non,biron2025automatic}), making these barriers  interesting objects of study in their own right. Moreover, these barriers can be efficiently approximated from the output of SMC algorithms.

Since the global barrier $\Lambda$ characterises problem complexity under our simplifying assumptions, we study the $\Lambda \to \infty$ limit in  \cref{sec:criticality} to understand the effects of the number of particles $N$, the number of intermediate distributions $T$, and the resampling frequency. This analysis reveals sharp cut-offs predicting which combinations of algorithmic parameters lead to normalising constant estimators with stable relative variance. These cut-offs yield practical tuning guidelines for ASMC/AIS algorithms taking into account computing architecture characteristics such as parallelism and memory use.

Our theoretical analysis motivates new methods in \cref{sec:methodology} addressing several practical limitations of existing AIS and ASMC algorithms. State-of-the-art SMC and AIS methods have random and hard-to-predict run times because their performance depends heavily on the schedule $0=\beta_0<\cdots<\beta_T=1$, which is tuned online. The prevailing approach, proposed in \citet{jasra_inference_2011,Zhou2016}, uses at each iteration $t$ a root-finding problem over $[\beta_{t}, 1]$ to obtain $\beta_{t+1}$ achieving a predetermined constant $\chi^2$-divergence $\delta$ between $\pi_{\beta_t}$ and $\pi_{\beta_{t+1}}$. This \emph{online schedule tuning} leads to a random number of iterations and hence random run time since the objective function is approximated by noisy particle weights. The random and unpredictable run time creates two problems: (1)~it is difficult to set \citeauthor{Zhou2016}'s tuning parameter $\delta$ to achieve a target approximation error and/or expected run time, which is undesirable since users often have a computational budget and want the best approximation within that budget; and (2)~when existing adaptive SMC algorithms are terminated before completion, they may not provide any usable output, in contrast to MCMC-based normalising constant estimation methods \citep{bennett_efficient_1976,xie_improving_2011}, which are \emph{anytime algorithms} that can be interrupted at any iteration and still provide useful output. We address both problems using a novel \emph{round-based schedule tuning} scheme, OASMC (\cref{alg:OASMC}), for \emph{optimising} ASMC. Our algorithm has deterministic run time and produces intermediate approximations at predictable intervals. Moreover, our normalising constant estimator is unbiased since our annealing schedule is tuned adaptively over rounds, rather than online.

As an additional benefit, the AIS variant of our algorithm performs schedule tuning with a memory cost that is constant in the number of particles. In contrast, existing schedule adaptation methods have memory costs that grow linearly with the number of particles. For high-dimensional target distributions with weak dependence between dimensions, our adaptive AIS algorithm requires $O(d)$ memory, compared with $O(d^{3/2})$ for optimally tuned ASMC samplers and PT algorithms, where $d$ is the problem dimensionality.

To demonstrate the generality and high performance of our proposed algorithms and make them accessible to practitioners, we provide two open-source implementations. The first implements both OASMC and OAIS as plug-ins to the Blang modelling language \citep{Bouchard2022Blang}, allowing users to specify statistical models using syntax similar to BUGS and Stan \citep{Lunn2000,stan2023stan}. We use this implementation to validate and benchmark our algorithms across 12 diverse models including Bayesian hierarchical models, spike-and-slab variable selection, estimation of ordinary differential equation (ODE) parameters, and phylogenetic inference (see Appendix~\ref{ap:models} for details). We also provide a Julia implementation of OAIS with GPU support, leveraging our algorithm's significantly reduced inter-thread communication requirements compared to previous AIS adaptation methods. Thanks to its limited communication overhead, our GPU-optimised OAIS algorithm achieves a 10--100$\times$ speedup compared with a state-of-the-art adaptive AIS algorithm with comparable variance of the normalising constant estimator.

\section{Annealed Sequential Monte Carlo samplers}\label{sec:background}

For $\beta\in[0,1]$, let $\pi_\beta$ be the \emph{annealing distribution} on $\mcX$ with unnormalised density $\gamma_\beta:\mcX\to\reals^{+}$ evaluable up to an unknown normalising constant $Z_\beta$:
\[
\pi_\beta(x)=\frac{\gamma_\beta(x)}{Z_\beta},\qquad Z_\beta=\int_\mcX \gamma_\beta(x)\,\dee x.
\]
We assume $\gamma_0(x)=\eta(x)$ is the normalised reference density with $Z_0=1$, and $\gamma_1(x)=\gamma(x)$ is the unnormalised target density with $Z_1=Z$. The \emph{annealing path} $\beta\mapsto \pi_\beta$ continuously interpolates between the tractable reference $\pi_0$ and the target $\pi_1$. The canonical choice is the \emph{geometric path} $\gamma_\beta(x)=\eta(x)^{1-\beta}\gamma(x)^\beta$, although other annealing paths can be used \citep{grosse2013annealing,masrani2021q,syed2021parallel,phillips2024particle}.

\subsection{Algorithm}
Given an \emph{annealing schedule} $\mathcal{B}_T=\beta_{0:T}$ of size $T+1$ partitioning $[0,1]$ with $0=\beta_0<\dots<\beta_T=1$, the ASMC algorithm generates weighted particles $\textbf{X}_{t}=X^{1:N}_t$ and $\textbf{w}_{t}=w^{1:N}_t$ to approximate $\pi_{\beta_t}$. At $t=0$, we initialize $N$ reference particles $X^n_0\iidsim\eta$ with uniform weights $w^n_0=1$. For each $t\in[T]$, we obtain $(\textbf{X}_{t},\textbf{w}_t)$ by: (1) propagating, (2) reweighting, and (optionally) (3) resampling the weighted particles $(\textbf{X}_{t-1},\textbf{w}_{t-1})$.\footnote{For $K\in\nats$ we denote $[K]=\{1,\dots,K\}$.} The algorithm is summarised in \cref{alg:AMCS}. See \cite{dai2022invitation} for a recent review.

\begin{algorithm}[t]
\caption{\texttt{ASMC}$(N, \mathcal{B}_T)$}
\label{alg:AMCS}
\begin{algorithmic}
\Require \# particles $N$, annealing schedule $\mathcal{B}_T$
\State $\textbf{X}_0 = X_0^{1:N},\quad X^n_0\iidsim \eta$  \Comment{Initialise particles}
\State $\textbf{w}_0 = w_0^{1:N},\quad w^n_0=1$  \Comment{Initialise weights}
\State $\hZ = 1$ \Comment{Initialise normalising constant estimator}
\For{$t\in [T]$} 
    \State $\textbf{X}_{t^-}=X^{1:N}_{t^-},\quad X^n_{t^-}\sim M_{\beta_{t-1},\beta_t}(X_{t-1}^{n},\dee x_{t^-})$\Comment{Propagation step}
    \State $\textbf{g}_t=g^{1:N}_t,\quad g^n_t= g_{\beta_{t-1},\beta_t}(X^n_{t-1},X^n_{t^-})$
    \State $\hat g_{t,i} = \sum_{n\in[N]}  w^n_{t-1}  (g^n_t)^i$ for $i\in \{0,1,2\}$ \Comment{Compute moments (see \cref{sec:schedule-update})}
    \State $\textbf{w}_{t^-}=w^{1:N}_{t^-},\quad w^n_{t^-} = w^n_{t-1} g^n_t$ \Comment{Reweight step}
    \If{$t=\tau_r$} \Comment{Resample step}
        \State $\hZ \gets  \hZ \frac{1}{N} \sum_{n \in [N]} w_{t^-}^n$\Comment{Update normalising constant}
        \State $\textbf{W}_{t^-}=W^{1:N}_{t^-},\quad W^n_{t^-}=w^n_{t^-}/\sum_{m\in[N]}w^m_{t^-}$
        \State Generate $\textbf{a}_t=a^{1:N}_t$ such that $\Pr[a^n_t=m]=W^m_{t^-}$\Comment{e.g., systematic resampling}
        \State $\textbf{X}_t=X^{1:N}_t,\qquad X^n_t= X^{a^n_t}_{t^-}$
        \State $\textbf{w}_t=w^{1:N}_t,\qquad w^n_t= 1$
    \Else
       \State $\textbf{X}_t= \textbf{X}_{t^-},\quad\textbf{w}_t= \textbf{w}_{t^-}$ \Comment{No resampling}
    \EndIf
\EndFor
\State \textbf{Return:} $\hZ, \hat{\textbf{g}}=\hat{g}_{1:T,0:2}$
\end{algorithmic}
\end{algorithm}

\subsubsection{Propagation step}
For $\beta,\beta'\in[0,1]$, the forward kernel $M_{\beta,\beta'}$ is a Markov transition kernel on $\mcX$ designed to propagate samples from $\pi_{\beta}$ towards $\pi_{\beta'}$. We require only that sampling from $M_{\beta,\beta'}(x,\cdot)$ is efficient for all $x\in\mathcal{X}$, without assuming ergodicity, irreducibility, or that the \emph{forward proposal} $\pi_{\beta}\otimes M_{\beta,\beta'}(\dee x,\dee x')=\pi_\beta(\dee x)M_{\beta,\beta'}(x,\dee x')$ admits a Lebesgue density. The propagation step generates $\textbf{X}_{t^-}$ by advancing particles $\textbf{X}_{t-1}$ using the forward kernel,
\[
X^n_{t^-}\sim M_{\beta_{t-1},\beta_t}(X^{n}_{t-1},\dee x_{t^-}).
\]
The canonical choice is an MCMC kernel $M_{\beta,\beta'}=K_{\beta'}$ that is $\pi_{\beta'}$-invariant, such as Hamiltonian Monte Carlo \citep{neal2011mcmc} or a slice sampler \citep{neal2003slice}. Other common choices include deterministic transports \citep{arbel2021annealed}, and unadjusted Langevin-based kernels are also possible \citep{heng2020controlled,doucet2022score}. See \cref{sec:analysis-forward-backward} and \cref{ap:explorer} for more information on the choice and adaptation of forward kernels. 

\subsubsection{Reweight step}
We introduce \emph{backward kernels} $L_{\beta',\beta}$ on $\mcX$ such that the \emph{backward target} $\pi_{\beta'}\otimes L_{\beta',\beta}$ is absolutely continuous with respect to the forward proposal $\pi_{\beta}\otimes M_{\beta,\beta'}$. The backward kernel accounts for distribution changes through changes in the \emph{incremental weight} $G_{\beta,\beta'}:\mcX\times\mcX\to\reals$ defined as the Radon--Nikodym derivative of the backward target with respect to the forward proposal. Any backward kernel suffices provided the unnormalised incremental weight function $g_{\beta,\beta'}(x,x')$ can be efficiently evaluated for all $x,x'\in\mathcal{X}$, 
\[
g_{\beta,\beta'}(x,x')=\frac{Z_{\beta'}}{Z_{\beta}}G_{\beta,\beta'}(x,x'),\qquad G_{\beta,\beta'}(x,x')
=\frac{\dee\pi_{\beta'}\otimes L_{\beta',\beta}}{\dee \pi_{\beta}\otimes M_{\beta,\beta'}}(x,x').
\]
At iteration $t$, the reweighting step generates $\textbf{w}_{t^-}=w^{1:N}_{t^-}$ by multiplying $\textbf{w}_{t-1}=w^{1:N}_{t-1}$ with incremental weights $\textbf{g}_t=g^{1:N}_t$:
\[
w^n_{t^-}=w^n_{t-1}g^n_t,\qquad g^n_t=g_{\beta_{t-1},\beta_t}(X^n_{t-1},X^n_{t^-}).
\]
For example, for the MCMC forward kernel $M_{\beta,\beta'}=K_{\beta'}$, the standard choice is the quasi-reversal $L_{\beta',\beta}=K_{\beta'}^*$ satisfying $\pi_{\beta'}(\dee x')K^*_{\beta'}(x',\dee x)=\pi_{\beta'}(\dee x)K_{\beta'}(x,\dee x')$, giving
\[
G_{\beta,\beta'}(x,x')=\frac{\dee\pi_{\beta'}}{\dee\pi_\beta}(x),\qquad g_{\beta,\beta'}(x,x')=\frac{\gamma_{\beta'}(x)}{\gamma_{\beta}(x)}.
\]
See \cref{sec:analysis-forward-backward} for additional examples and \cite{del_moral_sequential_2006, dai2022invitation} for more discussion on forward/backward kernels.

\subsubsection{Resample step}\label{sec:resampling}
After the propagation and reweighting steps, we can set $(\textbf{X}_t,\textbf{w}_t)=(\textbf{X}_{t^-},\textbf{w}_{t^-})$ and proceed to the next iteration. However, we may optionally trigger a \emph{resampling} event to stabilise weights and favour the propagation of particles with higher relative weights. Formally, we generate ancestral indices $\textbf{a}_{t}=(a^1_t,\dots,a^N_t)\in [N]^N$ such that $\Pr[a_t^n=m\mid\textbf{w}_{t^-}]=W^m_{t^-}$, where $W^n_{t^-}=w^n_{t^-}/\sum_{m\in[N]}w^m_{t^-}$ are the normalised weights. We then set
\[
X^n_t=X^{a^n_t}_{t^-},\qquad w^n_t=1.
\]
Many resampling schemes have been developed, e.g., systematic resampling is popular as it is easy to implement and performs well in theory and practice \citep{chopin2022resampling}.

\paragraph{Resampling strategies.}
Let $0=\tau_0<\dots<\tau_{R}= T$ denote the resampling times, where $\tau_r$ is the $r$-th resampling event and $R$ is the (possibly random) total number of resampling events. The resampling times $\tau_r$ are determined by the weighted particles generated up to time $t^-$. For notational convenience, we always trigger a final resampling at time $T=\tau_R$. 

Common strategies for determining the resampling times include \emph{annealed importance sampling} (AIS) \citep{neal_annealed_2001}, which uses no intermediate resampling ($R_{\AIS}=1$ with $\tau_1=T$); \emph{sequential importance resampling} (SIR) \citep{gordon_novel_1993}, which resamples at every iteration ($R_{\SIR}=T$ with $\tau_r=r$ for $r\in[T]$); and \emph{adaptive resampling} (AR) \citep{liu_blind_2012}, which resamples when the variability of $\textbf{w}_{t^-}$ exceeds a threshold $\nu\in(0,1)$. Formally, AR triggers resampling at $\tau_r=\inf\{\tau_{r-1}<t\leq T: N_\text{eff}(t)< \nu N\}$, where the \emph{effective sample size} (ESS) \citep{kong1994sequential} is
\[\label{eq:sample_size_IS}
N_\text{eff}(t)=\frac{\left(\sum_{n\in[N]}w^n_{t^-}\right)^2}{\sum_{n\in[N]} (w^n_{t^-})^2}\in [1,N].
\]
Unlike AIS and SIR, the total number of resampling events $R_{\AR}\in[T]$ for AR is random. 
\subsection{Estimators}
Using the weighted particles from \cref{alg:AMCS}, we can construct consistent estimators for expectations and the normalising constant. \cite{del_moral_sequential_2006} provide conditions under which the weighted particles before and after resampling at time $t$ are consistent estimators for expectations\footnote{Let $\E_{\beta}[f]$ and $\var_\beta[f]$ denote the expectation and variance of $f:\mcX\to\reals$ with respect to $\pi_\beta$. } with respect to the annealing distributions,
\[\label{eq:estimator_pi_t}
\E_{\beta_{t}}[f]&\stackrel{a.s.}{=}\lim_{N\to\infty}\sum_{n\in[N]}W^n_{t^-} f(X_{t^-}^n),\quad  W^n_{t^-}=\frac{w^n_{t^-}}{\sum_{m\in[N]}w^m_{t^-}}\\
&\stackrel{a.s.}{=}\lim_{N\to\infty}\sum_{n\in[N]}W^n_{t} f(X_{t}^n), \quad W^n_{t}=\frac{w^n_{t}}{\sum_{m\in[N]}w^m_{t}}.
\]
Similarly, we can use the propagated samples to obtain consistent estimators for expectations\footnote{For $\beta,\beta'\in[0,1]$, let $\E_{\beta,\beta'}[h]$ and $\var_{\beta,\beta'}[h]$ denote the expectation and variance of $h:\mcX\times\mcX\to\reals$ with respect to $\pi_\beta\otimes M_{\beta,\beta'}$.} with respect to the forward proposals,
\[\label{eq:estimator_forward_proposal}
\E_{\beta_{t-1},\beta_t}[h]\stackrel{a.s.}{=}\lim_{N\to\infty}\sum_{n\in[N]}W^n_{t-1}h(X^n_{t-1},X^n_{t^-}),\qquad W^n_{t-1}=\frac{w^n_{t-1}}{\sum_{m\in[N]}w^m_{t-1}}.
\]
Finally, \cref{alg:AMCS} outputs a consistent estimator $\hZ$ for the normalising constant $Z$ by multiplying the average unnormalised weights before resampling:
\[\label{eq:estimator_Z}
\hZ = \prod_{r\in[R]}\frac{1}{N}\sum_{n\in [N]}w^n_{\tau_r^-}=\prod_{r\in[R]}\frac{1}{N}\sum_{n\in [N]}g^n_{\tau_{r-1}+1}\cdots g^n_{\tau_{r}}.
\]
Like the expectation estimators, $\hZ$ is a consistent estimator for $Z$, i.e., $\hZ \stackrel{a.s.}{\to} Z$ as $N\to\infty$. Unlike the expectation estimators, however, $\hZ$ is unbiased for any finite $N$, even when the resampling times are adaptive. Although it is a commonly stated fact in the SMC literature (e.g.\ \citet[Exercise 16.4]{chopin2020introduction}), we formalise it in \cref{prop:estimators}, since the unbiasedness of $\hZ$ is a cornerstone of our theoretical analysis.

\bprop\label{prop:estimators}
Let $0=\tau_{0}<\cdot\cdot\cdot<\tau_{R}=T$ be a strictly increasing sequence of stopping times adapted to the filtration of weighted particles generated by the propagation and reweight steps at time $t$, then $\hat{Z}$ defined in \cref{eq:estimator_Z} is unbiased, i.e., $\mathbb{E}[\hat{Z}]=Z$.
\eprop

\section{Analysis of ASMC with a finite schedule}\label{sec:variance-finite-schedule} 
Since $\hZ$ is unbiased for any choice of forward and backward kernels and any resampling times, we can use its relative variance to analyse the performance of ASMC and guide algorithm design. This will be the central object of study for the remainder of this work.

\subsection{Performance Model Assumptions}
\label{sec:assumptions}

To analyse the variance of the normalising constant estimator $\hZ$ output by \cref{alg:AMCS}, we consider an idealised ``performance model'' for ASMC. In general, computing $\var[\hZ]$ is analytically intractable because of particle interactions introduced through resampling and temporal correlations induced by the forward kernels. We therefore impose the following  assumptions on the incremental weights.

\bassump[Integrability]\label{assump:integrability}
For all $\beta,\beta'\in[0,1]$, $\var_{\beta,\beta'}[g_{\beta,\beta'}]$ is finite.
\ebassump

\bassump[Temporal independence]\label{assump:independent_weights_time}
For $n\in[N]$, $g^n_{1},\dots, g^n_T$ are independent.
\ebassump

\bassump[Particle independence]\label{assump:independent_weights_particles}
For $t\in[T]$, $g^1_t,\dots,g^N_t$ are independent.
\ebassump

\bassump[Efficient local exploration]\label{assump:ELE}
For all $n\in[N]$ and $t\in[T]$, 
\[\label{eq:ELE_weights_law}
g^n_t\overset{d}{=} g_{\beta_{t-1},\beta_t}(X_{t-1},X_{t^-}), \qquad (X_{t-1},X_{t^-})\sim \pi_{\beta_{t-1}}\otimes M_{\beta_{t-1},\beta_t}.
\]
\ebassump

These assumptions are inspired by related performance models in the AIS/SMC \citep{neal_annealed_2001,grosse2013annealing,dai2022invitation} and PT \citep{syed2019non} literature. While not expected to hold exactly in practice, they can be approximated and provide algorithmic design guidelines that we validate empirically in \cref{sec:numerical-experiments}.

\cref{assump:integrability} ensures that the variance of $\hZ$ is well-defined and finite. \cref{assump:independent_weights_time} primarily concerns the quality of forward kernels, while \cref{assump:independent_weights_particles} relates to the quality of the resampling scheme: temporal interactions arise from propagating particles using the forward kernel, whereas particle interactions are induced by resampling. \cref{assump:independent_weights_particles} is exactly satisfied by AIS, and it can be approximated for SMC using discretisation-stable resampling schemes \citep{chopin2022resampling} with suitably dense schedules. \cref{assump:independent_weights_time,assump:ELE} can be approximated by introducing resample-move steps, i.e., $K$ iterations of an MCMC move targeting $\pi_{\beta_{t-1}}$ between each SMC iteration for sufficiently large $K$ \citep{gilks_following_2001}. In \cref{sec:inefficient-explorers}, we empirically demonstrate that the method derived from our analysis (see \cref{sec:methodology}) is robust to violations of our assumptions.

A limitation of our analysis is that \cref{assump:independent_weights_particles} does not distinguish between different resampling strategies (e.g., multinomial, stratified, or systematic resampling). Consequently, the variance analysis in \cref{sec:normalising_constant_analysis} does not account for the additional variance introduced by the resampling scheme itself, nor the correlations in the particles induced by the forward moves.

\subsection{Non-asymptotic variance of normalising constant estimator}\label{sec:normalising_constant_analysis}

Under \cref{assump:integrability,assump:independent_weights_time,assump:independent_weights_particles,assump:ELE}, the relative variance of $\hZ$ defined in \cref{eq:estimator_Z} decomposes in terms of the \emph{incremental discrepancy} $D(\beta,\beta')$, defined as the Rényi $2$-divergence between $\pi_\beta \otimes M_{\beta,\beta'}$ and $\pi_{\beta'} \otimes L_{\beta',\beta}$ measuring the relative variability of the incremental weights:
\begin{equation}\label{eq:incremental-discrepancy-def}
D(\beta,\beta') = \log \left(\frac{\E_{\beta,\beta'}[g_{\beta,\beta'}^2]}{\E_{\beta,\beta'}[g_{\beta,\beta'}]^2} \right)= \log(1 + \var_{\beta,\beta'}[G_{\beta,\beta'}]),
\end{equation}
where the expectation and variance are with respect to $\pi_{\beta}\otimes M_{\beta,\beta'}$.
For example, with MCMC kernels $M_{\beta,\beta'} = K_{\beta'}$ and $L_{\beta',\beta} = K_{\beta'}^*$, this simplifies to $D(\beta,\beta') = \log(1 + \var_\beta[\frac{\dee\pi_{\beta'}}{\dee\pi_\beta}])$. See \cref{sec:analysis-forward-backward} for expressions for $D(\beta,\beta')$ for other forward/backward kernels and \cref{sec:schedule-update} for an estimator for $D(\beta_{t-1},\beta_t)$ derived from \cref{alg:AMCS} output.

Intuitively, a larger discrepancy $D(\beta_{t-1},\beta_t)$ indicates a higher incremental weight variance when transitioning from $\pi_{\beta_{t-1}}$ to $\pi_{\beta_t}$ at iteration $t$ of \cref{alg:AMCS}. \cref{thm:variance_ERS} formalises this, showing the relative variance depends exponentially on the discrepancy accumulated between resampling events. Moreover, the relative variance admits a characterisation via a unique quantity $\rho \in [1, R]$, the \emph{effective resample size} (ERS), which measures efficiency relative to an idealised baseline ASMC algorithm that accumulates a constant discrepancy between the $R$ resampling events for a given $N$ and $\mcB_T$.

\bthm\label{thm:variance_ERS}
Suppose \cref{assump:integrability,assump:independent_weights_time,assump:independent_weights_particles,assump:ELE} hold with deterministic resampling times $\tau_{0:R}$. For $t \leq t'$, let $D(\mathcal{B}_T, t, t') = \sum_{s=t+1}^{t'} D(\beta_{s-1}, \beta_s)$ denote the accumulated discrepancy from time $t$ to $t'$, and let $D(\mathcal{B}_T) = \sum_{t\in[T]}D(\beta_{t-1},\beta_t)$ denote the total discrepancy.
\begin{enumerate}
    \item[(a)] The relative variance satisfies
    \begin{equation}\label{eq:ERS_identity_conditional}
    \var\left[\frac{\hZ}{Z}\right] = \prod_{r \in [R]} \left(1 + \frac{\exp(D(\mathcal{B}_T, \tau_{r-1}, \tau_r)) - 1}{N}\right) - 1.
    \end{equation}
    \item[(b)] If $N > 1$ and $D(\mathcal{B}_T) > 0$, there exists a unique $\rho \in [1, R]$ such that
    \begin{equation}\label{eq:ERS_identity}
    \var\left[\frac{\hZ}{Z}\right] = \left(1 + \frac{\exp(D(\mathcal{B}_T)/\rho) - 1}{N}\right)^\rho - 1.
    \end{equation}
    Moreover, $\rho = 1$ if and only if $D(\mathcal{B}_T, \tau_{r-1}, \tau_r) = D(\mathcal{B}_T)$ for some $r \in [R]$, and $\rho = R$ if and only if $D(\mathcal{B}_T, \tau_{r-1}, \tau_r) = D(\mathcal{B}_T)/R$ for all $r \in [R]$.
\end{enumerate}
\ethm
It follows from \cref{thm:variance_ERS} that, under our assumptions, the relative variance of $\hZ$ is determined entirely by the number of particles $N$, the annealing schedule through the total discrepancy $D(\mcB_T)$, and the resampling times through the ERS $\rho$. See \cref{sec:scaling_finite_T} for a discussion of how the relative variance scales with these quantities.

\subsection{Effective resample size for deterministic adaptive resampling}
We cannot directly apply \cref{thm:variance_ERS} to analyse AR, where resampling occurs stochastically at times when $N_\eff(t)<\nu N$. However, \citet[Theorems 2.2 and 2.3]{del2012adaptive} showed that, outside an event with probability exponentially small in $N$, the resampling times for AR are almost surely deterministic. We refer to this deterministic approximation of AR as DAR. Under our simplifying assumptions, DAR triggers a resampling event when the accumulated discrepancy from the previous resampling time exceeds $-\log \nu$:
\[\label{eq:DAR-times}
\tau_r = \inf\{\tau_{r-1} < t \leq T : \exp(-D(\mathcal{B}_T, \tau_{r-1}, t)) < \nu\}.
\]
Notably, DAR triggers no resampling for $t<T$ when $\nu \leq \exp(-D(\mcB_T))$, coinciding with AIS, and triggers resampling every iteration when $\nu > \max_{t\in[T]}\exp(-D(\beta_{t-1},\beta_t))$, coinciding with SIR. The threshold $\nu$ controls the amount of particle interaction, and DAR interpolates between AIS ($\nu = 0$) and SIR ($\nu = 1$). See \cref{sec:AR=DAR} for details.

Because DAR is deterministic, by \cref{thm:variance_ERS}, we can analyse the relative variance of $\hZ_\DAR$ through its ERS $\rho_\DAR(\nu)$, estimated in \cref{prop:adaptive_SMCS}.
\bprop\label{prop:adaptive_SMCS}
Suppose \cref{assump:integrability,assump:independent_weights_time,assump:independent_weights_particles,assump:ELE} hold. For all $N>1$ and $D(\mcB_T)>0$, the effective resample size $\rho_\DAR(\nu)$ for DAR with threshold $\nu \in (0,1)$ satisfies,
\begin{equation}\label{eq:ERS-bound-DAR}
\max\left\{1, \frac{D(\mathcal{B}_T)}{\max_{t \in [T]} D(\beta_{t-1}, \beta_t) - \log \nu}\right\} \leq \rho_{\DAR}(\nu) \leq \min\left\{T, 1 - \frac{D(\mathcal{B}_T)}{\log \nu}\right\}.
\end{equation}
\eprop

\cref{thm:variance_ERS}(a) recovers the variance estimate for AIS (no resampling) derived in \citet[Section 4]{neal_annealed_2001} and by \cref{thm:variance_ERS}(b), the ERS for AIS is $\rho_\AIS = 1$:
\[
\var\left[\frac{\hZ_\AIS}{Z}\right]=\frac{\exp(D(\mcB_T))-1}{N}.
\]
For SIR (resample every iteration), \cref{thm:variance_ERS}(a) recovers the variance estimate derived in \citet[Section 3.2]{dai2022invitation}:
\[
\var\left[\frac{\hZ_\SIR}{Z}\right]=\prod_{t\in[T]}\left(1+\frac{\exp(D(\beta_{t-1},\beta_t))-1}{N}\right)-1.
\]
Taking limits in \cref{eq:ERS-bound-DAR} as $\nu\to1^-$ yields estimates of the ERS $\rho_\SIR$:
\[
\frac{D(\mcB_T)}{\max_{t\in[T]}D(\beta_{t-1},\beta_t)}\leq \rho_\SIR\leq T,
\]
with equality on the right if and only if the incremental discrepancies are uniform, i.e., $D(\beta_{t-1},\beta_t)=D(\mcB_T)/T$ for all $t \in [T]$.

\section{The dense schedule limit for ASMC }\label{sec:dense_schedule_limit}
This section analyses how the total discrepancy and the asymptotic variance of $\hZ$ behave in the dense schedule limit, where  $\max_{t\in[T]}|\beta_t-\beta_{t-1}|\to 0$.

\subsection{Regularity assumptions}

To obtain tractable estimates of the incremental discrepancies $D(\beta,\beta')$ as $\beta'\to\beta$, we require the forward proposals $\pi_\beta\otimes M_{\beta,\beta'}$ and incremental weights $G_{\beta,\beta'}$ to be sufficiently smooth functions of $(\beta,\beta')$. Intuitively, this ensures that: (1) expectations under the forward proposal can be differentiated by passing derivatives under the integral, (2) incremental weights and their derivatives do not grow too rapidly, and (3) local changes in $\beta$ and $\beta'$ induce corresponding local changes in the distributions and weights. 

We formalize this through a nested hierarchy of function spaces $\mcF_0\supset\mcF_1\supset\mcF_2\supset\mcF_3$ that control the regularity and growth of functions $h:\mcX\times \mcX\to \reals$. Each $\mcF_i$ is a vector space containing constants and closed under domination (if $|h'|\leq |h|$ with $h\in\mcF_i$, then $h'\in\mcF_i$). The key requirement is that for $i\in\{0,1,2,3\}$ and $h\in\mcF_i$, then $(\beta,\beta')\mapsto \E_{\beta,\beta'}[h]$ is $i$-times continuously differentiable, and $\mcF_i$ contains all $(3-i)$-th order partial derivatives of $(\beta,\beta')\mapsto\log G_{\beta,\beta'}(x,x')$. The following three assumptions ensure these properties hold.

\bassump[Regular expectations]\label{assump:regular-proposal} 
For all $i\in\{0,1,2,3\}$ and $f\in\mcF_i$, $\E_{\beta,\beta'}[f]$ is $i$-times continuously differentiable in $(\beta,\beta')$. Moreover, for all $j,j'\in\nats$ with $j+j'\leq i$, there exist signed measures $\eta_{\beta,\beta',j,j'}$ over $\mcX\times\mcX$ integrable on $\mcF_i$ and the integral $\eta_{\beta,\beta',j,j'}[f]$ satisfies,
\[
f\in\mcF_i,\quad \eta_{\beta,\beta',j,j'}[f]=\frac{\partial^{j+j'}}{\partial\beta^{j}\partial\beta'^{j'}}\E_{\beta,\beta'}[f].
\] 
\ebassump

\bassump[Regular weights]\label{assump:regular-weights} 
For all $x,x'\in\mcX$, the incremental weight $G_{\beta,\beta'}(x,x')$ is positive and 3-times continuously differentiable in $(\beta,\beta')$. Moreover, for all $i\in\{0,1,2,3\}$ and $j_1,\dots,j_k\geq 1$ with $j_1+\dots+j_k=i$, we require $(1+\barG)^2 \bar{U}_{j_1}\cdots\bar{U}_{j_k}\in \mcF_{3-i}$, where,
\[
 \bar{G}(x,x')=\sup_{\beta,\beta'\in[0,1]}G_{\beta,\beta'}(x,x'),\quad \bar{U}_j(x,x')=\max_{i+i'=j}\sup_{\beta,\beta'\in [0,1]}\left|\frac{\partial^{i+i'}}{\partial\beta^i\partial\beta'^{i'}}\log G_{\beta,\beta'}(x,x')\right|.
\] 
\ebassump

\bassump[Non-degeneracy]\label{assump:non-degeneracy} 
$G_{\beta,\beta'}(x,x')=1$ for all $x,x'\in\mcX$, if and only if $\beta'= \beta$. Moreover, for all $\beta\in[0,1]$, the variance $\delta(\beta):=\var_{\beta,\beta}[\dot G_\beta]$ is strictly positive, where
\[
\dot{G}_\beta(x,x')=\left.\frac{\partial}{\partial \beta'} G_{\beta, \beta'}(x,x')\right|_{\beta' = \beta}.
\]\ebassump

\cref{assump:regular-proposal,assump:regular-weights} ensure that $D(\beta, \beta')=\log(1+\var_{\beta,\beta'}[G_{\beta, \beta'}])$ is 3-times continuously differentiable in $(\beta,\beta')$, with derivatives that can be explicitly computed from the partial derivatives of the forward proposal $\pi_{\beta}\otimes M_{\beta,\beta'}$ and incremental weights $G_{\beta,\beta'}$. \cref{assump:non-degeneracy} implies $D(\beta,\beta')\geq 0$, with equality if and only if $\beta' = \beta$, which allows us to interpret $D(\beta, \beta')$ as a statistical divergence between the annealing distributions that quantifies the distributional change from $\pi_\beta$ to $\pi_{\beta'}$ through the variance of the incremental weight with respect to the forward proposal.

\cref{assump:non-degeneracy} implies that, for all $\beta\in[0,1]$, as $\Delta\beta\to 0$, the incremental weight admits the asymptotic expansion $G_{\beta, \beta+\Delta\beta}(x,x') = 1 + \dot G_\beta(x,x') \Delta\beta+O(\Delta\beta^2).$ 
Thus, local changes in the incremental weights are encoded in $\dot G_\beta$ and its variance $\delta(\beta)$, which is strictly positive by \cref{assump:non-degeneracy}. $\delta(\beta)$ measures the sensitivity of incremental weights to infinitesimal changes in $\beta$. \cref{thm:incremental_discrepancy_estimate} shows that the incremental discrepancy is locally quadratic and controlled by $\delta(\beta)$.

\bthm\label{thm:incremental_discrepancy_estimate}
Suppose \cref{assump:regular-proposal,assump:regular-weights,assump:non-degeneracy} hold. Then $D(\beta,\beta')$ is $3$-times continuously differentiable in $\beta,\beta'$, and $\delta(\beta)$ is continuously differentiable in $\beta$. Moreover, for all $\beta,\beta'\in[0,1]$ and $\Delta\beta=\beta'-\beta$,
\[\label{eq:discrepancy-taylor}
\left|D(\beta,\beta+\Delta\beta)-\delta(\beta)\Delta\beta^2\right|\leq \frac{1}{6}\sup_{\beta,\beta'\in[0,1]}\left|\frac{\partial^3}{\partial\beta'^3}D(\beta,\beta')\right||\Delta\beta|^3.
\]
\ethm
We can compute $\dot G_\beta(x,x')$ and $\delta(\beta)$ explicitly for different kernel choices. For MCMC forward kernels with quasi-reversal backward kernels, $\dot G_\beta$ and $\delta(\beta)$ reduce to the Fisher score and Fisher information for the annealing distributions $(\pi_\beta)_{\beta\in[0,1]}$:
\[\label{eq:Fisher_information_example}
\dot G_\beta(x,x')=\frac{\dee}{\dee\beta}\log \pi_\beta(x),\qquad \delta(\beta)=\var_{\beta}\left[\frac{\dee}{\dee\beta}\log \pi_\beta\right].
\]
For the geometric path $\pi_\beta\propto \eta^{1-\beta}\pi^\beta$, \cref{assump:regular-proposal,assump:regular-weights,assump:non-degeneracy} hold provided $\pi\neq \eta$ and $(1+\exp(V))^2V^3$ is integrable with respect to both $\pi$ and $\eta$, where $V(x)=\log\gamma(x)-\log \eta(x)\neq 0$, $\dot G_\beta(x,x')=V(x)-\E_\beta[V]$ and $\delta(\beta)=\var_{\beta}[V]$.
See \cref{sec:analysis-forward-backward} for details and additional examples with different forward/backward kernel choices.

\subsection{Asymptotic variance in the dense schedule limit}\label{sec:dense-limit-variance}
We now explore how the performance of \cref{alg:AMCS} scales when the annealing schedule $\mcB_T$ becomes dense as $T\to\infty$. To obtain tractable limits of $D(\mcB_T)$, we compare schedules of different sizes by assuming the sequence $\{\mcB_T\}_{T\in\nats}$ is generated by a fixed function $\varphi$. 

\begin{assumption}[Schedule generator]\label{assump:schedule_generator}
There exists a continuously twice-differentiable function $\varphi:[0,1]\to [0,1]$ with $\varphi(0)=0$, $\varphi(1)=1$, and $\dot\varphi(u)=\frac{\dee}{\dee u}\varphi(u)>0$ such that for all $T\in\nats$ and $t\in[T]$, the schedule $\mathcal{B}_T=\beta_{0:T}$ satisfies $\beta_t=\varphi(u_t)$, where $u_t=t/T$.
\end{assumption}

Under \cref{assump:schedule_generator}, we have $\Delta\beta_t=\beta_t-\beta_{t-1}\approx\dot{\varphi}(u_t)\Delta u_t$ where $\Delta u_t=1/T$. Combining this with \cref{thm:incremental_discrepancy_estimate}, we obtain $D(\beta_{t-1},\beta_t)\approx  \delta(\varphi(u_t))\dot\varphi(u_t)^2\Delta u_t^2$ and $\sqrt{D(\beta_{t-1},\beta_t)}\approx \lambda(\beta_{t-1})\Delta\beta_t$, where $\lambda(\beta)=\sqrt{\delta(\beta)}$ is the \emph{local barrier}. Interpreting these as Riemann sum approximations yields the following integral approximations.

\begin{corollary}\label{cor:dense_schedule_limit_discrepancy}
Suppose \cref{assump:regular-proposal,assump:regular-weights,assump:non-degeneracy,assump:schedule_generator} hold. Then:
\begin{enumerate}
\item[(a)] There exists $C_\delta>0$ independent of $t\in[T]$ such that
\[
\left|D(\beta_{t-1},\beta_t)-\frac{1}{T}\int_{u_{t-1}}^{u_t}\delta(\varphi(u))\dot{\varphi}(u)^2\dee u\right|\leq \frac{C_\delta}{T^3}.\label{eq:accumulated-discrepancy-estimate}
\]

\item[(b)] There exists $C_\lambda>0$ independent of $t\in[T]$ such that
\[
\left|\sqrt{D(\beta_{t-1},\beta_t)}-\int_{\beta_{t-1}}^{\beta_t}\lambda(\beta)\dee \beta\right|\leq \frac{C_\lambda}{T^2}. \label{eq:accumulated-discrepancy-estimate-root}
\]
\end{enumerate}
\end{corollary}
Summing \cref{eq:accumulated-discrepancy-estimate} over $t\in[T]$ yields that the total discrepancy satisfies $\sum_{t\in[T]}D(\beta_{t-1},\beta_t)\approx E(\varphi)/T$, where $E(\varphi)=\int_{0}^{1}\delta(\varphi(u))\dot\varphi(u)^2\dee u$ is the \emph{kinetic energy} of $\varphi$. More remarkably, summing \cref{eq:accumulated-discrepancy-estimate-root} over $t$ shows that the sum of root discrepancies, $\sum_{t\in[T]}\sqrt{D(\beta_{t-1},\beta_t)}\approx\Lambda$, is approximately independent of the annealing schedule, where $\Lambda=\int_0^1\lambda(\beta)\dee \beta$ is the \emph{global barrier}. Moreover, by Jensen's inequality and the substitution $\beta=\varphi(u)$, we have $E(\varphi)\geq \Lambda^2$ for all $\varphi$:
\[
E(\varphi)
=\int_0^1\delta(\varphi(u))\dot\varphi(u)^2\dee u
\geq \left(\int_0^1\lambda(\varphi(u))\dot\varphi(u)\dee u\right)^2
=\left(\int_0^1\lambda(\beta)\dee\beta\right)^2
=\Lambda^2.
\]

Combining this with \cref{thm:variance_ERS} shows that the relative variance of $\hZ$ goes to zero as $T\to\infty$ at a rate controlled by $E(\varphi)$, independent of the resampling strategy. Thus, minimising the asymptotic variance is equivalent to minimising the kinetic energy $E(\varphi)$ in the dense schedule limit.

\bthm\label{thm:dense_limit_variance}
Suppose \cref{assump:integrability,assump:independent_weights_time,assump:independent_weights_particles,assump:ELE,assump:regular-proposal,assump:regular-weights,assump:non-degeneracy,assump:schedule_generator} hold, and let $\hZ$ be obtained from \cref{alg:AMCS} with deterministic resampling times. Then, as $T\to\infty$,
    \begin{equation}\label{eq:dense_limit_formula}
    \lim_{T\to\infty}T\var\left[\frac{\hZ}{Z}\right] = \frac{E(\varphi)}{N} \geq \frac{\Lambda^2}{N},
    \end{equation}
where the convergence is uniform over all choices of resampling times. Equality holds if and only if $\varphi=\varphi^*$, where the optimal schedule generator $\varphi^*=\argmin_{\varphi}E(\varphi)$ is twice differentiable and strictly increasing, with $\varphi^*(0)=0$, $\varphi^*(1)=1$, and for $u\in[0,1]$,
    \begin{equation}\label{eq:geodesic}
    \varphi^*(u) = \Lambda^{-1}(\Lambda u), \qquad \Lambda(\beta) = \int_0^\beta \lambda(\beta') \dee \beta'.
    \end{equation}
\ethm
\paragraph{Connection to KL divergence minimisation.}
Our variance-based analysis also applies to the common practice of choosing schedules to minimise the accumulated KL divergence $\KL(\mcB_T)=\sum_{t\in[T]}\KL(\beta_{t-1},\beta_t)$, where $\KL(\beta,\beta')=-\E_{\beta,\beta'}[\log G_{\beta,\beta'}]$ is the KL-divergence between $\pi_{\beta}\otimes M_{\beta,\beta'}$ and $\pi_{\beta'}\otimes L_{\beta',\beta}$ \citep{neal_annealed_2001,grosse2013annealing,arbel2021annealed}. In the dense schedule limit, $\lim_{T\to\infty}T\KL(\mcB_T)= E(\varphi)/2$ as $T\to\infty$, so minimising variance is equivalent to minimising the KL divergence. See \cref{sec:bias} for details.

\subsection{The global barrier for normalising constant estimation}\label{sec:global_barrier_discussion}

The global barrier $\Lambda$ depends on the annealing distributions and the forward/backward kernels, but not on the number of particles, annealing schedule, or resampling strategy. The local barrier $\lambda(\beta)$ measures the instantaneous rate of change in $\pi_\beta$ under small perturbations in $\beta$, while the global barrier measures the cumulative change along the entire annealing path from reference to target. The optimal schedule $\mathcal{B}_T^*=\beta^*_{0:T}$, with $\beta^*_t=\varphi^*(t/T)$, ensures that $\int_{\beta_{t-1}^*}^{\beta_t^*}\lambda(\beta)\dee\beta$ is constant for all $t\in[T]$, placing a higher density of annealing parameters where the path is most sensitive to changes in $\beta$. 
Expressions for $\varphi^*$ in the Gaussian case are derived in \citet[Sec.~3.3]{dai2022invitation}; see also \cref{sec:connection-online-cess} for connections to online schedule optimisation methods and \cref{sec:schedule-update} for a new empirical approximation method. See \cref{fig:vary-exploration,fig:gpus} in \cref{sec:numerical-experiments} for empirical examples.

\paragraph{Geometric interpretation}
Viewing the annealing distributions $\Pi=(\pi_\beta)_{\beta\in[0,1]}$ as a statistical manifold, a schedule generator $\varphi$ defines a curve $u\mapsto \pi_{\varphi(u)}$ in $\Pi$ between the reference $\pi_{\varphi(0)}=\eta$ and target $\pi_{\varphi(1)}=\pi$. \cref{thm:dense_limit_variance} reveals a deep connection between the asymptotic efficiency of normalising constant estimation and the Riemannian geometry of $\Pi$, equipped with the metric $\delta(\beta)$. The kinetic energy $E(\varphi)$ and global barrier $\Lambda$ are geometric quantities, and the optimal schedule $\varphi^*=\argmin_\varphi E(\varphi)$ is the geodesic between $\eta$ and $\pi$. Since $\Pi$ is one-dimensional, this geodesic coincides with the constant-speed reparameterisation of the annealing path of length $\Lambda$.

This geometric perspective extends classical results from the AIS literature. It has been well-known that with MCMC forward/backward kernels, the asymptotic efficiency depends on the Fisher information metric from \cref{eq:Fisher_information_example}; for example in \citet{gelman1998simulating,grosse2013annealing,rotskoff2017statistical,chopin2023connection,Omar2024,barzegar2024optimal}. \cref{thm:dense_limit_variance} extends this connection to general ASMC samplers without restricting the choice of resampling strategy or forward/backward kernels. 

We can overcome this barrier by expanding the annealing parameter space and optimising the annealing path  \citep{gelman1998simulating,grosse2013annealing,syed2021parallel,masrani2021q,barzegar2024optimal}, or by modifying the Riemannian metric $\delta(\beta)$ directly and optimising over the forward/backward kernels for example with a learnable flow or diffusion \citep{vaikuntanathan_escorted_2011,arbel2021annealed,barzegar2024optimal,phillips2024particle,chen2024sequential,skreta2025feynman,he2025rne}. See \cref{sec:barrier} for more details.

\section{The large barrier limit for ASMC}\label{sec:criticality}
We study how computational requirements scale with problem complexity. Consider an optimal  schedule $\mcB_T^*=\beta^*_{0:T}$, generated by $\varphi^*$ with $\beta^*_t=\varphi^*(u_t)$, for a problem with global barrier $\Lambda$. Even with this optimal schedule, the resources required for stable normalising constant estimation grow in $\Lambda$. We analyse how this scaling behaves as $\Lambda\to\infty$.

\subsection{Variance bounds with optimal schedule}
We first derive finite-sample bounds for the incremental discrepancy and particle requirements. \cref{thm:geodesics}(a) shows that, under the optimal schedule $\mcB_T^*$ of size $T$, $\kappa^{-1} \Lambda^2\leq TD(\mcB_T^*)\leq \kappa \Lambda^2$. The tightness of these bounds depends on a type of \emph{condition number} $\kappa=\sup_\beta\delta(\beta)/\inf_\beta\delta(\beta)\geq 1$, which measures the variability of the local barrier function $\delta(\beta)$ along the annealing path. A condition number close to one indicates uniform difficulty across the path, while $\kappa\gg 1$ reveals bottlenecks where the schedule must concentrate more steps. This provides a concrete link between the global barrier $\Lambda$ and the total discrepancy $D(\mcB_T^*)$ depending on a finite schedule of size $T$. 

Combined with \cref{thm:variance_ERS}, \cref{thm:geodesics}(b) establishes sharp particle requirements: it provides both lower bounds (below which the variance exceeds a tolerance $\epsilon$) and upper bounds (above which the variance falls below $\epsilon$). Together, these results fully characterise the computational cost of estimating $Z$ to a given accuracy using an optimal schedule.

\begin{theorem}\label{thm:geodesics}
\begin{enumerate}
\item[(a)] Suppose \cref{assump:regular-proposal,assump:regular-weights,assump:non-degeneracy} hold, then for all $t\in[T]$, 
\[\displaystyle\frac{\Lambda^2}{\kappa T^2}\leq D(\beta^*_{t-1},\beta^*_t)\leq \frac{\kappa\Lambda^2}{T^2}.\]

\item[(b)] Suppose additionally that \cref{assump:integrability,assump:independent_weights_time,assump:independent_weights_particles,assump:ELE} hold. Let $\hZ$ be the output of \cref{alg:AMCS} using an optimal schedule $\mcB_T^*$ of size $T$ generated by \cref{eq:geodesic}, with deterministic resampling times and effective resample size $\rho\in[1,T]$. Then for all $\epsilon>0$:
\begin{enumerate}
    \item[(i)] If $\displaystyle N < \frac{\rho}{\epsilon}\left(\exp\left(\frac{\Lambda^2}{\kappa \rho T}\right) -1\right)$, then $\displaystyle\var\left[\frac{\hZ}{Z}\right]>\epsilon$.
    \item[(ii)] If $\displaystyle N > \frac{\rho}{\log(1+\epsilon)}\left(\exp\left(\frac{\kappa\Lambda^2}{\rho T}\right) -1\right)$, then $\displaystyle\var\left[\frac{\hZ}{Z}\right]<\epsilon$.
\end{enumerate}
\end{enumerate}
\end{theorem}
\subsection{Particle stability analysis for ASMC}
\begin{figure}[t]
    \centering
    \includegraphics[width=1.0\textwidth]{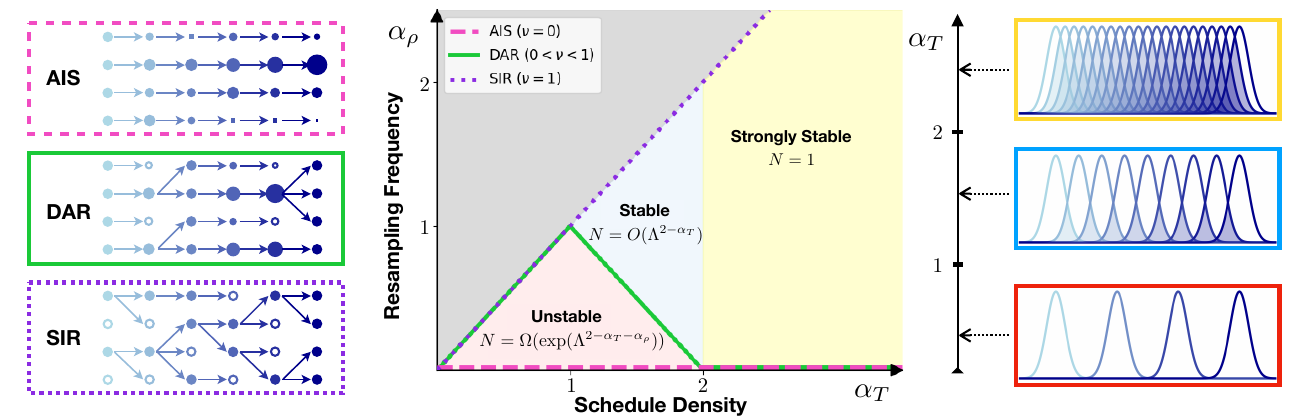}
    \caption{\textbf{Left:} Particle evolution for AIS (top) and DAR (middle) and SIR (bottom). \textbf{Middle:} Phase diagram identifying particle stability regimes when $T\sim \Lambda^{\alpha_T}$ and $\rho=\Theta(\Lambda^{\alpha_\rho})$. The unstable regime (red) requires exponentially many particles, the stable regime (green) requires a sublinear number, and the strongly stable regime (yellow) requires only a constant number. We also plot the stability regimes for AIS (dashed), DAR (solid), and SIR (dotted) as the schedule density $\alpha_T$ increases. \textbf{Right:} Discretised annealing paths between two Gaussian distributions at increasing schedule densities: $\alpha_T\in (0,1)$ (bottom), $\alpha_T\in (1,2)$ (middle), and $\alpha_T\in (2,\infty)$ (top). }
    \label{fig:phase_diagram}
\end{figure}

Suppose $\Lambda\to\infty$ with $T\sim\Lambda^{\alpha_T}$ and $\rho=\Theta(\Lambda^{\alpha_\rho})$,\footnote{We say $a(y)=O(b(y))$ if $|a(y)|\leq C|b(y)|$ for some $C>0$ for all $y$; $a(y)=\Omega(b(y))$ if $|a(y)|\geq C|b(y)|$; and $a(y)=\Theta(b(y))$ if both hold.} where $\alpha_T\geq0$ measures the schedule density and $0\leq \alpha_\rho\leq \alpha_T$ (since $1\le\rho\leq T$) measures the frequency of particle interaction through resampling. Using \cref{thm:geodesics}, we determine how many particles $N$ are required for the variance to remain stable as $\Lambda\to\infty$. Three distinct regimes emerge (\cref{fig:phase_diagram})---unstable, stable, and strongly stable---depending on the schedule density and particle interaction. These regimes identify sharp phase transitions where the scaling behaviour of \cref{alg:AMCS} changes discontinuously. Notably, this classification makes no assumptions on the state space, the choice of annealing distributions, or forward/backward kernels.

\paragraph{Unstable regime ($\alpha_\rho+\alpha_T<2$).} 
When the number of annealing distributions is sparse relative to $\Lambda$ ($\alpha_T<1$) or particles interact too infrequently ($\alpha_\rho<2-\alpha_T$), at least an exponential number of particles is required to obtain relative variance of $\hZ$ within $\epsilon>0$. By \cref{thm:geodesics}, $N=\Omega\left(\epsilon^{-1}\Lambda^{\alpha_\rho} \exp(\kappa^{-1}\Lambda^{2-\alpha_T-\alpha_\rho})\right)$. The exponential term is minimised when $\alpha_\rho=\alpha_T$, implying that frequent resampling is advantageous for sparse schedules.

\paragraph{Stable regime ($1\leq\alpha_T\leq 2$ and $\alpha_\rho\geq 2-\alpha_T$).} 
With sufficient overlap between distributions and adequate particle interaction, only $N=O(\log(1+\epsilon)^{-1}\Lambda^{2-\alpha_T})$ particles are needed as $\Lambda\to\infty$. This yields sublinear scaling ($2-\alpha_T\in[0,1]$), an exponential reduction compared to the unstable regime. The particle requirement does not decrease when $\alpha_\rho>2-\alpha_T$, showing that beyond a threshold, additional resampling provides no benefit. This regime demonstrates how one can maintain stable normalising constant estimates with fewer interacting particles by adding more annealing distributions.

\paragraph{Strongly stable regime ($\alpha_T>2$).} 
When the schedule is sufficiently dense relative to $\Lambda$, the relative variance of $\hZ$ converges to zero as $\Lambda\to\infty$, independently of $N$ and $\alpha_\rho$. In particular, it is possible to obtain stable estimates with $O(1)$ particles and no resampling.

\subsection{Particle stability of deterministic adaptive resampling}\label{sec:high-barrier-scaling-resampling}
Combining \cref{prop:adaptive_SMCS} and \cref{thm:geodesics}(a) under an optimal annealing schedule of size $T$, we obtain the following bounds on the ERS of DAR with resampling threshold $\nu\in [0,1]$, as defined in \cref{eq:DAR-times}, which are uniform in $N$:
\begin{equation}\label{eq:ERS-bounds-optimal}
\max\left\{1,\frac{\Lambda^2T}{\kappa^2\Lambda^2-\kappa T^2\log \nu }\right\}\leq \rho_\DAR(\nu)\leq \min\left\{T,1-\frac{\kappa\Lambda^2}{T\log \nu}\right\}.
\end{equation}
If $T\sim \Lambda^{\alpha_T}$ as $\Lambda\to\infty$, then \cref{eq:ERS-bounds-optimal} implies $\rho_\DAR(\nu)=\Theta(\Lambda^{\alpha_\rho})$, where $\alpha_\rho$ depends on both the schedule density (measured by $\alpha_T$) and the resampling threshold $\nu$. This allows us to determine the regime and particle requirements as a function of the density of the annealing schedule $\alpha_T$, summarised in \cref{fig:phase_diagram}.

\paragraph{Annealed importance sampling ($\nu=0$).}
With AIS, the ERS is $\rho_{\AIS}=\rho_\DAR(0)=1$, hence $\alpha_\rho=0$. As $\Lambda\to\infty$, AIS is in the unstable regime when $\alpha_T<2$, requiring  $N=\Omega(\exp(\kappa^{-1}\Lambda^{2-\alpha_T}))$ particles. It is in the strongly stable regime when $\alpha_T=2$, requiring $N=O(1)$ particles, and in the strongly stable regime when $\alpha_T>2$.

\paragraph{Sequential importance resampling ($\nu=1$).}
For SIR, we have $\rho_\SIR=\rho_{\DAR}(1)$ with $T/\kappa^2\leq \rho_{\SIR}\leq T$, hence $\rho_\SIR=\Theta(\Lambda^{\alpha_T})$. SIR is in the unstable regime when $\alpha_T<1$, requiring $N=\Omega(\Lambda^{\alpha_T}\exp(\kappa^{-1}\Lambda^{2-2\alpha_T}))$ particles; in the stable regime when $1\leq \alpha_T< 2$, requiring $N=O(\Lambda^{2-\alpha_T})$ particles; and in the strongly stable regime when $\alpha_T\geq 2$.

\paragraph{Deterministic adaptive resampling ($0<\nu<1$).}
When $\alpha_T<1$, we have $\rho_\DAR(\nu)=\Theta(\Lambda^{\alpha_T})$, placing DAR in the unstable regime where it scales like SIR. When $1\leq \alpha_T\leq 2$, we have $\rho_\DAR(\nu)=\Theta(\Lambda^{2-\alpha_T})$, placing DAR in the stable regime. Finally, when $\alpha_T>2$, we have $\rho_{\DAR}(\nu)\to 1$ as $\Lambda\to\infty$, placing DAR in the strongly stable regime where it scales like AIS with no resampling events triggered.

\begin{figure}[t]
\centering
\includegraphics[width=\textwidth]{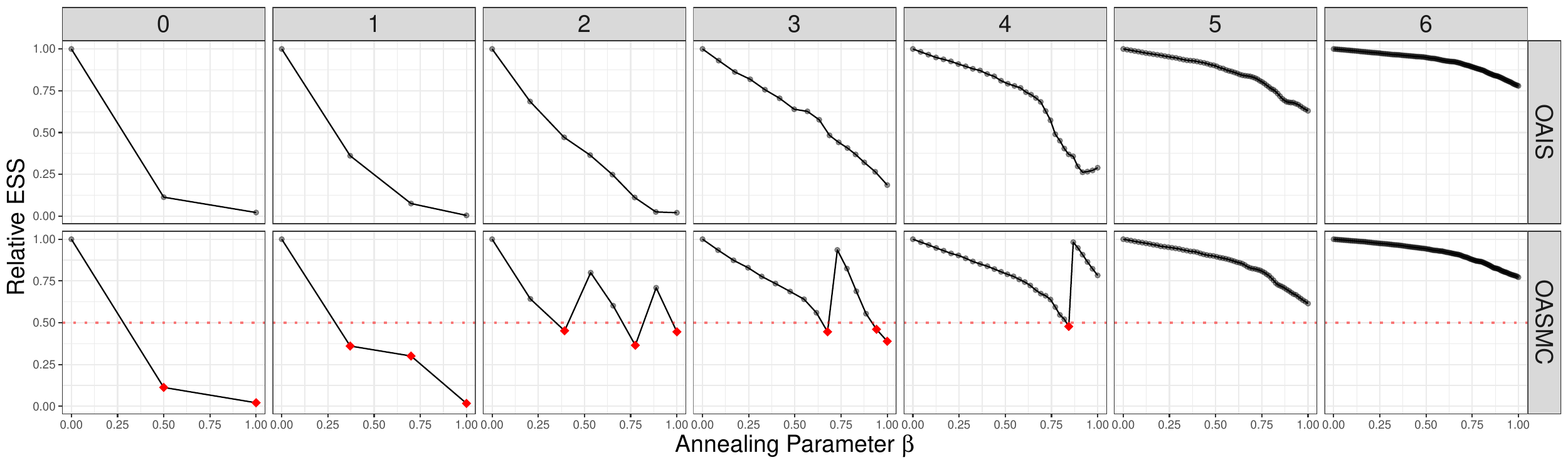}
\caption{Relative ESS $N_\text{eff}(t)/N$ as a function of iteration for OASMC targeting the Ising model. Facet columns denote adaptation round $k$ in Algorithm~\ref{alg:OASMC}. The top facet row shows OAIS, bottom facet row, OASMC. For OASMC, Adaptive resampling (AR) (\cref{sec:resampling}) is triggered when $N_\text{eff}(t)/N < \nu=1/2$ (dotted red line). As predicted by theoretical results for DAR in \cref{sec:high-barrier-scaling-resampling}, in the unstable schedule regime (small $k$), resampling occurs (red square) at every iteration; as $k$ increases, resampling becomes periodic and less frequent; in the strongly stable schedule regime (large $k$), no resampling occurs and OAIS and OASMC are equivalent in that regime.}
\label{fig:ising-ess}
\centering
\end{figure}

Notably, for any choice of $\alpha_T$, DAR triggers the minimal number of resampling events needed for stable estimation of the normalising constant (\cref{fig:phase_diagram}). Specifically, DAR automatically adapts to trigger minimal particle interaction in the stable regime whenever possible ($\alpha_T\geq 1$) and only enters the unstable or strongly stable regimes when unavoidable. Since AR coincides with DAR with high probability when $N$ is sufficiently large \citep{del2012adaptive}, this optimality explains why AR achieves robust empirical performance for large $N$ and is insensitive to the resampling threshold $\nu\in (0,1)$. \cref{fig:ising-ess} empirically demonstrates that AR automatically adapts to different regimes as $T$ increases.

\subsection{High-dimensional scaling}\label{sec:high-d-scaling}
Our analysis of the large barrier limit $\Lambda\to\infty$ extends naturally to high-dimensional scaling $d\to\infty$. Suppose we artificially expand the state space $\statespace^d$ to be $d$ independent copies of $\statespace$, with annealing distributions $\pi_\beta^{(d)}(x^{(1)},\dots,x^{(d)})=\pi_\beta(x^{(1)})\cdots\pi_\beta(x^{(d)})$ and product forward/backward kernels on $\statespace^d$. This yields $\delta^{(d)}(\beta)=d\delta(\beta)$, hence $E^{(d)}(\varphi)=d E(\varphi)$ and the global barrier $\Lambda^{(d)}=\sqrt{d} \Lambda$. Therefore, the large barrier analysis extends directly to high-dimensional scaling by substituting $\sqrt{d}\Lambda$ for $\Lambda$.

One advantage of formulating the analysis in terms of $\Lambda$ rather than dimensionality is that $\Lambda$ requires no structural assumptions on the annealing distributions or state space, and applies to complex low-dimensional regimes where dimensionality provides no insight. See \cref{ap:high-dim-details} for details.

\section{Methods}\label{sec:methodology}
In this section, we use the theory developed in the previous section to construct novel ASMC methods. As described in \cref{sec:intro}, our goal is to design an adaptive ASMC algorithm that produces a sequence of increasingly precise normalising constant estimates at deterministic and predictable times.

\subsection{Optimised Annealed Sequential Monte Carlo}
The key tuning parameter in the algorithm is the annealing schedule. To approximate the optimal schedule, we use \cref{thm:dense_limit_variance}, which reduces the problem to estimating the local barrier $\lambda(\beta)$. We first show how to estimate the local barrier and update the schedule using the output of an ASMC sampler. We then obtain Optimised ASMC (OASMC) by iterating these subroutines in a round-based scheme. Each round runs an increasingly precise ASMC algorithm using a schedule computed from the output of the previous round. We note that this round-based structure is also compatible with adaptive tuning of MCMC forward moves; see \cref{ap:explorer} for details.

\cref{alg:OASMC} shows the overall structure of OASMC. Notice that each round has a predictable run time by construction and that an unbiased estimator of the normalising constant is available at the end of each round. 
Figure~\ref{fig:rockets-schedules} shows an example of the sequence of schedules $\mcB_{T_{k}}$ produced by this algorithm, including comparisons with the schedules computed by the method of \citet{Zhou2016}.

\begin{algorithm}[t]
\caption{\texttt{OASMC}: Optimised Annealed Sequential Monte Carlo}
\label{alg:OASMC}
\begin{algorithmic}
\State $N_1 \gets $number of available threads 
\State $\mcB_{T_1} = (0, 1)$ \Comment{Initialise with the trivial 1-step schedule with $T_1=1$}
\For{$k=1, 2, \dots$}
    \State $(\hat Z_k, \hat{\textbf{g}}_k) \gets \texttt{ASMC}(N_k, \mcB_{T_k})$
    \State $N_{k+1}, T_{k+1} \gets \texttt{Budget}(N_k, \hat{\textbf{g}}_k)$
    \State $\mcB_{T_{k+1}} \gets \texttt{OptimiseSchedule}(T_{k+1}, \mcB_{T_k}, \hat{\textbf{g}}_k)$
\EndFor
\end{algorithmic}
\end{algorithm}

\subsubsection{Updating the schedule}\label{sec:schedule-update}
We now consider the following problem. Assume we have access to the output of \cref{alg:AMCS} based on schedule $\mcB_T =\beta_{0:T}$. We wish to compute $\mcB^*_{T'}=\beta_{0:T'}^*$ of size $T'$ approximating the schedule generated by $\varphi^*=\Lambda^{-1}(\Lambda  u)$. 
For all $t\in[T]$ and $i=0,1,2$, we can approximate $\E_{\beta_{t-1},\beta_t}[(g_{\beta_{t-1},\beta_t})^i]$ using \cref{eq:estimator_forward_proposal} in terms of the incremental weights $g^n_t=g_{\beta_{t-1},\beta_t}(X^n_{t-1},X^n_{t^-})$ computed in \cref{alg:AMCS}. We can then substitute them in \cref{eq:incremental-discrepancy-def}  to obtain a consistent estimator $\hD_t$ for $D(\beta_{t-1},\beta_t)$ as $N\to\infty$, 
\[\label{eq:CESS}
\hD_t
=\log \hg_{t,2}-2\log \hg_{t,1}+ \log \hg_{t,0},\quad \hg_{t,i}=\sum_{n\in [N]} w^n_{t-1}(g^n_t)^i.
\]
Here the term $ \hg_{t,0}$ normalises the weights $w^n_{t-1}$. 
\cref{cor:dense_schedule_limit_discrepancy} shows that we can approximate $\Lambda(\beta_t)$ using $\hLambda_t=\sum_{s=1}^t\sqrt{\hD_s}$ with an $O(t/T^2)$ error.
We can approximate $\Lambda^{-1}(\cdot)$ using a monotone spline $\hLambda^{-1}(\cdot)$ with knots $\{(\hLambda_t,\beta_t)\}_{t=0}^T$. Finally by substituting in $\hLambda_T$ and $\hLambda^{-1}(\cdot)$ for $\Lambda$ and $\Lambda^{-1}(\cdot)$ respectively, we obtain an approximation for $\hat\varphi^*(u)=\hat{\Lambda}^{-1}(\hat{\Lambda}_Tu)$, which we can use to generate $\mcB^*_{T'}$ (see Algorithm~\ref{alg:schedule-update}). The full output of ASMC is not needed and can instead be summarised using the sufficient statistics $\hat{\textbf{g}} = \hg_{1:T,0:2}$.

\begin{algorithm}[t]
\caption{\texttt{OptimiseSchedule}$(T', \mcB_T, \hat{\textbf{g}})$: update an annealing schedule}
\label{alg:schedule-update}
\begin{algorithmic}
\Require Current schedule $\mcB_T = \beta_{0:T}$, incremental weight statistics $\hat{\textbf{g}}=\hat{g}_{1:T,0:2}$ output from $\texttt{ASMC}(N, \mcB_{T})$, requested schedule size $T'$.
\State $\hLambda_0 = 0$ \Comment{Initialise $\Lambda(0)=0$.}
\For{$t\in [T]$}
    \State $\hD_t = \log \hg_{t,2} - 2 \log \hg_{t,1} + \log \hg_{t,0}$  \Comment{Approximate incremental discrepancy}
    \State $\hLambda_t = \hLambda_{t-1} + \sqrt{\hD_t}$     \Comment{Approximate $\Lambda(\beta_t)$}
\EndFor
\State $\hat \Lambda^{-1}(\cdot) = \texttt{Interpolate}(\{(\hLambda_0, \beta_0), (\hLambda_1,\beta_1),\dots,(\hLambda_T, \beta_T)\})$  
\State $\hat\varphi^*(u)= \hat{\Lambda}^{-1}(\hat{\Lambda}_Tu)$\Comment{Approximate optimal generator \cref{eq:geodesic}.}
\For{$t'\in [T']$}
    \State $\beta^*_{t'} = \hat \varphi^*(u_{t'}),\quad u_{t'}=t'/T'$.\Comment{Generate new schedule of size $T'$}
    \EndFor
\State \textbf{Return:} $\schedule^*_{T'}=\beta^*_{0:T'}$
\end{algorithmic}
\end{algorithm}

\subsubsection{Optimised AIS (OAIS)}\label{sec:constant-memory}

In OASMC, resampling required all $N$ particles to be stored in memory. 
However, when no resampling is used (i.e., the AIS case), we can permute in \cref{alg:AMCS} the order of the loop over the $N$ particles and the loop over the ASMC sampling iterations. 
After this transformation, the only quantities that need to be kept in memory are (1) a single particle and (2) partial sums needed to compute $\hat{\textbf{g}}$. 
We denote this scheme by OAIS. 

Note that these memory savings are not possible using existing adaptation methods such as \citet{Zhou2016}. The reason is that forming these previous methods' objective function for the line searches over schedule increments requires the $N$ particle weights to be computed before advancing to the next iteration. Hence, the same ``swap of the loops'' cannot be performed even in the AIS case with these previous adaptive methods.

\subsubsection{Increasing the number of particles and ASMC iterations}\label{sec:budget}

At the end of each round, our OASMC and OAIS algorithms have to determine the number of particles $N$ and iterations $T$ to use in the next round. 
This is the function denoted \texttt{Budget} in Algorithm~\ref{alg:OASMC}. 
The \texttt{Budget} function should account for the following factors: (1) as formalised in Section~\ref{sec:criticality}, $T$ and $N$ should be increased so as to avoid the unstable regime, and (2) $N$ should be at least the number of available compute cores, since parallelisation is possible only in $N$ and not in $T$. It is also useful to have $N$ not too small to ensure the accuracy of the estimates $\hat{\textbf{g}}$ used to compute the annealing schedule. For OAIS, it is possible to increase $N$ arbitrarily using the constant memory algorithm described in Section~\ref{sec:constant-memory}. For OASMC, memory constraints eventually prevent further increases in $N$. Fortunately, even in this case, the overall algorithm is consistent for $Z$ since we can still increase $T$ without increasing the memory requirements. 

A simple heuristic that satisfies the factors (1) and (2) is to start $N$ at the number of cores available and to double the budget at each round, e.g., by setting $N_{k+1} \gets \sqrt{2} N_k$ and $T_{k+1} \gets \sqrt{2} T_k$, until e.g., memory is exhausted, at which point set $N_{k+1} \gets N_k$ and $T_{k+1} \gets 2  T_k$. Theorem~\ref{cor:dense_schedule_limit_discrepancy} shows that, for sufficiently large $T$, the variance scales inversely in $N \cdot T$ and hence optimising preferential allocation of $N$ versus $T$ is not a dominant factor to consider for the purpose of normalising constant estimation. 

\subsection{Connection to online conditional ESS optimisation}\label{sec:connection-online-cess}

\begin{figure}[t]
\centering
\includegraphics[width=\textwidth]{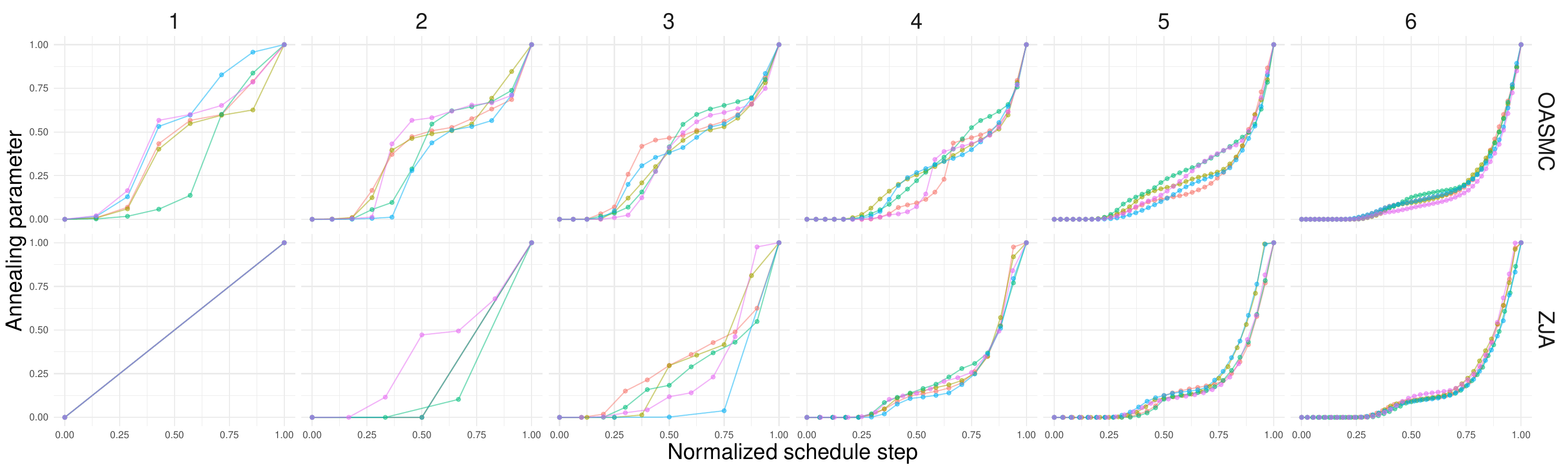}
\caption{\textbf{Top row:} Evolution of the schedule generators, plotting the annealing parameter $\beta_t$ as a function of the normalised schedule step $u_t=t/T$ over 6 rounds of OASMC (facet columns) with five random seeds (colours) on an ODE Bayesian parameter estimation problem for mRNA transfection data \citep{leonhardt_single-cell_2014}. \textbf{Bottom row:} method of \cite{Zhou2016} (labelled ZJA), each round here corresponds to a given budget, and we seek to compare that round's schedule to that produced by \citeauthor{Zhou2016} under the same budget. To achieve this, guided by our theoretical results we set the line search threshold in \citet{Zhou2016} using  $D(\beta_{t-1}^*, \beta_t^*) \approx \Lambda^2/T^2$, where $\Lambda \approx 16$ is estimated from a separate, large run (timing of that run not included for fairness of comparison).} 
\label{fig:rockets-schedules}
\centering 
\end{figure}

It is common to tune the schedule in an online fashion \citep{jasra_inference_2011,del_moral_adaptive_2012,Zhou2016}. 
To perform this tuning, \citet{Zhou2016} introduced the \emph{conditional ESS},
$\CESS_t=N \left(\hg_{t,1}\right)^2/(\hg_{t,2}\; \hg_{t,0}) \in[1,N]$, and proposed an algorithm that, at iteration $t-1$, selects the annealing parameter $\beta_t$ for iteration $t$ so that $\CESS$ remains constant across iterations. 
Since $\hat D_t = \log N - \log \CESS_t$ a.s.\ converges to $D(\beta_{t-1},\beta_t)$ as $N\to\infty$, we obtain that \citet{Zhou2016} and \cref{alg:schedule-update} optimise objective functions which are asymptotically equivalent in the limit of $T \to \infty$. 
As a result, our analysis provides theoretical justifications not only for our optimisation algorithm but also for the online method of \citet{Zhou2016}. 
An important practical advantage of our method is that its run time is deterministic and provides unbiased normalising constant estimates at predictable times. 
See \cref{fig:rockets-schedules} for an empirical comparison of the variance of $\hat Z$ computed by \citet{Zhou2016} and OASMC, and \cref{ap:numerical_schedules} for comparative plots of the schedules produced by the two methods and \cref{sec:gpus} for more details. 

\subsection{Computational duality with parallel tempering}\label{sec:theoretical-comparison-to-pt}
Suppose the annealing distributions $\pi_\beta\propto \eta^{1-\beta}\pi^\beta$, and $M_{\beta,\beta'}=K_{\beta'}$ is a $\pi_{\beta'}$-invariant Markov kernel corresponding to an MCMC algorithm. An alternative to ASMC to estimate the normalising constant $Z$ and expectations is the Parallel Tempering (PT) algorithm \citep{geyer1991markov,hukushima1996exchange}. Given an annealing schedule $\mcB_{N'}=\beta_{0:N'}$ of size $N'+1$, PT generates a Markov chain $\textbf{Y}_t=(Y^0_t,\dots,Y^{N'}_t)\in\mcX^{N'+1}$, jointly targeting the annealing distributions $\pi_{\beta_0}\otimes\cdots\otimes\pi_{\beta_{N'}}$. At iteration $t$, PT updates each chain in parallel using $K_{\beta_n}$ and then performs Metropolised swaps. Given the output of $T'$ iterations of PT, $\textbf{Y}_{1:T'}$, we can estimate $Z$ using the stepping stone estimator $\hat{Z}_\PT$, \citep{xie_improving_2011}, and $\E_{\pi}[f]$ using the Monte Carlo estimator $\hpi_\PT[f]$,
\[\label{eq:Z_PT}
\hZ_\PT = \prod_{n\in [N']}\frac{1}{T'}\sum_{t\in [ T']}\frac{\gamma_{\beta_{n}}(Y^{n-1}_t)}{\gamma_{\beta_{n-1}}(Y^{n-1}_t)},
\qquad 
\hpi_\PT[f] = \frac{1}{T'}\sum_{t\in[T']}f(Y^{N'}_t).
\]

Compare \cref{eq:Z_PT} to $\hat{Z}_\SIR$ and $\hpi_\SIR[f]$ output by \cref{alg:AMCS} using the annealing schedule $\mcB_T=\beta_{0:T}$ with SIR resampling at every iteration,
\[\label{eq:Z_SMC}
\hZ_\SIR = \prod_{t\in [T]}\frac{1}{N}\sum_{n\in [ N]}\frac{\gamma_{\beta_{t}}(X^{n}_{t-1})}{\gamma_{\beta_{t-1}}(X^{n}_{t-1})},
\qquad 
\hpi_\SIR[f] = \frac{1}{N}\sum_{n\in[N]}f(X^{n}_T).
\]
Note that the estimators in \cref{eq:Z_PT} and \cref{eq:Z_SMC} output by ASMC and PT have the
same functional form, provided the number of PT particles coincides with the number of 
ASMC iterations, $N' = T$, and the number of PT iterations, with the number of ASMC
particles, $T' = N$. The roles of parallelism and time are, therefore, interchanged when
going between PT and SIR. 

If, in addition, we assume the ``efficient local exploration'' assumptions of PT \citep[Section 3.3]{syed2019non}, and their SIR counterparts (\cref{sec:assumptions}), we have that $\hZ_\PT$ and $ \hZ_\SIR$ have the same law. This implies the theory and method developed to analyse the relative variance of $\hZ_\SIR$ can be applied to analyse the relative variance of $\hZ_\PT$ by simply interchanging the number of annealing distributions and number of particles. In particular, the computational duality between PT and SIR implies that the schedule to optimise the variance of $\hZ_\SIR$ also minimises the variance of $\hZ_\PT$.

Typically, NRPT is tuned not to minimise the variance of the normalising constant estimator, but rather to maximise a PT-specific notion of exchange efficiency called the round trip rate \citep{katzgraber2006feedback,lingenheil2009efficiency}. To maximise the round trip rate, 
based on simplifying assumptions, \citet{syed2019non} recommends using $N'=O(\Lambda_\PT)$ distributions and an annealing schedule generated by $\varphi^*(u)=\Lambda^{-1}(\Lambda u)$ but substituting $\lambda(\beta)$ and $\Lambda$ with $\lambda_\PT$ and $\Lambda_\PT$, the local and global barriers for PT,
\[
\lambda_\PT(\beta)=\frac{1}{2}\E|V_\beta-V_\beta'|,\quad \Lambda_\PT = \int_0^1 \lambda_\PT(\beta)\dee\beta,
\]
where $V_\beta=V(Y_\beta)$, $V'_\beta = V(Y'_\beta)$, with $V(y)=\log\pi(y)-\log \eta(y)$
and  
$Y_\beta, Y'_\beta \iidsim \pi_\beta$. We show in \Cref{sec:PT-details} that $\lambda_\PT(\beta)$ and $\Lambda_\PT$ scale linearly with $\lambda(\beta)$ and $\Lambda$ respectively, and are proportional in the high dimensional scaling limit from \cref{sec:high-d-scaling} (e.g., see \cref{fig:gpus} (right)). Therefore, the annealing schedule optimising the round trip rate is comparable to the annealing schedule optimising the variance of the normalising constant estimator, with equivalence in the high dimensional scaling limit. This motivated the use of the ``Parallel PT'' variant of NRPT introduced in \citet[Section 5.5]{syed2019non}, which maximises the round trips after the tuning rounds by running $K$-copies of PT in parallel with $N^*\approx 2\Lambda_\PT$ chains each as opposed to one copy of PT with $N=KN^*$ chains. This ensures that the recommendations based on the variance of $\hat Z_\PT$ and those based on round-trip rate align.

\subsection{Memory scaling analysis}\label{sec:memory-analysis}

As formalised in \cref{sec:criticality}, under our simplifying assumptions, both OASMC with adaptive resampling and PT require $T=T'\sim\Lambda$ iterations and $N=N'=O(\Lambda)$ particles to obtain stable normalising constant estimates. In contrast, OAIS (see \cref{sec:constant-memory}) requires $T\sim \Lambda^2$ with $N=O(1)$ particles. Despite the increased run-time for OAIS, it can still be an attractive alternative to OASMC and PT to estimate $Z$ when memory is constrained. For example, consider a weakly dependent $d$-dimensional target (formalised in Section~\ref{sec:high-d-scaling}), with $\Lambda = O(\sqrt{d})$ as $d\to\infty$. The particle interactions from the resampling moves in OASMC and swap NRPT, respectively, require storing $N=N'=O(\sqrt{d})$ $d$-dimensional particles in the stable regime, with a total memory cost of $O(d^{3/2})$. In contrast, OAIS only requires storing the adaptation statistics (namely, $\hat{\textbf{g}}$ consisting of three vectors of length $T \sim \Lambda^2=O(d)$) which is the same as the memory required to store a single particle. Hence, OAIS overall memory requirement is $O(d)$ for $N$ particles.

\subsection{Choosing an annealing algorithm}\label{sec:choosing}

\begin{figure}[t]
\includegraphics[width=1\textwidth]{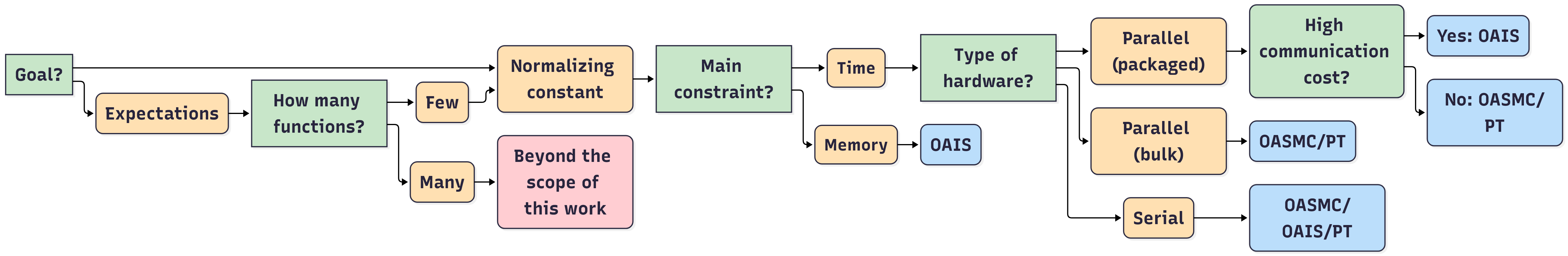}
\caption{A flowchart recommending an annealing algorithm based on goals and various information on the computing architecture. See Sections~\ref{sec:choosing} and \ref{sec:expectations}.}
\label{fig:flowchart}
\end{figure}

We have now described and analysed three algorithms: OASMC, OAIS, and PT.
An important practical question is which algorithm to use in a given context.
The choice depends on the computational goal: are we estimating the normalising constant or computing expectations (or both)? We focus first on the former and then provide comments on the latter in \cref{sec:expectations} (see also \cref{fig:flowchart} for a decision diagram).

The next consideration is memory cost. In certain scenarios such as phylogenetic inference (see, e.g., \citet{Jun2014Memory}), each particle takes a significant fraction of the memory available. In such a setting, as discussed in \cref{sec:memory-analysis}, OAIS may potentially be the only viable option. 
If memory is not constrained, then the key factor driving the choice of annealing algorithm is the type of computing hardware. 
We distinguish three situations: (1) serial or limited parallelism (e.g., a laptop's CPUs); (2) parallelism ``in bulk'' (e.g., MPI clusters), where cost scales linearly with the number of cores; and (3) ``packaged'' parallelism (e.g., a workstation's GPU), where a block of parallel cores is available and the user would like to fully utilise these cores. 

\paragraph{(1) Serial:}
For all three algorithms, our theory predicts that OASMC, OAIS and PT (with their respective optimal tuning) will perform similarly. Indeed, in the serial regime, the total run time is given by $TN$, so that OASMC and PT can use $T = N = \Theta(\Lambda)$ and provide a similar accuracy compared to OAIS with $T = \Theta(\Lambda^2)$ and $N = O(1)$. 

\paragraph{(2) Parallel (in bulk):}
Since the algorithms presented in this work parallelise over particles (in $N$) but not over iterations (in $T$) (and the other way around for PT), OAIS will be slower than OASMC and an optimised PT.

\paragraph{(3) Parallel (packaged):}
This is the main case where it may be useful to run algorithms in the unstable regime. Recall that the number of particles needed to achieve high accuracy is exponential in the unstable regime, but in a GPU context, the number of cores might be sufficiently large to support such a large number of particles. In such case, OAIS might be competitive since it avoids communication between cores (see \cref{sec:gpus} for an example). 

\subsection{Computing expectations}\label{sec:expectations}

So far, we have focussed on normalising constants. If the user is interested in computing expectations---either in addition to, or instead of, estimating $Z$---to what extent are our methods relevant? The answer depends on several factors: the computing architecture, the nature of the target, if we are computing expectations under one or several test functions, and whether we are interested in only computing expectations, or both expectations and $Z$. We discuss some specific examples here. 

If the target is regular (including, e.g., log-concave), and we are only interested in estimating expectations under the target (and not $Z$), and using a serial computer, \emph{and} many test functions are needed, then it is likely preferable to use straightforward MCMC. 

In the other cases (parallel architectures, or a single test function, or irregular targets), SMC methods (including OASMC) can be an attractive option. 
First, if the user is interested in a single or a few test functions, then it is possible to cast the problem of computing an expectation into normalising constant estimation; see \citet{rainforth_targetaware_2020}. This reduction is particularly useful when the test function puts emphasis on low mass regions of the posterior distribution (e.g., rare events). In such cases, all the theory studied in this paper applies directly.

Second, we consider the problem of computing expectations as a ``by-product'': if the user is interested in both $Z$ and expectations of test functions, it is useful to obtain expectations as a side product of (OA)SMC. The standard approach for doing this is to use the particles at the last SMC iteration (in the case of OASMC, the particles at the last SMC iteration of each round). A limitation of this standard approach is that only a small fraction of the $TN$ particles are used to compute the expectations. Fortunately, a simple workaround is possible: after the last SMC iteration, resample the final particle population and simulate $N$ parallel MCMC chains for $T$ steps, each initialised at one of the resampled final particles. The $TN$ final samples can then be used for estimating expectations. SMC acts as a parallelised burn-in strategy in that context. By design this ``final iteration rejuvenation'' only increases the computational cost by a factor two compared to standard ASMC. In the case of OASMC, we perform final iteration rejuvenation at the end of each round, again incurring only a computational cost increase by a factor of two compared to basic OASMC. See \cref{ap:rejuvenation} for empirical results. 

Finally, computing expectations under irregular targets (e.g., multi-modal distributions) warrants a cautionary note. On the one hand, annealing is often successful at approximating irregular targets (for example, many of the numerical examples explored in Section~\ref{sec:numerical-experiments} are multi-modal \citep[Appendix I.5]{syed2019non}). However, we want to highlight a limitation of our work which is especially prominent when estimating expectations under an irregular target. Given the simplifications imposed by our performance model, we do not take into account that repeated resampling may kill the diversity of certain test function evaluations, e.g., when test functions are sensitive to multi-modality. As a result, it may be advantageous to use our OAIS algorithm rather than OASMC  to estimate expectations under irregular targets. The cost is a potentially larger $T$ to obtain stable particle weights ($\Lambda^2$ instead of $\Lambda$), but the benefit is that the multi-modality might be better preserved. We leave the investigation of this to future work.

\section{Numerical experiments}\label{sec:numerical-experiments}

\begin{figure}[t]
\centering
\includegraphics[width=0.95\textwidth]{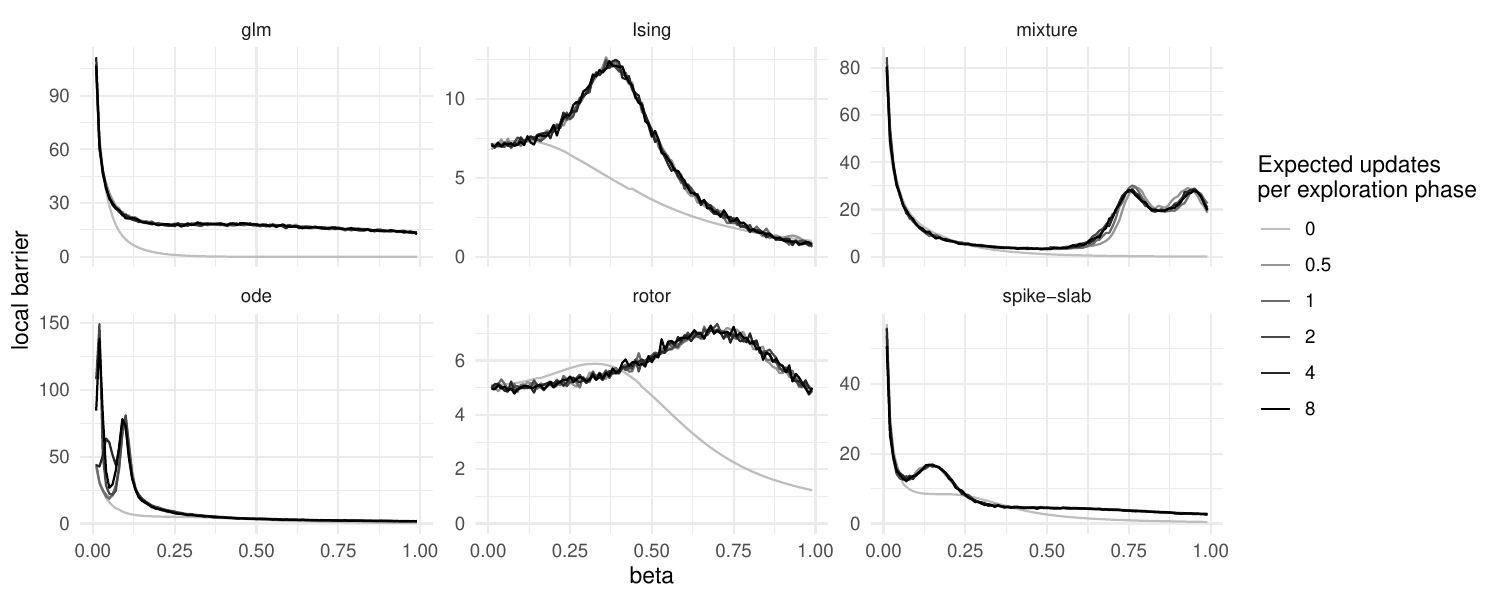}
\caption{Estimates of the local barrier $\lambda$ for six inference problems (facets) obtained with OASMC for 20 rounds. Each curve is obtained by differentiating the cubic spline  computed by Algorithm~\ref{alg:schedule-update} in the last round. The different lines denote an increasing expected number of updates per component of the target during the propagation step of \cref{alg:AMCS}. For example, the curve with label $0.5$ means that only a random half of the variables are updated by $M_{\beta, \beta'}$.}
\label{fig:vary-exploration}
\centering
\end{figure}

\begin{figure}[t]
        \centering
	\includegraphics[width=\textwidth]{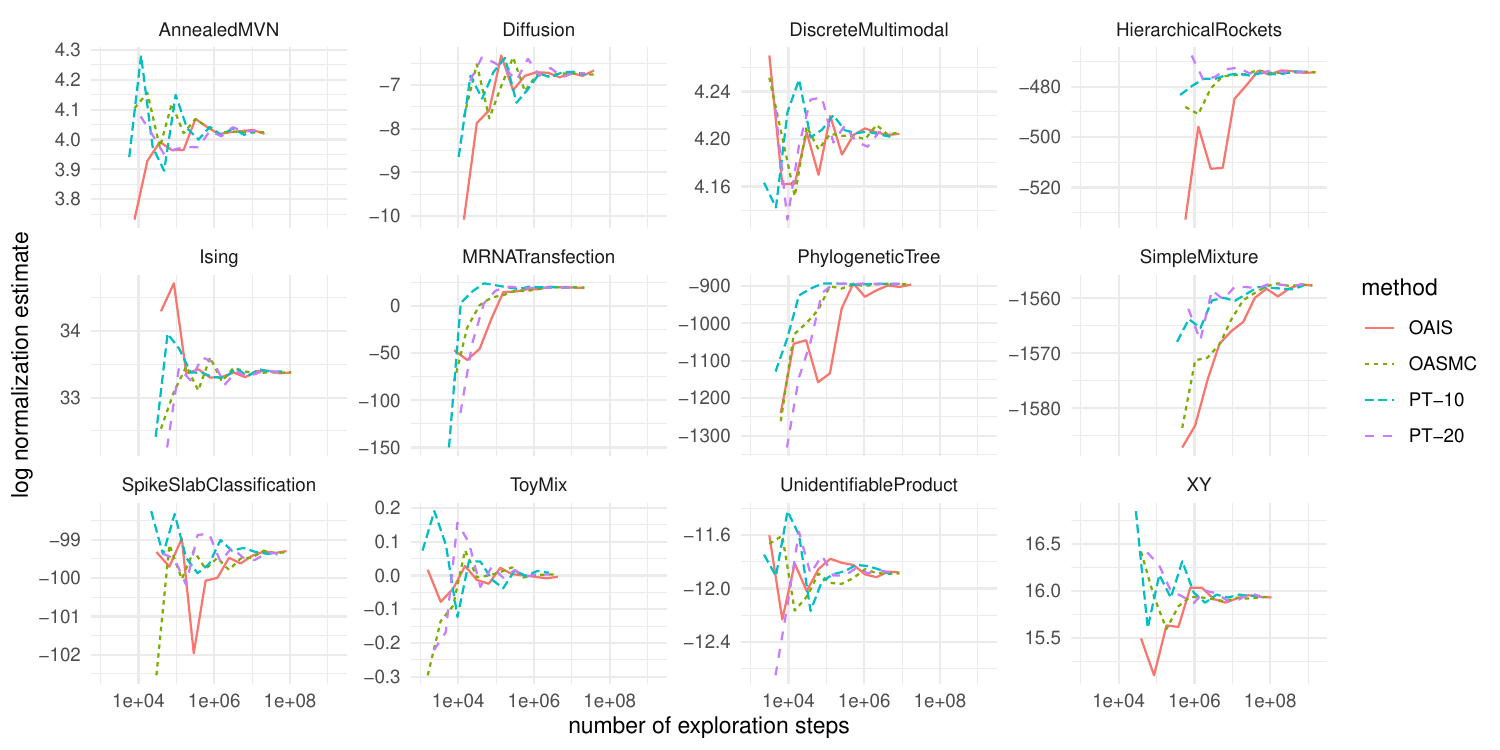}
	\caption{Log normalising constant estimates as a function of the number of MCMC steps (used in the propagation step of OASMC/OAIS and exploration step of PT), a proxy for the CPU computational cost. PT-$i$ refers to Non-Reversible PT with $i$ chains. Each facet represents a distinct inference problem. }
 \label{fig:logz-estimates}
\end{figure}

\begin{figure}[t]
\centering
\includegraphics[height=7.0cm]{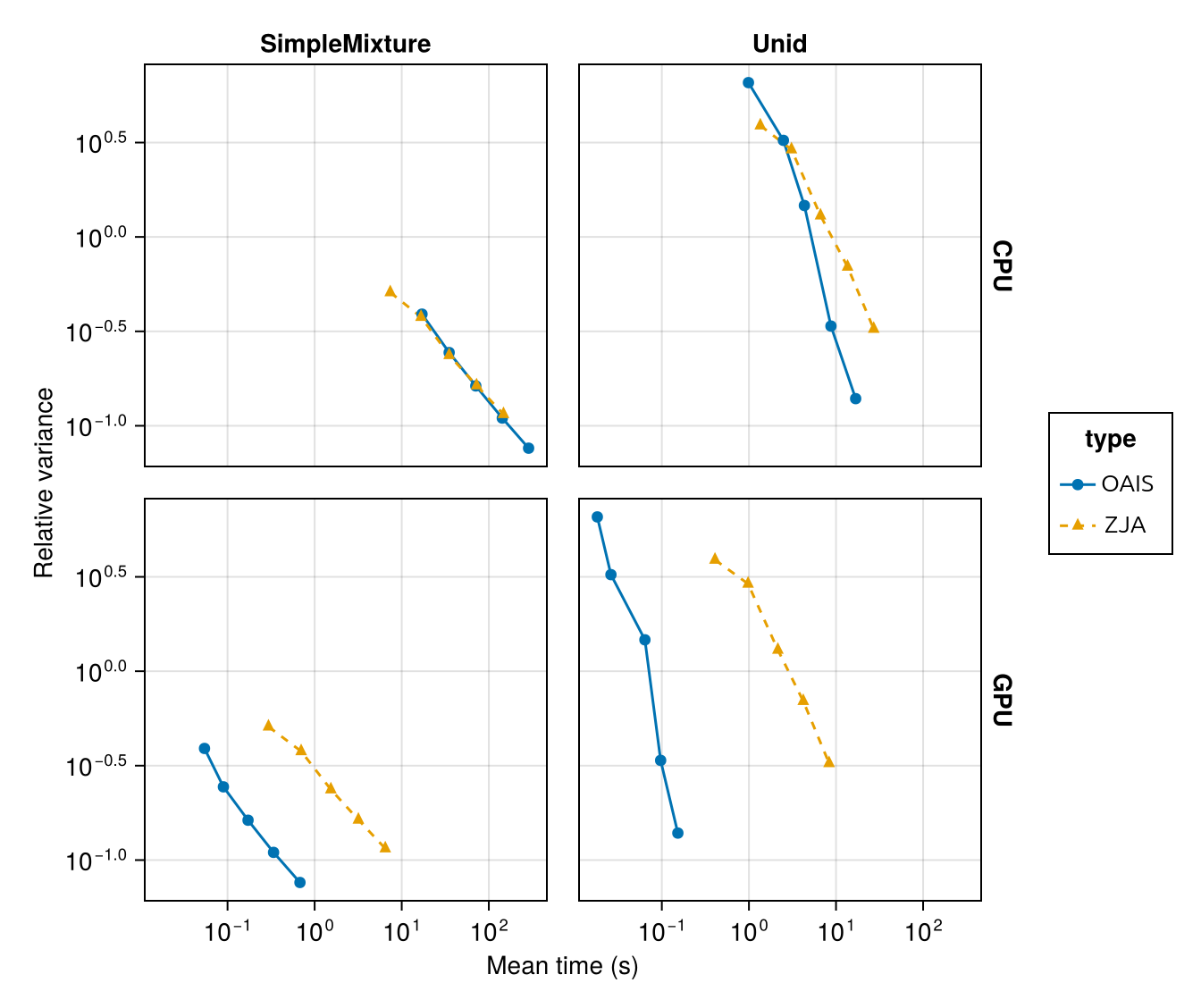}
\includegraphics[height=7.0cm]{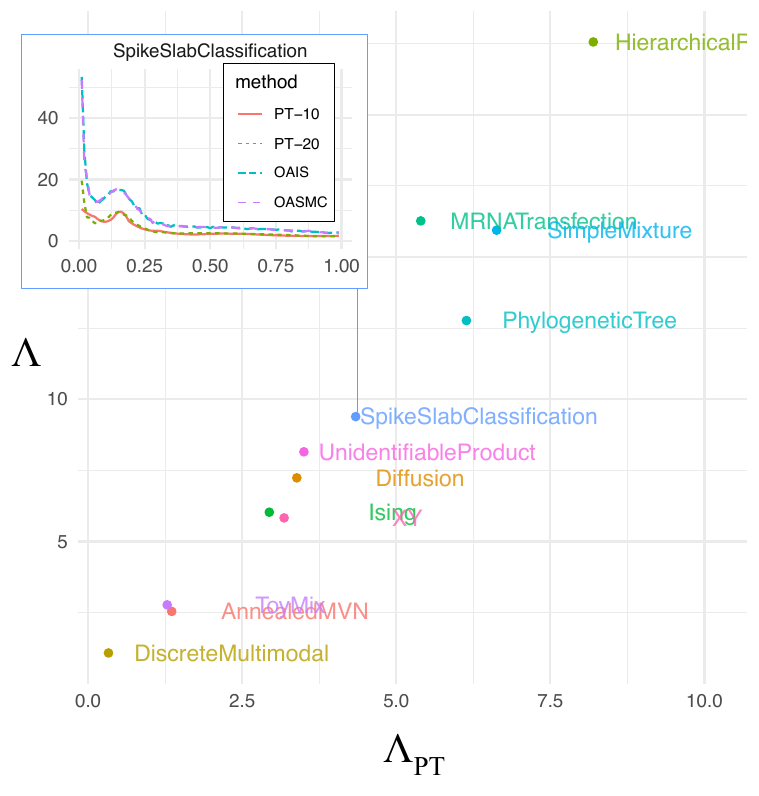}
\caption{{\bf Left}: comparative performance of our OAIS algorithm and of the previous state-of-the-art adaptive method of \cite{Zhou2016} (labelled ZJA). Each dot is a replication of 1000 independent executions with different random seeds. The number of particles $N$ is held fixed to $2^{14}$ while the number of annealing distributions $T$ increases. Each dot's $x$ coordinate is the mean wall clock time averaged over the 1000 independent seeds, and its $y$-axis, the sample variance of $\hat Z$ relative to $Z$, also over the 1000 seeds (both $x$ and $y$ axes are in log-scale). Facet rows are GPU/CPU backends, and facet columns, results on two representative models. {\bf Right}: comparison of the PT global communication barrier $\Lambda_\text{PT}$ ($x$-axis) and AIS/SMC's global barrier $\Lambda$ for 12 different models. Linear regression with intercept set to zero gives a coefficient of $0.4$. Inset: example of local barriers for one of the models (plots for the other barriers can 
be found in \cref{ap:numerical}).}
\label{fig:gpus}
\centering
\end{figure}

\subsection{Robustness to violation of performance model assumptions}\label{sec:inefficient-explorers}

OASMC is based on performance model assumptions, in particular on \cref{assump:ELE}, which we do not expect to describe the details of the inefficiency of real-world samplers. However, we can approximate \cref{assump:ELE} by increasing the quality of the forward kernel. This section investigates how this gap between theory and implementation affects the estimation of the local barrier, the key quantity used in OASMC.

\Cref{fig:vary-exploration} shows the estimates of the local barriers for a variety of inference problems obtained using \cref{alg:OASMC} with slice-within-Gibbs sampler MCMC kernels \citep{neal2003slice}. For each target, we consider a sequence of experiments where the forward kernel progressively performs more and more variable updates. The local barrier is well approximated, provided that the forward kernel is non-degenerate (the line with label $0$ corresponds to $M_{\beta, \beta'}$ being the identity, i.e., where the only movement comes from sampling at $\eta$). Hence, the schedules obtained in \cref{alg:schedule-update} are expected to be close approximations to the optimal schedule, even when our performance model assumptions are violated.

\subsection{Empirical comparison of annealing algorithms}\label{sec:empirical-comparison}

In this section, we empirically compare the normalising constant estimates obtained by OASMC and OAIS methods with those obtained from the Non-Reversible PT algorithm (NRPT) \citep{syed2019non} combined with the stepping stone method (labelled PT for short in the plots). See also \cref{ap:wfsmc} for additional results and connections with Waste-Free SMC \citep{dau2022waste}. 
To make run times comparable, we use the same MCMC exploration kernels  for both NRPT and PT. Specifically, we measure time complexity in terms of the total number of MCMC exploration steps taken. 
Whereas OASMC and OAIS have no free tuning parameters, NRPT requires the user to specify the number of chains. To illustrate the effect of this free parameter, we ran NRPT with 10 and 20 chains based on pilot runs and the recommendation of \cite{syed2019non} to use a number of chains on the order of the PT global communication barrier.   

Figure~\ref{fig:logz-estimates} shows the normalising constant estimates as a function of the total number of exploration steps. As expected, all methods converge to the same value with a sufficiently large time budget. In the 12 problems we considered, the performance of the two algorithms is similar when running on a CPU, with a slight advantage to properly tuned NRPT in some of the examples. 
However, the OASMC and OAIS methods still offer three practical advantages over NRPT even in the CPU context, namely, they do not require tuning the number of schedule discretisation points, they provide an unbiasedness guarantee on the normalising constant estimator, and they can achieve lower memory cost. 

Next, we show in Figure~\ref{fig:gpus} (right) and \cref{ap:numerical} Figure~\ref{fig:barriers} estimates of the global and local barriers for both OASMC and NRPT, $\lambda(\beta), \Lambda$ and $\lambda_\text{PT}(\beta), \Lambda_\text{PT}$ respectively. 
For both OASMC and NRPT, the cumulative barrier is fitted using a monotone cubic spline, and automatic
differentiation is applied to the resulting piecewise polynomial function to obtain an estimate of 
the local barriers. 
As predicted in Section~\ref{sec:theoretical-comparison-to-pt}, the two local barriers are in 
close agreement, with the PT barrier lower than the OASMC barrier by a small multiplicative constant. 

\subsection{High-performance GPU implementation of adaptive schemes}\label{sec:gpus}

The numerical results so far have focused on CPU-based implementations. 
We now turn to GPU-based implementations, a context where our OAIS method 
is particularly attractive. We show that a GPU implementation of OAIS has a runtime up to 100 times faster than a GPU implementation of the previous state-of-the-art adaptive schedule method, 
\cite{Zhou2016} (abbreviated ZJA in the following). Additional information on the GPU implementation can be found in \cref{ap:GPU-details}.

We hypothesised that OAIS would be advantageous compared to ZJA applied to AIS in the GPU context, since OAIS requires markedly less communication between particles/threads compared 
to ZJA (once per round rather than once per iteration). 
To test this hypothesis, we set up an experiment with five factors:
(1) GPU vs.\ CPU backends, (2) OAIS vs.\ ZJA applied to AIS, (3) two representative Bayesian models 
(Bayesian inference over mixture model parameters, `SimpleMixture'; and 
a model with unidentifiable likelihood creating concentration on a sub-manifold, `Unid'; 
see \cref{ap:simulations} for more details), (4) a sequence of 5 levels of sampling effort 
(for OAIS, obtained by increasing the number of rounds, starting at 5 rounds for `Unid' and 
3 for `SimpleMixture', for ZJA, the relationship $D(\beta^*_{t-1}, \beta^*_t) \approx \Lambda^2 / T^2$ from \cref{sec:dense-limit-variance} was used to obtain matching 
conditional ESS thresholds), (5) 1,000 random seeds. 
All experiments used $2^{14}$ particles (selected to reach full GPU utilisation) and  Float64 precision. 
For each of the 40,000 factorial treatments, we measured the wall clock time and $\hat{Z}$.  

The results, summarised in \cref{fig:gpus}, support our hypothesis: on the GPU backend, to obtain a given level of accuracy 
on the normalising constant (horizontal line, measured using the variance since the estimator is unbiased), OAIS is always at least 10 times faster than ZJA, and up to 100 times faster on the `Unid' model. This contrasts to the situation on the CPU backend, 
where the performance of OAIS and ZJA are nearly identical in the `SimpleMixture' model and OAIS has 
a modest advantage over ZJA on the `Unid' model.

\bibliographystyle{rss}
\bibliography{sources}
\clearpage

\begin{appendix}

\section{Supplement for \cref{sec:background}}
\subsection{Feynman-Kac operator for ASMC}

Consider a sequence of target distributions $\pi_t$ (corresponding to $\pi_{\beta_t}$) on a measurable space $\mathcal{X}$, with normalising constants $Z_t$ (corresponding to $Z_{\beta_t}$). We aim to estimate $Z=Z_T$, assuming $Z_0=1$. The ASMC uses forward kernels $M_t$ (corresponding to $M_{\beta_{t-1},\beta_{t}}$) and unnormalised incremental weight functions $g_t$ (corresponding to $g_{\beta_{t-1},\beta_t}$).
The evolution of the unnormalized measures $\gamma_t = Z_t \pi_t$ is governed by the linear Feynman-Kac (FK) operator $\Psi_t$:
\begin{equation}
\Psi_t(\mu)[f] = \int_{\mathcal{X}} \mu(\dee x_{t-1}) \int_\mathcal{X} M_t(x_{t-1}, \dee x_t) g_t(x_{t-1}, x_t) f(x_t).
\end{equation}
This operator satisfies $\Psi_t(\gamma_{t-1})[f] = \gamma_t[f]$, where $\gamma_t[f]=\int_\mathcal{X}f(x_t)\gamma_t(\dee x_t)$ is the expectation of $f$ with respect to $\gamma_t$.

We use $N$ particles $\{X_t^n, w_t^n\}_{n=1}^N$. We define the filtration capturing the history:
\begin{itemize}
    \item $\mathcal{F}_{t-1}$: History up to the end of iteration $t-1$.
    \item $\mathcal{F}_{t^-}$: History after the Mutation (Propagation/Reweight) step at $t$, but before the Selection (Resample) step.
\end{itemize}

Resampling occurs at predictable stopping times $0 = \tau_0 < \dots < \tau_R = T$. The decision to resample at time $t$, $R_t = \mathbb{I}(t \in \{\tau_r\})$, depends on the state just before resampling. Therefore, $R_t$ is $\mathcal{F}_{t^-}$-measurable. When invoked, the resampling scheme must be unbiased: the expected number of offspring of particle $n$ must equal $N W_{t^-}^n$ conditional on $\mathcal{F}_{t^-}$ (where $W_{t^-}^n$ are normalised weights).

Let $\tau(t)$ be the time of the most recent resampling up to time $t$. The particle approximation of the unnormalized measure is:
\begin{equation}
\hat{\gamma}_t[f] = \hat{Z}_{\tau(t)} \cdot \left( \frac{1}{N} \sum_{n=1}^N w_t^n f(X_t^n) \right),
\end{equation}
where $\hat{Z}_{\tau(t)}$ is the accumulated normalising constant estimator up to $\tau(t)$; i.e. the product over resampling times $r\leq \tau(t)$ of the empirical average incremental weights as in \eqref{eq:estimator_Z}. The final estimator is $\hat{Z} = \hat{\gamma}_T[1]$.

\subsection{Proof of \cref{prop:estimators}}
We prove by induction that $\mathbb{E}[\hat{\gamma}_t[f]] = \gamma_t[f]$ for any bounded measurable function $f$. 

\paragraph{Base Case:}
At $t=0$, we have  $X_0^n \sim \pi_0$ i.i.d., $\hat{Z}_0=1, w_0^n=1$ so 
\begin{equation}
\mathbb{E}[\hat{\gamma}_0[f]] = \mathbb{E}\left[ \frac{1}{N} \sum_{n=1}^N f(X_0^n) \right] = \pi_0[f] = \gamma_0[f].
\end{equation}

\paragraph{Inductive Step:}
Assume $\mathbb{E}[\hat{\gamma}_{t-1}[f]] = \gamma_{t-1}[f]$. We analyse the transition from time $t-1$ to $t$.

\paragraph{Step 1: Mutation (Propagation and Reweighting)}

We analyse the intermediate measure $\hat{\gamma}_{t^-}$. Note that $\tau(t^-) = \tau(t-1)$.
\begin{equation}
\hat{\gamma}_{t^-}[f] = \hat{Z}_{\tau(t-1)} \frac{1}{N} \sum_{n=1}^N w_{t-1}^n g_t(X_{t-1}^n, X_{t^-}^n) f(X_{t^-}^n).
\end{equation}
We take the expectation conditioned on $\mathcal{F}_{t-1}$ (over the randomness of $M_t$).
\begin{align}
\mathbb{E}[\hat{\gamma}_{t^-}[f] | \mathcal{F}_{t-1}] &= \hat{Z}_{\tau(t-1)} \frac{1}{N} \sum_{n=1}^N w_{t-1}^n \int M_t(X_{t-1}^n, dx_t) g_t(X_{t-1}^n, x_t) f(x_t).
\end{align}
Recognizing the definition of the FK operator $\Psi_t$ applied to $\hat{\gamma}_{t-1}$:
\begin{equation}
\mathbb{E}[\hat{\gamma}_{t^-}[f] | \mathcal{F}_{t-1}] = \Psi_t(\hat{\gamma}_{t-1})[f]. \label{eq:mutation_unbiased}
\end{equation}

\paragraph{Step 2: Selection (Adaptive Resampling)}

We analyse the transition from $\hat{\gamma}_{t^-}$ to $\hat{\gamma}_t$. We condition on the filtration $\mathcal{F}_{t^-}$.
\begin{equation}
\hat{\gamma}_t[f] = R_t \cdot \hat{\gamma}_t^{RS}[f] + (1-R_t) \cdot \hat{\gamma}_t^{Id}[f].
\end{equation}
Since $R_t$ is $\mathcal{F}_{t^-}$-measurable, it is treated as a constant within the conditional expectation:
\begin{equation}
\mathbb{E}[\hat{\gamma}_t(f) | \mathcal{F}_{t^-}] = R_t \cdot \mathbb{E}[\hat{\gamma}_t^{RS}[f] | \mathcal{F}_{t^-}] + (1-R_t) \cdot \mathbb{E}[\hat{\gamma}_t^{Id}[f] | \mathcal{F}_{t^-}].
\end{equation}

\emph{Case 2a: Identity Operation ($R_t=0$)}
$\hat{\gamma}_t^{Id} = \hat{\gamma}_{t^-}$. Thus, $\mathbb{E}[\hat{\gamma}_t^{Id}[f] | \mathcal{F}_{t^-}] = \hat{\gamma}_{t^-}[f]$.

\emph{Case 2b: Resampling Operation ($R_t=1$)}
The estimator is updated: $\hat{Z}_t = \hat{\gamma}_{t^-}(1)$. New particles $X_t^n$ are drawn from the normalized empirical distribution $\hat{\pi}_{t^-} = \hat{\gamma}_{t^-} / \hat{\gamma}_{t^-}[1]$ using a resampling scheme
\begin{equation}
\hat{\gamma}_t^{RS}[f] = \hat{Z}_t \cdot \left( \frac{1}{N} \sum_{n=1}^N f(X_t^n) \right).
\end{equation}
The property of unbiased resampling ensures
\begin{equation}
\mathbb{E}\left[ \frac{1}{N} \sum_{n=1}^N f(X_t^n) | \mathcal{F}_{t^-} \right] = \hat{\pi}_{t^-}[f].
\end{equation}
Therefore:
\begin{align}
\mathbb{E}[\hat{\gamma}_t^{RS}[f] | \mathcal{F}_{t^-}] &= \hat{Z}_t \cdot \hat{\pi}_{t^-}[f] = \hat{\gamma}_{t^-}[1] \cdot \frac{\hat{\gamma}_{t^-}[f]}{\hat{\gamma}_{t^-}[1]} = \hat{\gamma}_{t^-}[f].
\end{align}

\textit{Combining Adaptively:}
Substituting the results from both cases:
\begin{equation}
\mathbb{E}[\hat{\gamma}_t[f] | \mathcal{F}_{t^-}] = R_t \cdot \hat{\gamma}_{t^-}[f] + (1-R_t) \cdot \hat{\gamma}_{t^-}[f] = \hat{\gamma}_{t^-}[f]. \label{eq:selection_unbiased}
\end{equation}
The adaptive selection step preserves the expectation of the measure.

\paragraph{Step 3: Completing the Induction}
We use the tower property of expectations.
\begin{align}
\mathbb{E}[\hat{\gamma}_t[f]] &= \mathbb{E}[ \mathbb{E}[\hat{\gamma}_t[f] | \mathcal{F}_{t^-}] ] \\
&= \mathbb{E}[\hat{\gamma}_{t^-}[f]] \quad (\text{from Eq. \ref{eq:selection_unbiased}}) \\
&= \mathbb{E}[ \mathbb{E}[\hat{\gamma}_{t^-}[f] | \mathcal{F}_{t-1}] ] \\
&= \mathbb{E}[\Psi_t(\hat{\gamma}_{t-1})[f]] \quad (\text{from Eq. \ref{eq:mutation_unbiased}}).
\end{align}
By the linearity of $\Psi_t$ and the inductive hypothesis ($\mathbb{E}[\hat{\gamma}_{t-1}] = \gamma_{t-1}$):
\begin{equation}
\mathbb{E}[\hat{\gamma}_t[f]] = \Psi_t(\mathbb{E}[\hat{\gamma}_{t-1}])[f] = \Psi_t(\gamma_{t-1})[f] = \gamma_t[f].
\end{equation}

By induction, $\mathbb{E}[\hat{\gamma}_T[f]] = \gamma_T[f]$. Setting $f=1$, we obtain $\mathbb{E}[\hat{Z}] = Z$.

\section{Supplement for \cref{sec:variance-finite-schedule}}
\subsection{Proof of \cref{thm:variance_ERS}}

\benum

\item [(a)]
We begin by expressing the relative normalising constant estimator $\hZ/Z$ in terms of the relative incremental weights $G^n_t = (Z_{\beta_{t-1}}/Z_{\beta_{t}}) g^n_t$:
\[
\frac{\hZ}{Z}
= \prod_{r\in[R]} \frac{1}{N} \sum_{n\in[N]} \tilde{w}^n_{\tau_r^-},
\qquad \text{where} \quad
\tilde{w}^n_{\tau_r^-} = G^n_{\tau_{r-1}+1} \cdots G^n_{\tau_r}.
\]
By \cref{assump:integrability,assump:ELE}, $G^n_t$ has finite second moments with $\E[G^n_t]=1$ and
\[
\var[G^n_t] = \E[(G^n_t)^2] - 1 = \exp(D(\beta_{t-1},\beta_t)) - 1.
\]
By \cref{assump:independent_weights_time}, $G^n_{\tau_{r-1}+1}, \ldots, G^n_{\tau_r}$ are independent, so $\tilde{w}^n_{\tau_r^-}$ has mean
\[
\E[\tilde{w}^n_{\tau_r^-}] = \E[G^n_{\tau_{r-1}+1}] \cdots \E[G^n_{\tau_r}] = 1
\]
and variance
\begin{align}
\var[\tilde{w}^n_{\tau_r^-}]
&= \E[(G^n_{\tau_{r-1}+1} \cdots G^n_{\tau_r})^2] - 1 \\
&= \E[(G^n_{\tau_{r-1}+1})^2] \cdots \E[(G^n_{\tau_r})^2] - 1 \\
&= \exp(D(\beta_{\tau_{r-1}}, \beta_{\tau_{r-1}+1})) \cdots \exp(D(\beta_{\tau_{r}-1}, \beta_{\tau_{r}})) - 1 \\
&= \exp(D(\mcB_T, \tau_{r-1}, \tau_r)) - 1.
\end{align}
By \cref{assump:independent_weights_particles}, $\tilde{w}^n_{\tau_{r}^-}$ are independent across $n \in [N]$, so
\[
\E\left[\frac{1}{N}\sum_{n\in[N]} \tilde{w}^n_{\tau_r^-}\right] = 1,
\qquad
\var\left[\frac{1}{N}\sum_{n\in[N]} \tilde{w}^n_{\tau_r^-}\right] = \frac{\exp(D(\mcB_T, \tau_{r-1}, \tau_r)) - 1}{N}.
\]
Finally, by \cref{assump:independent_weights_time,assump:independent_weights_particles}, the sums $\sum_{n\in[N]} \tilde{w}^n_{\tau_r^-}$ are independent across $r$, and hence
\[
\var\left[\frac{\hZ}{Z}\right]
&= \prod_{r\in[R]} \left(1 + \var\left[\frac{1}{N}\sum_{n\in[N]} \tilde{w}^n_{\tau_r^-}\right]\right) - 1 \\
&= \prod_{r\in[R]} \left(1 + \frac{\exp(D(\mcB_T, \tau_{r-1}, \tau_r)) - 1}{N}\right) - 1.
\]
\item [(b)]
By (a) the relative variance of $\hZ$ satisfies,
\[
\log\left(1+\var\left[\frac{\hZ}{Z}\right]\right)=\sum_{r\in[R]}v(u_r),
\]
where $u_r=D(\mcB_T,\tau_{r-1},\tau_r)/D(\mcB_T)$ and for $u\in [0,1]$,
\[
v(u)=\log \left(1+\frac{\exp(u D(\mcB_T))-1}{N}\right).
\]
Since $u\mapsto v(u)$ is convex and $\sum_{r\in[R]}u_r=1$, Jensen's inequality gives
\[
R v\left(\frac{1}{R}\right)\leq
\sum_{r\in[R]} v(u_r)\leq v(1).
\]
The upper bound holds with equality if and only if $u_r=1$ for some $r\in [R]$, which occurs when $D(\mcB_T,\tau_{r-1},\tau_r)=D(\mcB_T)$ for some $r$ (i.e., all discrepancy is concentrated in one interval). The lower bound holds with equality if and only if $u_r=1/R$ for all $r\in[R]$, which occurs when the discrepancy is evenly distributed across intervals.

Rearranging, we obtain bounds on the conditional relative variance:
\[\label{eq:variance_bound_conditional}
V(R)\leq \var\left[\frac{\hZ}{Z}\right]
\leq V(1),
\]
where for $r\geq 1$,
\[
V(r)=\exp(rv(1/r))-1=\left(1+\frac{\exp(D(\mcB_T)/r)-1}{N}\right)^r-1.
\]
When $N>1$ and $D(\mcB_T)>0$, the function $V(r)$ is strictly decreasing in $r$. Therefore, by the intermediate value theorem, there exists a unique $\rho\in [1,R]$ such that
\[
\var\left[\frac{\hZ}{Z}\right]
= V(\rho)
= \left(1+\frac{\exp(D(\mcB_T)/\rho)-1}{N}\right)^{\rho}-1.
\]

\eenum

\subsection{Asymptotic variance with a fixed annealing schedule}\label{sec:scaling_finite_T}

We now analyse the asymptotic properties of the relative variance through \cref{thm:variance_ERS}.

\subsubsection{Scaling with particles}

Suppose $D(\mcB_T)$ and the resampling times $0=\tau_0<\cdots<\tau_{R}=T$ are fixed. By \cref{thm:variance_ERS}(a), the relative variance is a decreasing function of $N$, equalling $\exp(D(\mathcal{B}_T))-1$ when $N=1$ and decaying to zero as $N\to\infty$ at a rate depending exponentially on the discrepancy accumulated between resampling steps:
\[
\lim_{N\to\infty} N\var\left[\frac{\hZ}{Z}\right] = \sum_{r\in[R]} \exp(D(\mcB_T,\tau_{r-1},\tau_r)) - R.
\]
In the special cases of AIS and SIR, this recovers the asymptotic variance estimates derived in \citet[Proposition 2]{del_moral_sequential_2006} under comparable simplifying assumptions.

We can use \cref{thm:variance_ERS}(b) to equivalently characterise the asymptotic variance in terms of the effective resample size. Let $\rho_1=1$ and for each $N>1$ let $\rho_N$ be the ERS satisfying
\[
\var\left[\frac{\hZ}{Z}\right] = \left(1+\frac{\exp(D(\mcB_T)/\rho_N)-1}{N}\right)^{\rho_N}-1.
\]
Taking limits as $N\to\infty$ gives
\begin{equation}\label{eq:scaling_particle}
\lim_{N\to \infty} N\var\left[\frac{\hZ}{Z}\right] = \rho\left(\exp\left(\frac{D(\mathcal{B}_T)}{\rho}\right)-1\right),
\end{equation}
where $\rho=\lim_{N\to\infty}\rho_N\in[1,R]$ is the unique solution to
\[
\rho\left(\exp\left(\frac{D(\mathcal{B}_T)}{\rho}\right)-1\right) = \sum_{r\in[R]} \exp(D(\mcB_T,\tau_{r-1},\tau_r)) - R.
\]
\cref{eq:scaling_particle} shows that the asymptotic variance grows exponentially with the total discrepancy per effective resample, $D(\mathcal{B}_T)/\rho$. The exponential dependence on $N$ aligns with guidelines in \citep{chatterjee2018sample,agapiou2017importance} for importance sampling.

\subsubsection{Scaling with resample size}
If $N$ and $D(\mathcal{B}_T)$ are fixed, the relative variance is a monotonically decreasing function of the effective resample size $\rho\in[1,T]$ and satisfies the following bounds for any annealing schedule $\mcB_T$ and $N$,
\[
\exp\left(\frac{D(\mathcal{B}_T)}{N}\right)-1\leq \var\left[\frac{\hZ}{Z}\right]\leq \frac{\exp(D(\mathcal{B}_T))-1}{N}.
\]
This implies that frequent resampling can exponentially reduce the relative variance compared to AIS when $D(\mathcal{B}_T)$ is large, however, there are marginal gains if $D(\mathcal{B}_T)$ is small. Frequent resampling will still result in a high variance if the number of particles is small relative to the total discrepancy but may force unnecessary particle interaction.
However, we note that monotonicity in $\rho$ holds under our idealised performance model assumptions, in practice, this may not hold due to added path degeneracy and variance from inefficiencies in the resampling kernel that our simplifying assumptions cannot capture. 

\subsubsection{Scaling with total discrepancy}
For fixed $N$ and $\rho$, the relative variance increases monotonically in $D(\mcB_T)$, growing exponentially for large $D(\mcB_T)$:
\[\label{eq:scaling_discrepancy_large}
\var\left[\frac{\hZ}{Z}\right]\sim \frac{\exp(D(\mathcal{B}_T))}{N^\rho}=\left(\frac{\exp(D(\mcB_T)/\rho)}{N}\right)^\rho\;\;\text{ as }D(\mathcal{B}_T)\to \infty.
\]
Stabilising the relative variance in this regime requires $N$ exponentially large in the total discrepancy per effective resample, $D(\mathcal{B}_T)/\rho$. Conversely, as $D(\mathcal{B}_T)\to 0^+$, the relative variance decays linearly in $D(\mathcal{B}_T)/N$, independently of $\rho$:
\[\label{eq:scaling_discrepancy_small}
\var\left[\frac{\hZ}{Z}\right] \sim \frac{D(\mathcal{B}_T)}{N}\;\;\text{ as }D(\mathcal{B}_T)\to 0^+.
\]
Thus, when the total discrepancy is small, the number of particles required to stabilise the relative variance is only linear in $D(\mcB_T)$.

\subsection{Adaptive resampling in the large particle limit}\label{sec:AR=DAR}
Let $0=\tau^N_0<\cdots<\tau^N_{R^N}=T$ denote the resampling schedule for adaptive resampling (AR) with threshold $\nu$, satisfying $\tau^N_0=0$, $\tau^N_{R^N}=T$, and for $r=[R^N-1]$,
\[
\tau^N_{r}=\inf\{\tau^N_{r-1}<t\leq T: N_\text{eff}(t)<\nu N\}.
\]
Similarly, let $\tau_{0:R}$ denote the resampling schedule for stabilised adaptive resampling (DAR) with threshold $\nu$, satisfying $\tau_0=0$, $\tau_{R}=T$, and for $r=1,\dots,R-1$,
\[
\tau_r=\inf\left\{t>\tau_{r-1}:D(\mcB_T,\tau_{r-1},t)>-\log\nu\right\}.
\]
Define $\tau^N(t)$ and $\tau(t)$ as the most recent resampling time by iteration $t$:
\[
\tau^N(t)=\max\{\tau^N_r: \tau^N_r\leq t\},\qquad
\tau(t)=\max\{\tau_r: \tau_r\leq t\}.
\]
We will say that AR equals DAR if and only if $\tau^N(t)\stackrel{a.s}{=}\tau(t)$ for all $t\in[T]$.

\bprop\label{prop:AR_convergence}
Suppose \cref{assump:integrability,assump:independent_weights_time,assump:independent_weights_particles,assump:ELE} hold. If $\nu\neq \exp(-D(\mcB_T,t,t'))$ for all $t,t'\in[T]$, then AR converges to DAR almost surely as $N\to\infty$.
\eprop

\bprf
We prove by induction on $t$ that $\tau^N(t)\xrightarrow{a.s.}\tau(t)$ as $N\to\infty$. 

\paragraph{Base case ($t=0$).} 
Trivially, $\tau^N(0)=\tau(0)=0$.

\paragraph{Inductive step.} 
Suppose $\tau^N(t)\xrightarrow{a.s.}\tau(t)$. By the induction hypothesis, almost surely there exists $N(t)$ large enough such that for all $N\geq N(t)$, we have $\tau^N(t)=\tau(t)$. 

Let $w^n_{t+1^-}$ denote the weight of particle $n$ after the propagation and reweighting step at iteration $t+1$. By \cref{assump:independent_weights_time,assump:independent_weights_particles}, the weights $(w^n_{t+1^-})_{n\in[N]}$ are \iid. The first moment satisfies
\[
\E[w^n_{t+1^-}]
=\E\left[\prod_{s=\tau(t)+1}^{t+1} g^n_s\right]
=\prod_{s=\tau(t)+1}^{t+1}\E[g^n_s]
=\prod_{s=\tau(t)+1}^{t+1}\frac{Z_{\beta_s}}{Z_{\beta_{s-1}}}
=\frac{Z_{\beta_{t+1}}}{Z_{\beta_{\tau(t)}}}.
\] 
Similarly, the second moment satisfies
\[
\E[(w^n_{t+1^-})^2]
=\prod_{s=\tau(t)+1}^{t+1}\E[(g^n_s)^2]
=\frac{Z_{\beta_{t+1}}^2}{Z_{\beta_{\tau(t)}}^2}\prod_{s=\tau(t)+1}^{t+1}\exp(D(\beta_{s-1},\beta_s)).
\]
By the strong law of large numbers,
\[
\frac{N_\text{eff}(t+1)}{N}
=\frac{\left(\frac{1}{N}\sum_{n\in[N]}w^n_{t+1^-}\right)^2}{\frac{1}{N}\sum_{n\in[N]}(w^n_{t+1^-})^2}
\xrightarrow[N\to\infty]{a.s.} \frac{\E[w_{t+1^-}]^2}{\E[(w_{t+1^-})^2]}
=\exp(-D(\mcB_T,\tau(t),t+1)).
\]

Since $\nu\neq \exp(-D(\mcB_T,\tau(t),t+1))$ by assumption, let $\epsilon>0$ satisfy
\[
0<\epsilon<\left|\exp(-D(\mcB_T,\tau(t),t+1))-\nu\right|.
\]
There exists $N(t+1)\geq N(t)$ such that for all $N>N(t+1)$,
\[
\left|\frac{N_\text{eff}(t+1)}{N}-\exp(-D(\mcB_T,\tau(t),t+1))\right|<\frac{\epsilon}{2}.
\] 

\textbf{Case 1: $\tau(t+1)=t+1$.} By the DAR criterion, $\exp(-D(\mcB_T,\tau(t),t+1))<\nu-\epsilon$. By the triangle inequality,
\begin{align*}
\frac{N_\text{eff}(t+1)}{N}
&\leq \exp(-D(\mcB_T,\tau(t),t+1))+ \left|\frac{N_\text{eff}(t+1)}{N}-\exp(-D(\mcB_T,\tau(t),t+1))\right|\\
&<\nu-\epsilon+\epsilon/2
=\nu-\epsilon/2.
\end{align*}
Therefore, $N_\text{eff}(t+1)<N\nu$, hence $\tau^N(t+1)=t+1=\tau(t+1)$. 

\textbf{Case 2: $\tau(t+1)\neq t+1$.} Then $\tau(t+1)=\tau(t)$ and $\exp(-D(\mcB_T,\tau(t),t+1))>\nu+\epsilon$. By the reverse triangle inequality,
\begin{align*}
\frac{N_\text{eff}(t+1)}{N}
&\geq \exp(-D(\mcB_T,\tau(t),t+1))- \left|\frac{N_\text{eff}(t+1)}{N}-\exp(-D(\mcB_T,\tau(t),t+1))\right|\\
&>\nu+\epsilon-\epsilon/2
=\nu+\epsilon/2.
\end{align*}
Therefore, $N_\text{eff}(t+1)>N\nu$, hence $\tau^N(t+1)=\tau^N(t)=\tau(t)=\tau(t+1)$ by the induction hypothesis.

In both cases, for all $N\geq N(t+1)$, we have $\tau^N(t+1)=\tau(t+1)$, which implies $\tau^N(t+1)\xrightarrow{a.s.}\tau(t+1)$. By induction, $\tau^N(t)\xrightarrow{a.s.}\tau(t)$ for all $t\in[T]$, establishing that AR converges to DAR almost surely.
\eprf
\subsection{Proof of \cref{prop:adaptive_SMCS}}

\blem\label{lem:adaptive_SMCS_2}
Suppose \cref{assump:integrability,assump:independent_weights_time,assump:independent_weights_particles,assump:ELE} hold. If there exists $\epsilon>0$ such that almost surely $D(\mcB_T,\tau_{r-1},\tau_r) \leq \epsilon D(\mcB_T)$ for all $r\in[R]$, then $\rho>1/\epsilon$.
\elem

\bprf[Proof of \cref{lem:adaptive_SMCS_2}]
Summing over $r\in[R]$ gives $D(\mcB_T)\leq R\epsilon D(\mcB_T)$, hence $R\geq 1/\epsilon$ almost surely. By \cref{thm:variance_ERS}(a), the relative variance satisfies
\[
\log\left(1+\var\left[\frac{\hZ}{Z}\right]\right)=\sum_{r\in[R]}v(u_r),
\]
where $u_r=D(\mcB_T,\tau_{r-1},\tau_r)/D(\mcB_T)$ and 
\[
v(u)=\log \left(1+\frac{\exp(u D(\mcB_T))-1}{N}\right).
\]

Since $u\mapsto v(u)$ is strictly convex, the maximum of $\sum_{r\in[R]}v(u_r)$ subject to the constraints $0\leq u_r\leq \epsilon$ and $\sum_{r\in[R]}u_r=1$ is attained when $u_r=\epsilon$ for $r=1,\dots,R_{\epsilon}$, where $R_\epsilon=\lfloor 1/\epsilon\rfloor< R$, $u_{R_\epsilon+1}=1-\epsilon R_\epsilon$, and $u_r=0$ for $r=R_{\epsilon}+2,\dots,R$. Therefore,
\begin{align*}
\sum_{r\in[R]}v(u_r)
&=\sum_{r=1}^{R_{\epsilon}} v(\epsilon)+v\left(1-\epsilon R_\epsilon\right) +\sum_{r=R_{\epsilon}+2}^R v(0)\\
&=R_\epsilon v(\epsilon)+v\left(1-\epsilon R_\epsilon\right)\\
&\leq \frac{1}{\epsilon} v(\epsilon),
\end{align*}
where the last inequality follows from convexity of $v(u)$. Rearranging yields
\[
\var\left[\left.\frac{\hZ}{Z}\right|\tau_{0:R}\right]
\leq \exp( v(\epsilon)/\epsilon)-1
=\left(1+\frac{\exp(\epsilon D(\mcB_T))-1}{N}\right)^{1/\epsilon}-1.
\]
Taking expectations and applying \cref{thm:variance_ERS} gives
\[
\var\left[\frac{\hZ}{Z}\right]
\leq \left(1+\frac{\exp(\epsilon D(\mcB_T))-1}{N}\right)^{1/\epsilon}-1,
\]
hence $\rho\geq 1/\epsilon$.  
\eprf

\bprf[Proof of \cref{prop:adaptive_SMCS}]
We prove the upper and lower bounds separately.

\paragraph{Upper bound.}
For all $r\in [R]$, the accumulated discrepancy between $\tau_{r-1}$ and $\tau_r$ satisfies
\begin{align*}
D(\mcB_T,\tau_{r-1},\tau_r)
&\leq  -\log\nu +\max_{t\in[T]} D(\beta_{t-1},\beta_t)\\
&= \frac{-\log\nu +\max_{t\in[T]} D(\beta_{t-1},\beta_t)}{D(\mcB_T)}D(\mcB_T).
\end{align*}
By \cref{lem:adaptive_SMCS_2},
\[
\rho_\DAR(\nu) \geq \frac{D(\mcB_T)}{-\log\nu +\max_{t\in[T]} D(\beta_{t-1},\beta_t)}. 
\]
Since $\rho_\DAR(\nu)\geq 1$, we obtain
\[
\rho_\DAR(\nu)\geq\max\left\{1,\frac{D(\mcB_T)}{\max_{t\in[T]} D(\beta_{t-1},\beta_t)-\log\nu}\right\}.
\]

\paragraph{Lower bound.}
When $\nu\leq \nu_{\min}$, no resampling events occur, so $R=1$ and hence $\rho_\DAR(\nu)=1$ by \cref{thm:variance_ERS}, making the lower bound trivial. 

When $\nu>\nu_{\min}$, there exists $t<T$ such that $D(\mcB_T,0,t)>-\log\nu$, hence $R>1$. For each $r\in[R-1]$, the adaptive resampling criterion ensures $D(\mcB_T,\tau_{r-1},\tau_r)> -\log \nu$. Summing over $r\in[R-1]$ gives
\[
D(\mcB_T)\geq \sum_{r=1}^{R-1}D(\mcB_T,\tau_{r-1},\tau_r)> -(R-1)\log \nu.
\]
Rearranging yields an upper bound on $R$:
\[
R< 1-\frac{D(\mcB_T)}{\log \nu}.
\]
Taking expectations and applying \cref{thm:variance_ERS} gives
\[
\rho_\DAR(\nu)\leq \E[R]< 1-\frac{D(\mcB_T)}{\log \nu}.
\]
Since $\rho_\DAR(\nu)\leq T$, we obtain
\[
\rho_\DAR(\nu)\leq \min\left\{T,1-\frac{D(\mcB_T)}{\log \nu}\right\}.
\]
\eprf

\section{Supplement for \cref{sec:dense_schedule_limit}}
\subsection{Regularity of paths and measures}\label{sec:regularity}
\subsubsection{Path regularity of functions and measures}
Let $\mcM(\mcX)$ be the set of finite signed measures over $\mcX$. Given $\mu\in \mcM(\mcX)$, let $L^1(\mu)$ denote the set of integrable functions with respect to $\mu$, and for $f\in L^1(\mu)$, let $\mu[f]=\int_\mcX f(x)\mu(\dee x)$ be the integral.
\paragraph{Regularity class.}
A \emph{regularity class}, $\mcF=(\mcF_i)_{i\in \nats}$ is a sequence of subspaces of measurable functions $f:\mcX\mapsto \reals$ such that:
\bitems
\item For all $i\in\nats$, $\mcF_i$ is a vector space 
and $\mcF_{i+1}\subseteq \mcF_i$.
\item If a measurable $g:\mcX\mapsto\reals$ such that, $|g(x)|\leq |f(x)|$ for $f\in\mcF_i$, then $g\in \mcF_i$. 
\item For all $i\in\nats$, $\mcF_i$ contains the constant function $1(x)=1$.
\eitems

\paragraph{Regularity of real functions.}
Let $B$ be a $d$-dimensional compact manifold parameterised by $\beta=(\beta_1,\dots, \beta_d)$. Given $h:B\mapsto\reals$, we will say $\beta\mapsto h_\beta\in C^k(B)$ if for all $\textbf{i}=(i_1,\dots, i_d)$ with $|\textbf{i}|=i_1+\dots+i_d\leq k$, the $\textbf{i}$-th order partial derivative of $h_\beta$, denoted $h_{\beta,\textbf{i}}$ exists and is continuous in $\beta$, where
\[
h_{\beta,\textbf{i}}=\frac{\partial^{|\textbf{i}|}}{\partial\beta^{\textbf{i}}}h_\beta.
\] 
Here we use $\partial\beta^{\textbf{i}}=\partial\beta_1^{i_1}\cdots\partial\beta_d^{i_d}$ for convenience.

\paragraph{Regularly measurable functions.}
Given measurable functions $f_\beta:\mcX\mapsto\reals$ parametrised by $\beta\in B$, and regularity class $\mcF$, we will say $\beta\mapsto f_\beta\in C^k(B,\mcF)$ if and only if
\bitems
\item For all $x\in\mcX$, $\beta\mapsto f_\beta(x)\in C^k(B)$.
\item For all $i=0,\dots,k$,and $\textbf{i}=(i_1,\dots,i_d)$ with $|\textbf{i}|= i$, we have $\sup_{\beta\in B} |f_{\beta,\textbf{i}}|\in \mcF_{k-i}$, where $f_{\beta,\textbf{i}}$ is the partial derivatives of $f_\beta$ of order $\textbf{i}$ evaluated pointwise,  
\[
 x\in \mcX,\quad f_{\beta,\textbf{i}}(x)=\frac{\partial^{|\textbf{i}|}}{\partial\beta^{\textbf{i}}}f_\beta(x).
\]
\eitems
Note that if $\beta\mapsto f_\beta$ is $C^k(B,\mcF)$ then $f_{\beta,\textbf{i}}\in C^{k-|\textbf{i}|}(B,\mcF)$.

\paragraph{Regularity signed-measures.}
Given signed-measures $\mu_\beta\in \mcM(\mcX)$ parametrised $\beta\in B$, and regularity class $\mcF$, we will say $\mu_\beta \in C^k(B,\mcF)$ if and only if
\bitems
\item For all $f\in \mcF_k$,  $\beta\mapsto \mu_\beta[f]\in C^k(B)$.
\item For all $i=0,\dots,k$,and $\textbf{i}=(i_1,\dots,i_d)$ with $|\textbf{i}|\leq i$, there exists a signed measure $\mu_{\beta,\textbf{i}}$ such that $\mcF_i\subset L^1(\mu_{\beta,\textbf{i}})$ and,
\[\label{eq:MVD}
f\in \mcF_i,\quad \mu_{\beta,\textbf{i}}[f] = \frac{\partial^{|\textbf{i}|}}{\partial\beta^{\textbf{i}}}\mu_\beta[f].
\]
\eitems
If $\beta\mapsto \mu_\beta$ is $C^k(B,\mcF)$ then the measure-valued derivative $\mu_{\beta,\textbf{i}}$ are uniquely defined only on functions of regularity $\mcF_{|\textbf{i}|}$, but in general they are not unique.

\cref{lem:product-rule} shows that if $f_\beta$ and $\mu_\beta$ are $C^k(\mcB,\mcF)$ functions and measures respectively, with respect to some regularity class $\mcF$, then the expectations $\beta\mapsto \mu_\beta[f_\beta]$ is $C^k(\mcB,\reals)$, and the derivatives satisfy a product rule. In particular, when $B=[0,1]$ and $k=1$, we have,
\[\label{eq:product_rule}
\frac{\dee}{\dee\beta}\mu_\beta[f_\beta]=\mu_{\beta,1}[f_\beta]+\mu_\beta\left[f_{\beta,1}\right].
\]

\blem\label{lem:product-rule}

Suppose $\beta\mapsto f_{\beta}\in C^k(B,\mcF)$ and $\beta\mapsto \mu_\beta\in C^k(B,\mcF)$, then:

\benum
\item $\beta\mapsto \mu_\beta[f_\beta]\in C^k(B)$, and  all $\textbf{i}=(i_1,\dots, i_d)$ and $\textbf{i}'=(i_1',\dots, i_d')$ with $0\leq |\textbf{i}|+|\textbf{i}|'< k$, for $j=1,\dots,d$ the partial derivative with respect to $\beta_j$ exists and satisfies,
\[
 \frac{\partial}{\partial\beta_j}\mu_{\beta,\textbf{i}}[f_{\beta,\textbf{i}'}]=\mu_{\beta,\textbf{i}+\textbf{e}_j}[f_{\beta,\textbf{i}'}]+\mu_{\beta,\textbf{i}}[f_{\beta,\textbf{i}'+\textbf{e}_j}]
\]
where $\textbf{e}_j=(\delta_{1,j},\dots, \delta_{d,j})$ where $\delta_{j,j}=1$ and $\delta_{i,j}=0$ otherwise.

\item If $B=[0,1]$, then for $i\leq k$, 
\[
\frac{\dee^i}{\dee\beta^i}\mu_\beta[f_\beta]=\sum_{j=0}^i{i\choose j}\mu_{\beta,j}\left[f_{\beta,i-j}\right].
\]
\eenum
\elem

\bprf[Proof of \cref{lem:product-rule}]
\benum
\item 
Given $\beta\in B$ and $\beta'=\beta+ h \textbf{e}_j$, we have 
\[
\frac{\mu_{\beta',\textbf{i}}[f_{\beta',\textbf{i}'}]-\mu_{\beta,\textbf{i}}[f_{\beta,\textbf{i}'}]}{h}
&=\frac{\mu_{\beta',\textbf{i}}[f_{\beta',\textbf{i}'}]-\mu_{\beta,\textbf{i}}[f_{\beta',\textbf{i}'}]}{h}+\frac{\mu_{\beta,\textbf{i}}[f_{\beta',\textbf{i}'}]-\mu_{\beta,\textbf{i}}[f_{\beta,\textbf{i}'}]}{h}\\
&=\frac{\mu_{\beta',\textbf{i}}-\mu_{\beta,\textbf{i}}}{h}[f_{\beta',\textbf{i}'}]
+\mu_{\beta,\textbf{i}}\left[\frac{f_{\beta',\textbf{i}'}-f_{\beta,\textbf{i}'}}{h}\right]\\
&=\mu_{\beta,\textbf{i}+\textbf{e}_j}[f_{\beta,\textbf{i}}]+\mu_{\beta,\textbf{i}}\left[f_{\beta,\textbf{i}'+\textbf{e}_j}\right]
+\eps_1+\eps_2+\eps_3,
\]
where
\[
\eps_1& =(\mu_{\beta',\textbf{i}}-\mu_{\beta,\textbf{i}})\left[\frac{f_{\beta',\textbf{i}'}-f_{\beta,\textbf{i}'}}{h}\right]\\
\eps_2& =\left(\frac{\mu_{\beta',\textbf{i}}-\mu_{\beta,\textbf{i}}}{h}-\mu_{\beta,\textbf{i}+\textbf{e}_j}\right)[f_{\beta,\textbf{i}'}]\\
\eps_3&=\mu_{\beta,\textbf{i}}\left[\frac{f_{\beta',\textbf{i}}-f_{\beta,\textbf{i}}}{h}-f_{\beta,\textbf{i}'+\textbf{e}_j}\right].
\]
By the mean value theorem
\[
|\eps_1|
\leq \left|\mu_{\beta',\textbf{i}}-\mu_{\beta,\textbf{i}}\right|\left[\left|\frac{f_{\beta',\textbf{i}'}-f_{\beta,\textbf{i}'}}{h}\right|\right]
\leq |\mu_{\beta',\textbf{i}}-\mu_{\beta,\textbf{i}}|\left[\sup_{\beta\in B} |f_{\beta,\textbf{i}'+\textbf{e}_j}|\right].
\]
Since $\sup_{\beta\in B} |f_{\beta,\textbf{i}'+\textbf{e}_j}|\in \mcF_{k-|\textbf{i}'|-1}\subseteq \mcF_{|\textbf{i}|}$ , and $\mu_{\beta,\textbf{i}}[f]$ is continuous for all $f\in \mcF_{|\textbf{i}|}$, we have $\eps_1\to 0$ as $h\to 0$.

By definition of $\mu_{\beta,\textbf{i}}$, for all $f\in \mcF_{|\textbf{i}|+1}$, we have,
\[
\lim_{h\to 0}\frac{\mu_{\beta',\textbf{i}}-\mu_{\beta,\textbf{i}}}{h}[f]
&=\lim_{h\to 0}\frac{1}{h}\left(\frac{\partial^{|\textbf{i}|}}{\partial\beta^\textbf{i}}\mu_{\beta'}[f]-\frac{\partial^{|\textbf{i}|}}{\partial\beta^\textbf{i}}\mu_{\beta'}[f]\right) \\
&= \frac{\partial^{|\textbf{i}|+1}}{\partial\beta^{\textbf{i}+\textbf{e}_j}}\mu_{\beta'}[f]\\
&=\mu_{\beta,\textbf{i}+\textbf{e}_j}[f].
\]
Since $f_{\beta,\textbf{i}'}\in \mcF_{k-|\textbf{i}|}\subseteq \mcF_{|\textbf{i}|+1}$ and have $\eps_2\to 0$ as $h\to 0$.

Finally $\eps_3$ satisfies,
\[
|\eps_3|\leq |\mu_{\beta,\textbf{i}}|\left[\left|\frac{f_{\beta',\textbf{i}'}-f_{\beta,\textbf{i}'}}{h}-f_{\beta,\textbf{i}'+\textbf{e}_j}\right|\right].
\]
By the mean value theorem, the integrand is uniformly bounded by $2 \sup_{\beta\in B}|f_{\beta,\textbf{i}'+\textbf{e}_j}|$ and converges to zero almost surely as $\beta'\to\beta$. By the dominated convergence theorem, $\eps_3\to 0$ as $h\to\infty$. 

Since $\mu_{\beta}[f_\beta]=\mu_{\beta,\textbf{i}}[f_{\beta,\textbf{i}'}]$ with $\textbf{i}=\textbf{i}'=(0,\dots,0)$, we can continuously differentiate $\mu_{\beta}[f_\beta]$ $k$ times using the project rule derived above. Hence $\beta\mapsto \mu_{\beta}[f_\beta]$ is in $C^k(B)$.

\item We proceed with induction on $i=0,\dots,k$. When $i=0$, the result trivially holds. Now suppose for all $i< k$. 
\[
\frac{\dee^i}{\dee\beta^i}\mu_\beta[f_\beta]
=\sum_{j=0}^i{i\choose j}\mu_{\beta,j}\left[f_{\beta,i-j}\right].
\]
Then for $i+1$ have
\[
\frac{\dee^{i+1}}{\dee\beta^{i+1}}\mu_\beta[f_\beta]
&=\der{}{\beta}\frac{\dee^{i}}{\dee\beta^i}\mu_\beta[f_\beta]\\
&=\sum_{j=0}^i{i\choose j}\der{}{\beta}\mu_{\beta,j}\left[f_{\beta,i-j}\right]\\
&=\sum_{j=0}^i{i\choose j}\left(\mu_{\beta,j+1}\left[f_{\beta,i+1-j+1}\right]+\mu_{\beta,j)}\left[f_{\beta,i+1-j}\right]\right)\\
&=\mu_{\beta,0}[f_{\beta,k+1}]+\sum_{j=1}^{i}\left({i\choose j-1}+{i\choose j}\right)\mu_{\beta,j}\left[f_{\beta,i+1-j}\right]+\mu_{\beta,i+1}[f_{\beta,0}]\\
&=\mu_{\beta,0}[f_{\beta,i+1}]+\sum_{j=1}^{i}{i+1 \choose j}\mu_{\beta,j}\left[f_{\beta,i+1-j}\right]+\mu_{\beta,i+1}[f{\beta,0}]\\
&=\sum_{j=0}^{i+1}{i+1\choose j}\mu_{\beta,j}\left[f_{\beta,i+1-j)}\right].
\]
The second line used the induction hypothesis, and the third line used the product rule from (a). 

\eenum
\eprf
\subsubsection{Regularity of measure-valued paths with densities}
Suppose $\mu_\beta(\dee x)=\mu_\beta(x)\dee x$ is absolutely continuous with respect to $\dee x$ for all $\beta\in B$, and for all $x\in\mcX$ we have $\mu_\beta(x)>0 , \beta\mapsto \mu_\beta(x)\in C^k(B)$. For $i=0,\dots, k$, define $\mcF_i$ as the set of measurable functions $f:\mcX\mapsto\reals$, such that for all $\textbf{i}_1,\dots,\textbf{i}_j$ with $|\textbf{i}_1|,\dots,|\textbf{i}_j|\geq 1$ and $|\textbf{i}_1|+\dots+|\textbf{i}_j|\leq i$,
\[
\int_{\mcX}f(x)\bar{V}_{\textbf{i}_1}(x)\cdots \bar{V}_{\textbf{i}_j}(x)\bar\mu(x)\dee x<\infty.
\]
Here we define $\bar{\mu}(x)=\sup_{\beta\in B}\mu_\beta(x)$ as the pointwise supremum of the $\mu_\beta(x)$ over $B$, and $\bar{V}_{\textbf{i}}(x)=\sup_{\beta\in B} |V_{\beta,\textbf{i}}(x)|$ is the pointwise supremum of the $\textbf{i}$-th order partial derivative of the log-density $V_\beta(x)=\log\mu_\beta(x)$. For $i>k$ let $\mcF_i$ be the set of constant functions, then $\mcF=(\mcF_i)_{i\in\nats}$ defines a regularity class.

\cref{lem:regularity-measures} shows that $\beta\mapsto \mu_\beta$ is $C^k(B,\mcF)$ and the measure-valued derivatives $\mu_{\beta,\textbf{i}}$ are absolutely continuous with respect to $\dee x$ as well. In particular, in the special case where $B=[0,1]$, we have for all $f\in \mcF_1$,
\[
\frac{\dee}{\dee\beta}\mu_\beta[f] =\mu_{\beta,1}[f]= \mu_\beta\left[f \frac{\dee}{\dee\beta}\log \mu_\beta\right],
\]
where $\frac{\dee}{\dee\beta}\log \mu_\beta(x)$ is the Fisher score function for $\mu_\beta$. If $\beta\mapsto f_\beta\in C^1([0,1],\mcF)$ then  \cref{lem:product-rule} implies,
\[
\der{}{\beta}\mu_\beta[f_\beta] = \mu_\beta\left[f \der{}{\beta}\log \mu_\beta + \der{f_\beta}{\beta}\right].
\]
\blem\label{lem:regularity-measures}
Suppose $\mu_\beta(\dee x)$ is absolutely continuous with respect to $\dee x$ for all $\beta\in B$, and for all $x\in\mcX$, the density $\mu_\beta(x)>0$ and $\beta\mapsto \mu_\beta(x)\in C^k(B)$, and let $\mcF$ be the regularity class defined above.
\benum 
\item  $\beta\mapsto \mu_\beta\in C^k(B,\mcF)$, and for all $|\textbf{i}|\leq i\leq k$,
\[
f\in \mcF_i,\quad  \frac{\partial^{|\textbf{i}|}}{\partial\beta^\textbf{i}}\mu_\beta[f]=\mu_{\beta,\textbf{i}}[f],
\]
where $\mu_{\beta,\textbf{i}}\in \mcM(\mcX)$ is absolutely continuous with respect to $\dee x$, 
\[\label{eq:RN-derivative_mu}
\mu_{\beta,\textbf{i}}(\dee x)
=\frac{\partial^{|\textbf{i}|}}{\partial\beta^\textbf{i}}\mu_\beta(x)\dee x.
\]
\item  If $B=[0,1]$, then $\mu_{\beta,0}(x)=\mu_\beta(x)$, and for $i=1,\dots,k$, 
\[
\der{\mu_{\beta,i}}{\mu_\beta}(x)=\der{\mu_{\beta,i-1}}{\mu_\beta} (x)\der{}{\beta}\log \mu_\beta(x)+\der{}{\beta}\der{\mu_{\beta,i-1}}{\mu_\beta}(x).
\]
\eenum
\elem

\bprf[Proof of \cref{lem:regularity-measures}]
\benum
\item  
For $i\leq k$ define, $V_{\beta,\textbf{i}}(x)$ as the $\textbf{i}$-th derivative of $V_\beta(x)=\log \mu_\beta(x)$ with respect to $\beta$. We have the $\textbf{i}$-th derivative of $\mu_\beta(x)=\exp(\log \mu_\beta(x))$ with respect to $\beta$ can be expressed as $\mu_\beta(x)$ times a linear combination of $V_{\beta,\textbf{i}_1}\cdots V_{\beta,\textbf{i}_1}$ over multi-indices $\textbf{i}_1,\dots,\textbf{i}_j$ with $|\textbf{i}_1|,\dots,|\textbf{i}_j|\geq 1$ and $|\textbf{i}_1|+\dots+|\textbf{i}_j|= |\textbf{i}|$. Therefore for some $C(\textbf{i}_1,\dots,\textbf{i}_j)\geq 0$, 
\[
\frac{\partial^{|\textbf{i}|}}{\partial\beta^\textbf{i}}\mu_\beta(x)
&=\sum_{|\textbf{i}_1|+\dots+|\textbf{i}_1|=|\textbf{i}|} C(\textbf{i}_1,\dots,\textbf{i}_j) V_{\beta,\textbf{i}_1}(x)\cdots V_{\beta,\textbf{i}_j}(x)\mu_\beta(x).
\]
By taking the supremum, we have the following upper bound uniformly in $\beta$,
\[
\sup_{\beta\in B}\left|\frac{\partial^{|\textbf{i}|}}{\partial\beta^\textbf{i}}\mu_\beta(x)\right|
\leq
\sum_{|\textbf{i}_1|+\dots+|\textbf{i}_1|=|\textbf{i}|} C(\textbf{i}_1,\dots,\textbf{i}_j)  \bar{V}_{\textbf{i}_1}(x)\cdots \bar{V}_{\textbf{i}_j}(x)\bar\mu(x).
\]
By definition of $\mcF_i$ we can apply the Leibniz integration rule for all $f\in \mcF_i$, 
\[
 \frac{\partial^{|\textbf{i}|}}{\partial\beta^\textbf{i}}\mu_\beta[f]
 = \frac{\partial^{|\textbf{i}|}}{\partial\beta^\textbf{i}} \int f(x)  \mu_\beta(x)\dee x
 =  \int  f(x)\frac{\partial^{|\textbf{i}|}}{\partial\beta^\textbf{i}}\mu_\beta (x)\dee x
 =\mu_{\beta,\textbf{i}}[f].
 \] 
Here $\mu_{\beta,\textbf{i}}\in\mcM(\mcX)$ is a signed measure absolutely continuous with respect to $\dee x$,
\[
\mu_{\beta,\textbf{i}}(\dee x)=\frac{\partial^{|\textbf{i}|}}{\partial\beta^\textbf{i}}\mu_\beta (x)\dee x.
\]
For $i> k$, $\mcF_i$ is the set of constant functions, so for all $f\in \mcF_i$, we have $\mu_\beta[f]$ is constant, and for all $|\textbf{i}|>k$, 
\[
\frac{\partial^{|\textbf{i}|}}{\partial\beta^\textbf{i}}\mu_\beta[f]=0=\mu_{\beta,\textbf{i}}[f],
\]
where $\mu_{\beta,\textbf{\textit{i}}}=0\in \mcM(\mcX)$ is the zero measure. 

\item Since $\mu_\beta(x)>0$, we have for all $i\in \nats$ that
\[
 \mu_{\beta,i}[f] = \mu_\beta\left[f\der{\mu_{\beta,i}}{\mu_\beta}\right],
\]
where
\[
\der{\mu_{\beta,i}}{\mu_\beta}
= \frac{1}{\mu_\beta(x)} \frac{\dee^i}{\dee\beta^i}\mu_\beta(x).
\]
For all $i=1,\dots, k$, and $f\in \mcF_i$,
\[
\mu_{\beta,i}[f]
=\der{}{\beta}\mu_{\beta,i-1}[f]
=\der{}{\beta}\mu_\beta\left[f\der{\mu_{\beta,i-1}}{\mu_\beta}\right].
\]
By the product rule,
\[
\mu_{\beta,i}[f]
&=\mu_{\beta,1}\left[f\der{\mu_{\beta,i-1}}{\mu_\beta}\right]+
\mu_\beta\left[\der{}{\beta}\left(f\der{\mu_{\beta,i-1}}{\mu_\beta}\right)\right].
\]
By (a) we have $\der{\mu_{\beta,1}}{\mu_\beta}=\der{}{\beta}\log\mu_\beta$, which implies,
\[
\mu_{\beta,i}[f]
&=\mu_\beta\left[f\der{\mu_{\beta,i-1}}{\mu_\beta} \der{}{\beta}\log\mu_\beta\right]+
\mu_\beta\left[f\der{}{\beta}\der{\mu_{\beta,i-1}}{\mu_\beta}\right]\\
&=\mu_\beta\left[f\left(\der{\mu_{\beta,i-1}}{\mu_\beta} \der{}{\beta}\log\mu_\beta + \der{}{\beta}\der{\mu_{\beta,i-1}}{\mu_\beta}\right)\right].
\]
Therefore 
\[
\der{\mu_{\beta,i}}{\mu_\beta}=\der{\mu_{\beta,i-1}}{\mu_\beta} \der{}{\beta}\log\mu_\beta + \der{}{\beta}\der{\mu_{\beta,i-1}}{\mu_\beta}.
\]
\eenum
\eprf
\subsection{Proof of \cref{thm:incremental_discrepancy_estimate}}\label{sec:proof_discrepancy_estimate}

For $\beta,\beta'$ we define $U_{\beta,\beta'},H_{\beta,\beta'}:\mcX\times\mcX\mapsto\reals_+$ as,
\[
G_{\beta,\beta'}(x,x')&=\frac{\dee \pi_{\beta'}\otimes L_{\beta,\beta'}}{\dee \pi_\beta\otimes M_{\beta,\beta'}}(x,x'),\\
U_{\beta,\beta'}(x,x')&=\log G_{\beta,\beta'}(x,x'),\\
H_{\beta,\beta'}(x,x')&= G_{\beta,\beta'}(x,x')^2.
\]
For $\textbf{i}=(i,i')$ where $i,i'\in \nats$ and $x,x'\in \mcX$ define the following partial derivatives when they exist,
\[
G_{\beta,\beta',\textbf{i}}(x,x')&=\frac{\partial^{i+i'}}{\partial\beta^i\partial\beta'^{i'}}G_{\beta,\beta'}(x,x'),\quad 
\bar{G}_{\textbf{i}}(x,x')=\sup_{\beta,\beta'\in[0,1]}|G_{\beta,\beta',\textbf{i}}(x,x')|,\\
U_{\beta,\beta',\textbf{i}}(x,x')&=\frac{\partial^{i+i'}}{\partial\beta^i\partial\beta'^{i'}}U_{\beta,\beta'}(x,x'),\quad 
\bar{U}_{\textbf{i}}(x,x')=\sup_{\beta,\beta'\in[0,1]}|U_{\beta,\beta',\textbf{i}}(x,x')|,\\
H_{\beta,\beta',\textbf{i}}(x,x')&=\frac{\partial^{i+i'}}{\partial\beta^i\partial\beta'^{i'}}H_{\beta,\beta'}(x,x'), \quad 
\bar{H}_{\textbf{i}}(x,x')=\sup_{\beta,\beta'\in[0,1]}|H_{\beta,\beta',\textbf{i}}(x,x')|.
\]
For $\beta,\beta'$, we also denote the forward proposal as $\eta_{\beta,\beta'}=\pi_\beta\otimes M_{\beta,\beta'}\in \mcM(\mcX\times\mcX)$ and for multi-index $\textbf{i}=(i,i')$ of order $|\textbf{i}|=i+i'$ define the measure-valued derivatives,
\[
f\in \mcF_{|\textbf{i}|},\quad \eta_{\beta,\beta',\textbf{i}}[f]=\frac{\partial^{i+i'}}{\partial\beta^i\partial\beta'^{i'}}\eta_{\beta,\beta'}[f].
\]

\blem\label{lem:regularity}
\benum

\item  If \cref{assump:regular-weights} holds, then $\beta,\beta'\mapsto\log U_{\beta,\beta'},G_{\beta,\beta'}, H_{\beta,\beta'}$ are in $C^3([0,1]^2,\mcF)$,
\item  If \cref{assump:regular-proposal,assump:regular-weights,assump:non-degeneracy} hold, then for all $\beta\in [0,1]$,
\[
\eta_{\beta,\beta,0,0}\left[G_{\beta,\beta,0,1}\right]=0, \quad
\eta_{\beta,\beta,0,1}\left[G_{\beta,\beta,0,1}\right]
=-\frac{1}{2}\eta_{\beta,\beta,0,0}\left[G_{\beta,\beta,0,2}\right].
\]
\eenum
\elem
\bprf[Proof of \cref{lem:regularity}]

\benum
\item By \cref{assump:regular-weights}, for all $x,x'\in \mcX$ we have $U_{\beta,\beta'}(x,x')$, $G_{\beta,\beta'}(x,x')$, and $H_{\beta,\beta'}(x,x')$ are in $C^3([0,1]^2)$. Let $\textbf{i}=(i,i')$ with $\textbf{i}=i+i'$, we are done if we can show that $\bar{U}_\textbf{i},\bar G_\textbf{i}$ and $\bar{H}_\textbf{i}$ are in $\mcF_{3-|\textbf{i}|}$ for $|\textbf{i}|\leq 3$. Note that,
\[
\bar{U}_\textbf{i}\leq (1+\bar{G})^2\bar{U}_{|\textbf{i}|}\in \mcF_{3-i}.
\]
Since $\mcF_{3-k}$ is closed under domination, we have $\bar{U}_\textbf{i}\in \mcF_{3-k}$. 

Since $G_{\beta,\beta'}=\exp(U_{\beta,\beta'}), H_{\beta,\beta'}=\exp(2U_{\beta,\beta'})$, 
\[
G_{\beta,\beta',\textbf{i}}
&=\sum_{|\textbf{i}_1|+\dots+|\textbf{i}_j|=|\textbf{i}|} C_G(\textbf{i}_1,\dots,\textbf{i}_j) G_{\beta,\beta'}U_{\beta,\beta',\textbf{i}_1}\cdots U_{\beta,\beta',\textbf{i}_j}\\
H_{\beta,\beta',\textbf{i}}
&=\sum_{|\textbf{i}_1|+\dots+|\textbf{i}_j|=|\textbf{i}|} C_H(\textbf{i}_1,\dots,\textbf{i}_j)H_{\beta,\beta'}U_{\beta,\beta',\textbf{i}_1}\cdots U_{\beta,\beta',\textbf{i}_j}
\]
where $C_G(\textbf{i}_1,\dots,\textbf{i}_j)$ and $C_H(\textbf{i}_1,\dots,\textbf{i}_j)\geq 0$ are constants, and $\textbf{i}_1=(i_1,i_1'),\dots, \textbf{i}_j=(i_j,i_j')$ are multi-indices.
Therefore we have,
\[
\bar{G}_\textbf{i}
&\leq \sum_{|\textbf{i}_1|+\dots+|\textbf{i}_j|=|\textbf{i}|} C_G(\textbf{i}_1,\dots,\textbf{i}_j) \bar{G}\bar{U}_{\textbf{i}_1}\cdots \bar{U}_{\textbf{i}_j}\\
&\leq \sum_{|\textbf{i}_1|+\dots+|\textbf{i}_j|=|\textbf{i}|} C_G(\textbf{i}_1,\dots,\textbf{i}_j) (1+\bar{G})^2\bar{U}_{|\textbf{i}_1|}\cdots \bar{U}_{|\textbf{i}_j|}.
\]
Similarly $\bar{\textbf{H}_\textbf{i}}$ satisfies
\[
\bar{H}_\textbf{i}
&\leq \sum_{|\textbf{i}_1|+\dots+|\textbf{i}_j|=|\textbf{i}|} C_G(\textbf{i}_1,\dots,\textbf{i}_j) \bar{G}^2\bar{U}_{\textbf{i}_1}\cdots \bar{U}_{\textbf{i}_j}\\
&\leq \sum_{|\textbf{i}_1|+\dots+|\textbf{i}_j|=|\textbf{i}|} C_G(\textbf{i}_1,\dots,\textbf{i}_j) (1+\bar{G})^2\bar{U}_{|\textbf{i}_1|}\cdots \bar{U}_{|\textbf{i}_j|}.
\]
Since $\mcF_{3-k}$ is closed under domination, we have $\bar{G}_\textbf{i}$ and $H_{\textbf{i}}\in\mcF_{3-|\textbf{i}|}$.

\item  Let $\beta$ be fixed. By \cref{lem:regularity}(a) and (b) we have $\beta'\mapsto \eta_{\beta,\beta'}$ and $\beta'\mapsto G_{\beta,\beta'}$ are in $C^3([0,1],\mcF)$, and hence by \cref{lem:product-rule}, $\beta'\mapsto \eta_{\beta,\beta'}[G_{\beta,\beta'}]$ is in $C^3([0,1])$. Since $G_{\beta,\beta'}$ is the Radon--Nikodym derivative between the backward target $\pi_{\beta'}\otimes L_{\beta,\beta'}$ and the forward proposal $\eta_{\beta,\beta'}=\pi_{\beta}\otimes M_{\beta,\beta'}$, for all $\beta,\beta'$, $\eta_{\beta,\beta'}[G_{\beta,\beta'}]=1$. By taking the first and second derivatives of both sides and applying \cref{lem:product-rule}, we obtain the following identities,
\[
0&=\eta_{\beta,\beta',0,0}[G_{\beta,\beta',0,1}]+\eta_{\beta,\beta',0,1}[G_{\beta,\beta'}]\\
0&=\eta_{\beta,\beta',0,0}[G_{\beta,\beta',0,2}]
+ 2\eta_{\beta,\beta',0,1}[G_{\beta,\beta',0,1}]
+\eta_{\beta,\beta',0,2}[G_{\beta,\beta'}].
\]
By setting $\beta'=\beta$ and using $G_{\beta,\beta}=1$,
\[
0&=\eta_{\beta,\beta,0,0}[G_{\beta,\beta',0,1}]+\eta_{\beta,\beta,0,1}[1]\\
0&=\eta_{\beta,\beta,0,0}[G_{\beta,\beta',0,2}]
+ 2 \eta_{\beta,\beta,0,1}[G_{\beta,\beta',0,1}]
+\eta_{\beta,\beta,0,2}[1].
\]
We have $\eta_{\beta,\beta,0,k}[1]=0$ for $k\geq 1$ since,
\[
\eta_{\beta,\beta',0,k}[1]
=\frac{\partial^k}{\partial\beta'^k}\eta_{\beta,\beta'}[1]
=\frac{\partial^k}{\partial\beta'^k}1
=0.
\]

\eenum
\eprf

\bprf[Proof of \cref{thm:incremental_discrepancy_estimate}]
We have $\beta,\beta'\mapsto H_{\beta,\beta'}$ and $\beta,\beta'\mapsto \eta_{\beta,\beta'}$ are $C^3([0,1]^2,\mcF)$ by \cref{lem:regularity} (a) and (b) respectively. Therefore, by \cref{lem:product-rule}, the incremental discrepancy $\beta,\beta'\mapsto D(\beta,\beta')=\log(\eta_{\beta,\beta'}[H_{\beta,\beta'}])$ is in $C^3([0,1]^2)$. Moreover, for $i=0,\dots,3$,
\[\label{eq:chisquare-derivative}
\frac{\partial^i}{\partial\beta'^i}\eta_{\beta,\beta'}[H_{\beta,\beta'}]
=\sum_{j=0}^i {i\choose j}\eta_{\beta,\beta',0,j}[H_{\beta,\beta',0,i-j}].
\]
Notably for $\beta'=\beta$, we have using \cref{assump:non-degeneracy},
\[
H_{\beta,\beta,0,0}(x,x')&=1\\
H_{\beta,\beta,0,1}(x,x')&=2G_{\beta,\beta,0,1}(x,x')\\
H_{\beta,\beta,0,2}(x,x')&= 2G_{\beta,\beta,0,1}(x,x')^2 +2G_{\beta,\beta,0,2}(x,x').
\]
We will now compute the first and second derivatives of $D(\beta,\beta')$ with respect to $\beta'$. The first derivative equals,
\[
\frac{\partial}{\partial\beta'}D(\beta,\beta')
&=\frac{1}{\eta_{\beta,\beta'}[H_{\beta,\beta'}]}\frac{\partial}{\partial\beta'}\eta_{\beta,\beta'}[H_{\beta,\beta'}]\\
&=\frac{\eta_{\beta,\beta',0,0}[H_{\beta,\beta',0,1}]+\eta_{\beta,\beta',0,1}[H_{\beta,\beta',0,0}]}{\eta_{\beta,\beta'}[H_{\beta,\beta'}^2]}.
\]
By setting $\beta'=\beta$ and using \cref{lem:regularity}(c) we have,
\[
\frac{\partial}{\partial\beta'}D(\beta,\beta)
=\frac{\eta_{\beta,\beta,0,0}[G_{\beta,\beta',0,1}]+\eta_{\beta,\beta,0,1}[1]}{\eta_{\beta,\beta}[1]}
=\frac{0+0}{1}=0.
\]
We can similarly compute the second derivative: 
\[
\frac{\partial^2}{\partial\beta'^2}D(\beta,\beta')
&=\frac{1}{\eta_{\beta,\beta'}[H_{\beta,\beta'}]}\frac{\partial^2}{\partial\beta'^2}\eta_{\beta,\beta'}[H_{\beta,\beta'}]-\left(\frac{\partial}{\partial\beta'}D(\beta,\beta')\right)^2\\
&=\frac{\eta_{\beta,\beta',0,0}[H_{\beta,\beta',0,2}]
+2\eta_{\beta,\beta',0,1}[H_{\beta,\beta',0,1}]
+\eta_{\beta,\beta',0,2}[H_{\beta,\beta',0,0}]}{1+\eta_{\beta,\beta'}[H_{\beta,\beta'}]}\\&\quad-\left(\frac{\partial}{\partial\beta'}D(\beta,\beta')\right)^2.
\]
Again, by setting $\beta'=\beta$, using \cref{lem:regularity} combined with $\frac{\partial}{\partial\beta'}D(\beta,\beta)=0$, we get
\[
\frac{\partial^2}{\partial\beta'^2}D(\beta,\beta)
&=\frac{\eta_{\beta,\beta,0,0}[2G_{\beta,\beta,0,1}^2+2G_{\beta,\beta,0,2}]
+2\eta_{\beta,\beta,0,1}[2G_{\beta,\beta,0,1}]
+\eta_{\beta,\beta,0,2}[1]}{1+\eta_{\beta,\beta}[1]}\\
&=\frac{2\eta_{\beta,\beta,0,0}[G_{\beta,\beta,0,1}^2]+2(\eta_{\beta,\beta,0,0}[G_{\beta,\beta,0,2}]
+2\eta_{\beta,\beta,0,1}[G_{\beta,\beta,0,1}])
+0}{1+0}\\
&=2\eta_{\beta,\beta}[G_{\beta,\beta,0,1}^2]\\
&=2\delta(\beta).
\]
The last line follows since $\dot G_\beta=G_{\beta,\beta,0,1}$ and $\eta_{\beta,\beta}[\dot G_\beta]=0$ by \cref{lem:regularity}(c). Finally, by Taylor's theorem,
\[
\left|D(\beta,\beta+\Delta\beta)-\delta(\beta)\Delta\beta^2\right|
&\leq \frac{1}{6}\sup_{\beta'\in[0,1]}\frac{\partial^3}{\partial\beta'^3}D(\beta,\beta')|\Delta\beta|^3\\
&\leq \frac{1}{6}\sup_{\beta,\beta'\in[0,1]}\frac{\partial^3}{\partial\beta'^3}D(\beta,\beta')|\Delta\beta|^3
\]
\eprf

\subsection{Proof of \cref{cor:dense_schedule_limit_discrepancy}}

Let $\beta_t = \varphi(u_t)$ where $u_t = t/T$ for $t \in [T]$. We have $\Delta u_t = u_t - u_{t-1} = 1/T$ and $\Delta\beta_t = \beta_t - \beta_{t-1}$.

\paragraph{Part (a): Estimate for $D(\beta_{t-1},\beta_t)$.}
By the triangle inequality,
\[
\left|D(\beta_{t-1},\beta_t)-\frac{1}{T}\int_{u_{t-1}}^{u_t}\delta(\varphi(u))\dot{\varphi}(u)^2\dee u\right|
\leq  \sum_{i=1}^3\epsilon_i,
\]
where
\[
\epsilon_1&=\left|D(\beta_{t-1},\beta_t)-\delta(\beta_{t-1}) \Delta\beta_t^2\right|,\\
\epsilon_2&=\left|\delta(\beta_{t-1}) \Delta\beta_t^2- \delta(\beta_{t-1})\dot{\varphi}(u_{t-1})^2 \Delta u_t^2\right|,\\
\epsilon_3 &=\left|\delta(\beta_{t-1})\dot{\varphi}(u_{t-1})^2 \Delta u_t^2 - \frac{1}{T}\int_{u_{t-1}}^{u_t}\delta(\varphi(u))\dot{\varphi}(u)^2\dee u\right|.
\]

For $\epsilon_1$, by \cref{thm:incremental_discrepancy_estimate} and Taylor's theorem,
\[
\epsilon_1
\leq\frac{1}{6} \sup_{\beta,\beta'\in[0,1]}\left|\frac{\partial^3 D(\beta,\beta')}{\partial\beta'^3}\right||\Delta\beta_t|^3.
\]
Using $|\Delta\beta_t|\leq \sup_{u\in[0,1]}|\dot{\varphi}(u)||\Delta u_t| = \sup_{u\in[0,1]}|\dot{\varphi}(u)|/T$, we obtain
\[
\epsilon_1 \leq \frac{1}{6} \sup_{\beta,\beta'\in[0,1]}\left|\frac{\partial^3 D(\beta,\beta')}{\partial\beta'^3}\right|\sup_{u\in[0,1]}|\dot{\varphi}(u)|^3 \frac{1}{T^3}.
\]

For $\epsilon_2$, using the continuity of $\delta$,
\[
\epsilon_2
&= \sup_{\beta\in[0,1]}\delta(\beta)\left|\Delta\beta_t^2-\dot{\varphi}(u_{t-1})^2\Delta u_t^2\right|\\
&\leq \sup_{\beta\in[0,1]}\delta(\beta) |\Delta\beta_t+\dot{\varphi}(u_{t-1})\Delta u_t||\Delta\beta_t-\dot{\varphi}(u_{t-1})\Delta u_t|.
\]
By the mean value theorem, $|\Delta\beta_t+\dot{\varphi}(u_{t-1})\Delta u_t|\leq 2\sup_{u\in[0,1]}|\dot{\varphi}(u)|/T$. By Taylor's theorem,
\[
\left|\Delta\beta_t-\dot{\varphi}(u_{t-1}) \Delta u_t\right|\leq \frac{1}{2} \sup_{u\in[0,1]}|\ddot{\varphi}(u)| \frac{1}{T^2}.
\]
Therefore,
\[
\epsilon_2\leq \sup_{\beta\in[0,1]}\delta(\beta)\sup_{u\in[0,1]}|\dot{\varphi}(u)|\sup_{u\in[0,1]}|\ddot{\varphi}(u)| \frac{1}{T^3}.
\]
For $\epsilon_3$, we use the error bound for the left Riemann sum approximating the integral of $\delta(\varphi(u))\dot{\varphi}(u)^2$ from $u_{t-1}$ to $u_t$ with one rectangle. The function $s \mapsto \delta(\varphi(u))\dot{\varphi}(u)^2$ is continuously differentiable, giving
\[
\epsilon_3
&\leq \frac{1}{2}\sup_{u\in[0,1]}\left|\frac{\dee}{\dee u}\left[\delta(\varphi(u))\dot{\varphi}(u)^2\right]\right||\Delta u_t|^2 \\
&= \frac{1}{2}\sup_{u\in[0,1]}\left( |\dot{\delta}(\varphi(u))||\dot{\varphi}(u)|^3+2\delta(\varphi(u))|\dot{\varphi}(u)||\ddot{\varphi}(u)|\right)\frac{1}{T^3}.
\]

Combining these three estimates yields
\[
\left|D(\beta_{t-1},\beta_t)-\frac{1}{T}\int_{u_{t-1}}^{u_t}\delta(\varphi(u))\dot{\varphi}(u)^2\dee u\right|\leq \frac{C_\delta}{T^3},
\]
where $C_\delta>0$ depends only on the suprema of derivatives of $\delta$, $\varphi$, and compositions thereof, but is independent of $t$.

\paragraph{Part (b): Estimate for $\sqrt{D(\beta_{t-1},\beta_t)}$.}
By the triangle inequality,
\[
\left|\sqrt{D(\beta_{t-1},\beta_t)}-\int_{\beta_{t-1}}^{\beta_t} \lambda(\beta)\dee \beta\right|
\leq  \epsilon_1+\epsilon_2,
\]
where
\[
\epsilon_1&=\left|\sqrt{D(\beta_{t-1},\beta_t)}-\lambda(\beta_{t-1})\Delta\beta_t\right|,\\
\epsilon_2&=\left|\lambda(\beta_{t-1})\Delta\beta_t-\int_{\beta_{t-1}}^{\beta_t} \lambda(\beta)\dee \beta\right|.
\]

For $\epsilon_1$, we write
\[
\epsilon_1
=\frac{|D(\beta_{t-1},\beta_t)-\lambda(\beta_{t-1})^2\Delta\beta_t^2|}{\sqrt{D(\beta_{t-1},\beta_t)}+\lambda(\beta_{t-1})\Delta\beta_t}.
\]
By Taylor's theorem and \cref{thm:incremental_discrepancy_estimate}, the numerator satisfies
\[
|D(\beta_{t-1},\beta_t)-\lambda(\beta_{t-1})^2\Delta\beta_t^2|\leq \frac{1}{6} \sup_{\beta,\beta'\in[0,1]}\left|\frac{\partial^3 D(\beta,\beta')}{\partial\beta'^3}\right||\Delta\beta_t|^3.
\]
Since $\lambda(\beta)=\sqrt{\delta(\beta)}$ is continuous and strictly positive by \cref{thm:incremental_discrepancy_estimate,assump:non-degeneracy}, the denominator satisfies
\[
\sqrt{D(\beta_{t-1},\beta_t)}+\lambda(\beta_{t-1})\Delta\beta_t\geq \inf_{\beta\in[0,1]}\lambda(\beta)|\Delta\beta_t|>0.
\]
Therefore,
\[
\epsilon_1\leq \frac{1}{6}\sup_{\beta,\beta'\in[0,1]}\left|\frac{\partial^3 D(\beta,\beta')}{\partial\beta'^3}\right|\frac{1}{\inf_{\beta\in[0,1]}\lambda(\beta)}|\Delta\beta_t|^3.
\]

For $\epsilon_2$, we use the error bound for the left Riemann sum approximation of $\int_{\beta_{t-1}}^{\beta_t}\lambda(\beta)\dee\beta$ with one rectangle. Since $\lambda(\beta)=\sqrt{\delta(\beta)}$ is continuously differentiable,
\[
\epsilon_2\leq \frac{1}{2}\sup_{\beta\in[0,1]}\dot{\lambda}(\beta)|\Delta\beta_t|^2=\frac{1}{2}\sup_{\beta\in[0,1]}\frac{|\dot{\delta}(\beta)|}{2\sqrt{\delta(\beta)}}|\Delta\beta_t|^2.
\]

By the mean value theorem and \cref{assump:schedule_generator}, $|\Delta\beta_t| = |\dot{\varphi}(\tilde{u})||\Delta u_t|$ for some $\tilde{u} \in [u_{t-1}, u_t]$, hence $|\Delta\beta_t| \leq \sup_{u\in[0,1]}|\dot{\varphi}(u)|/T$. Combining the bounds for $\epsilon_1$ and $\epsilon_2$ yields
\[
\left|\sqrt{D(\beta_{t-1},\beta_t)}-\int_{\beta_{t-1}}^{\beta_t}\lambda(\beta)\dee\beta\right|\leq C_\lambda |\Delta\beta_t|^2 \leq \frac{C_\lambda}{T^2},
\]
where $C_\lambda>0$ depends only on the suprema of derivatives of $\delta$, $\lambda$, and $\varphi$, but is independent of $t$.

\subsection{Proof of \cref{thm:dense_limit_variance}}
For each $T$, by \cref{thm:variance_ERS}(b) there exists $\rho_T\in[1,T]$ such that 
\[
\var\left[\frac{\hZ}{Z}\right]=\left(1+\frac{\exp(D(\mcB_T)/\rho_T)-1}{N}\right)^{\rho_T}-1.
\]
Since the variance is monotonically decreasing in $\rho_T$, we obtain the following upper and lower bounds independent of $\rho_T$:
\[
\left(1+\frac{\exp(D(\mcB_T)/T)-1}{N}\right)^{T}-1\leq\var\left[\frac{\hZ}{Z}\right]\leq \frac{\exp(D(\mcB_T))-1}{N}.
\]
By \cref{cor:dense_schedule_limit_discrepancy}, we have $\lim_{T\to\infty}TD(\mcB_T)=E(\varphi)$.
Using $\exp(d)\sim 1+d$ as $d\to 0$ and $(1+d/T)^T\sim 1+d$ as $T\to\infty$, we obtain
\[
\liminf_{T\to\infty}T\var\left[\frac{\hZ}{Z}\right]
&\geq \lim_{T\to\infty}T\left[\left(1+\frac{\exp(D(\mcB_T)/T)-1}{N}\right)^{T}-1\right]\\
&= \lim_{T\to\infty}T\left[\left(1+\frac{D(\mcB_T)}{TN}\right)^{T}-1\right]\\
&= \lim_{T\to\infty}T\left[1+\frac{D(\mcB_T)}{N}-1\right]\\
&=\lim_{T\to\infty}\frac{TD(\mcB_T)}{N}\\
&=\frac{E(\varphi)}{N},
\]
and similarly,
\[
\limsup_{T\to\infty}T\var\left[\frac{\hZ}{Z}\right]
&\leq \lim_{T\to\infty}T\left(\frac{\exp(D(\mcB_T))-1}{N}\right)\\
&=\lim_{T\to\infty}\frac{TD(\mcB_T)}{N}\\
&=\frac{E(\varphi)}{N}.
\]
Therefore the limit exists and equals uniformly in the resampling times,
\[
\lim_{T\to\infty}T\var\left[\frac{\hZ}{Z}\right]=\frac{E(\varphi)}{N}.
\]
By Jensen's inequality, using the substitution $\beta=\varphi(u)$, we have
\[
E(\varphi)
&=\int_0^1\delta(\varphi(u))\dot\varphi(u)^2\dee u\\
&\geq \left(\int_0^1\lambda(\varphi(u))\dot\varphi(u)\dee u\right)^2\\
&=\left(\int_0^1\lambda(\beta)\dee\beta\right)^2\\
&=\Lambda^2,
\]
where the second line uses $\lambda(\beta)=\sqrt{\delta(\beta)}$ and the third uses $\varphi(0)=0$ and $\varphi(1)=1$.
We have 
\[
E(\varphi)=\Lambda^2\quad
&\Longleftrightarrow \quad
\delta(\varphi(u))\dot\varphi(u)^2=\Lambda^2\\
&\Longleftrightarrow \quad
\lambda(\varphi(u))\dot\varphi(u)=\Lambda\\
&\Longleftrightarrow \quad
\int_0^u\lambda(\varphi(u'))\dot\varphi(u')\dee u'=\Lambda u\\
&\Longleftrightarrow \quad
\int_0^{\varphi(u)}\lambda(\beta)\dee\beta=\Lambda u\\
&\Longleftrightarrow \quad
\Lambda(\varphi(u))=\Lambda u\\
&\Longleftrightarrow \quad
\varphi(u)=\Lambda^{-1}(\Lambda u),
\]
where we recall that $\Lambda(\beta)=\int_0^\beta\lambda(\beta')\dee\beta'$. Since $\Lambda$ is strictly increasing with $\Lambda(0)=0$ and $\Lambda(1)=\Lambda$, we conclude that $\varphi^*(u)=\Lambda^{-1}(\Lambda u)$ is strictly increasing with $\varphi^*(0)=0$ and $\varphi^*(1)=1$.
By \cref{thm:incremental_discrepancy_estimate}, since $\delta(\beta)$ is positive and continuously differentiable, $\lambda(\beta)=\sqrt{\delta(\beta)}$ is also positive and continuously differentiable. Hence $\Lambda(\beta)$ is twice continuously differentiable, which implies $\varphi^*$ is as well.

\subsection{Asymptotic equivalence of $Z$ estimation variance and $\log Z$ estimation bias}\label{sec:bias}
For any schedule $\mcB_T=\beta_{0:T}$, we have $\log Z=\ELBO(\mcB_T)+\KL(\mcB_T)$, where $\ELBO(\mcB_T)$ is the \emph{evidence lower bound} and $\KL(\mcB_T)$ is the KL divergence between the forward proposal and backward target accumulated over $\mcB_T$:
\[
\ELBO(\mcB_T) = \sum_{t \in [T]} \E_{\beta_{t-1}, \beta_t}[\log g_{\beta_{t-1}, \beta_t}],\quad  
\KL(\mcB_T)=-\sum_{t\in[T]}\E_{\beta_{t-1},\beta_t}[\log G_{\beta_{t-1},\beta_t}].
\]
We can approximate the ELBO consistently as $N\to\infty$ using the output of \cref{alg:AMCS} via the estimator $\frac{1}{N}\sum_{n \in [N]} \sum_{t \in [T]} W^n_{t-1} \log g^n_t$.

It is common in ASMC to choose the annealing schedule $\mcB_T$ to minimise $\KL(\mcB_T)$ rather than the asymptotic variance of $\hZ$ \citep{neal_annealed_2001,grosse2013annealing,arbel2021annealed}. Under our regularity assumptions, \cref{prop:bias} shows that $\KL(\mcB_T)\sim E(\varphi)/(2T)$ decays to zero as $T \to \infty$, generalising \citet[Theorem 1]{grosse2013annealing}. Comparing to \cref{eq:dense_limit_formula}, we see that minimising the variance of $\hZ$ is equivalent to minimising $\KL(\mcB_T)$ in the dense schedule limit.

\bprop\label{prop:bias}
Suppose \cref{assump:regular-proposal,assump:regular-weights,assump:non-degeneracy,assump:schedule_generator} hold. There exists $C_\KL>0$ such that for all $T\in\nats$,
\[
\left|\KL(\mcB_T)-\frac{E(\varphi)}{2T}\right|\leq \frac{C_\KL}{T^2}.
\]
\eprop

\bprf
For $\beta,\beta'\in[0,1]$, let $\KL(\beta,\beta')$ denote the KL divergence between the forward proposal $\pi_{\beta}\otimes M_{\beta,\beta'}$ and backward target $\pi_{\beta'}\otimes L_{\beta',\beta}$:
\[
\KL(\beta,\beta')=-\E_{\beta,\beta'}[\log G_{\beta,\beta'}].
\]
By \cref{lem:regularity}(a) and (b), both $\eta_{\beta,\beta'}$ and $U_{\beta,\beta'}=-\log G_{\beta,\beta'}$ are in $C^3([0,1]^2,\mcF)$. Therefore, by \cref{lem:product-rule}, $\KL(\beta,\beta')$ is $C^3([0,1]^2)$ with derivatives
\[\label{eq:KL-derivative}
\frac{\partial^i}{\partial\beta'^i}\KL(\beta,\beta')
=-\sum_{j=0}^i {i\choose j}\eta_{\beta,\beta',0,j}[U_{\beta,\beta',0,i-j}]
\]
for $i=0,\dots, 3$. Recall,
\[
U_{\beta,\beta',i,j}(x,x')=\frac{\partial^{i+i'}}{\partial\beta^i\partial\beta'^{i'}}U_{\beta,\beta'}(x,x').
\]
We now compute the derivatives at $\beta'=\beta$. By \cref{lem:regularity}, we have
\[
U_{\beta,\beta}(x,x')&=0,\\
U_{\beta,\beta,0,1}(x,x')&=G_{\beta,\beta,0,1}(x,x'),\\
U_{\beta,\beta,0,2}(x,x')&=G_{\beta,\beta,0,2}(x,x')-G_{\beta,\beta,0,1}(x,x')^2.
\]
For the first derivative of $\KL(\beta,\beta')$ at $\beta'=\beta$,
\[
\frac{\partial}{\partial\beta'}\KL(\beta,\beta)
&=-\eta_{\beta,\beta}[U_{\beta,\beta,0,1}]-\eta_{\beta,\beta,0,1}[U_{\beta,\beta}]\\
&=-\eta_{\beta,\beta}[G_{\beta,\beta,0,1}]-0=0,
\]
where we used $\eta_{\beta,\beta}[G_{\beta,\beta,0,1}]=0$ from \cref{lem:regularity}.
For the second derivative,
\[
\frac{\partial^2}{\partial\beta'^2}\KL(\beta,\beta)
&=-\eta_{\beta,\beta}[U_{\beta,\beta,0,2}]
-2\eta_{\beta,\beta,0,1}[U_{\beta,\beta,0,1}]
-\eta_{\beta,\beta,0,2}[U_{\beta,\beta}]\\
&=-\eta_{\beta,\beta}[G_{\beta,\beta,0,2}-G_{\beta,\beta,0,1}^2]
-2\eta_{\beta,\beta,0,1}[G_{\beta,\beta,0,1}]\\
&=\eta_{\beta,\beta}[G_{\beta,\beta,0,1}^2]
=\delta(\beta),
\]
where we used $\eta_{\beta,\beta,0,1}[G_{\beta,\beta,0,1}]=0$ and the definition of $\delta(\beta)$.

We have shown that $\KL(\beta,\beta')$ is $C^3([0,1]^2)$ and satisfies
\[
\KL(\beta,\beta)=0,\quad \frac{\partial}{\partial\beta'}\KL(\beta,\beta)=0,\quad \frac{\partial^2}{\partial\beta'^2}\KL(\beta,\beta)=\delta(\beta).
\]
Since the cumulative KL divergence satisfies $\KL(\mcB_T)=\sum_{t\in[T]} \KL(\beta_{t-1},\beta_t)$, following proof of \cref{cor:dense_schedule_limit_discrepancy} applies verbatim with $\KL(\beta,\beta')$ instead of $D(\beta,\beta')$ to yield
\[
\left|\KL(\mcB_T)-\frac{1}{T}E(\varphi)\right|\leq \frac{C_\KL}{T^2},
\]
for some $C_\KL>0$ independent of $T$. 
\eprf

\subsection{Analysis of forward/backward kernels} 
\label{sec:analysis-forward-backward}
\subsubsection{Regularity class for forward proposal}
Let $\pi_\beta$ be the annealing distribution and $M_{\beta,\beta'}$ be the forward proposal. Our goal is to identify sufficient conditions and a regularity class $\mcF$ such that the forward proposal on $\eta_{\beta,\beta'}=\pi_\beta\otimes M_{\beta,\beta'}$ is $C^k([0,1]^2,\mcF)$.

\paragraph{Prior regularity class}\label{sec:prior_class}
For all $x\in\mcX$, suppose $\beta\mapsto \pi_\beta(x)$ is in $C^3([0,1])$ and strictly positive. Define $\bar\pi:\mcX\mapsto\reals_+$ as follows:
\[
 \bar{\pi}(x)=\sup_{\beta\in[0,1]} \pi_\beta(x),
\]
and for $i\leq k$ define:
\[
V_{\beta,i}(x)&=\frac{\dee^i}{\dee\beta^i}\log \pi_\beta(x),\quad \bar{V}_i(x)=\sup_{\beta\in[0,1]}|V_{\beta,i}(x)|.
\]
Define the \emph{prior regularity class} $\mcF^\pre=(\mcF^\pre_i)_{i\in\nats}$ as follows: for $i\leq k$, let $\mcF^\pre_i$ be the set of functions $f:\mcX\times \mcX\mapsto\reals$ such that $f(x,x')=\tilde f(x)$ for some $\tilde{f}:\mcX\mapsto\reals$ and for all $i_1,\dots,i_j\geq 1$ with $i_1+\dots+i_j\leq i$,
\[
 \int_{\mcX}\tilde{f}(x)\bar{V}_{i_1}(x)\cdots \bar{V}_{i_j}(x)\bar{\pi}(x)\dee x<\infty.
\]
For $i>k$ let $\mcF_i$ be the set of constant functions.

\blem\label{lem:prior_class}
If for all $x\in\mcX$, the mapping $\beta\mapsto \pi_\beta(x)$ is in $C^3([0,1])$ and strictly positive, then $\beta,\beta'\mapsto \eta_{\beta,\beta'}$ is in $C^3([0,1]^2,\mcF^\pre)$.
\elem
\bprf[Proof of \cref{lem:prior_class}]
Suppose $i\leq k$ and $f\in \mcF^\pre_i$, with $f(x,x')=\tilde{f}(x,x')$ we have for all $\beta,\beta'$,
\[
\eta_{\beta,\beta'}[f]=\pi_\beta[\tilde{f}].
\]
By \cref{lem:regularity-measures}, the mapping $\beta\mapsto \pi_{\beta}[\tilde{f}]$ is $C^i([0,1])$, and
\[
\frac{\partial^i}{\partial\beta^i}\eta_{\beta,\beta'}[f]
=\frac{\dee^i}{\dee\beta^i}\pi_\beta[\tilde{f}]
=\pi_{\beta,i}[\tilde f]
=\eta_{\beta,\beta',i,0}[f],
\]
where $\eta_{\beta,\beta',i,0}$ is a signed measure on $\mcX\times\mcX$ with density
\[
\eta_{\beta,\beta',i,0}(\dee x,\dee x')=\frac{\dee^i}{\dee\beta^i}\pi_\beta(x)\dee x \dee x'.
\]
Since $\eta_{\beta,\beta'}[f]$ is independent of $\beta'$, we have for all $i\in\nats$ and $i'\geq 1$, 
\[
\frac{\partial^{i+i'}}{\partial\beta^i\partial\beta'^{i'}}\eta_{\beta,\beta'}[f]=0=\eta_{\beta,\beta,i,i'}[f],
\]
where $\eta_{\beta,\beta',i,i'}=0$ is the trivial measure.
\eprf
\paragraph{Posterior regularity class}\label{sec:posterior_class}
For $\beta,\beta'$ define $m_{\beta,\beta'}\in \mcM(\mcX)$ as 
\[
m_{\beta,\beta'}(\dee x')=\int_\mcX \pi_\beta(x)M_{\beta,\beta'}(x,\dee x')\in \mcM(\mcX).
\]
For all $x'\in\mcX$, suppose $\beta,\beta'\mapsto m_{\beta,\beta'}(x')$ is in $C^3([0,1]^2)$ and strictly positive with density $m_{\beta,\beta'}(x')$ with respect to $\dee x'$. For all $x'\in\mcX$, define $\bar m:\mcX\mapsto\reals_+$ as
\[
 \bar m(x') =\sup_{\beta,\beta'\in[0,1]} m_{\beta,\beta'}(x'),
\]
and for $\textbf{i}=(i,i')\leq k$ with $\textbf{i}=i+i'\leq k$, define $W_{\beta,\beta',\textbf{i}},\bar{W}_{\beta,\beta',\textbf{i}}:\mcX\mapsto \reals$ as
\[
W_{\beta,\beta',\textbf{i}}(x')&=\frac{\partial^{i+i'}}{\partial\beta^i\partial\beta'^{i'}}\log m_{\beta,\beta'}(x'),\quad \bar{W}_\textbf{i}(x')=\sup_{\beta\in[0,1]}|W_{\beta,\beta',\textbf{i}}(x')|.
\]
Define the \emph{posterior regularity class} as $\mcF^\post=(\mcF^\post_i)_{i\in\nats}$ as follows: for $i\leq k$, let $\mcF^\post_i$ be the set of functions $f:\mcX\times \mcX\mapsto\reals$ such that $f(x,x')=\tilde f(x')$ for some $\tilde{f}:\mcX\mapsto\reals$ and for all $\textbf{i}_1=(i_1,i_1'),\dots,\textbf{i}_j=(i_j,i_j')$ with $|\textbf{i}_1,\dots,|\textbf{i}_j|\geq 1$ with $|\textbf{i}_1|+\dots+|\textbf{i}_j|\leq i$,
\[
 \int_{\mcX}\tilde{f}(x')\bar{W}_{\textbf{i}_1}(x')\cdots \bar{W}_{\textbf{i}_j}(x')\bar{m}(x')\dee x'<\infty.
\]
For $i>k$ let $\mcF_i$ be the set of constant functions.

\blem\label{lem:posterior_class}
If for all $x'\in\mcX$, $\beta,\beta'\mapsto m_{\beta,\beta'}(x')$ is in $C^3([0,1])$ and strictly positive, then $\beta,\beta'\mapsto \eta_{\beta,\beta'}$ is in $C^k([0,1]^2,\mcF^\post)$.
\elem
\bprf[Proof of \cref{lem:posterior_class}]
Suppose $i\leq k$ and $f\in \mcF^\pre_i$, with $f(x,x')=\tilde{f}(x,x')$ we have for all $\beta,\beta'$,
\[
\eta_{\beta,\beta'}[f]=m_{\beta,\beta'}[\tilde{f}].
\]
By \cref{lem:regularity-measures}, $\beta,\beta'\mapsto \pi_{\beta}[\tilde{f}]$ is $C^i([0,1]^2)$, and
\[
\frac{\partial^{i+i'}}{\partial\beta^i\partial\beta'^{i'}}\eta_{\beta,\beta'}[f]
=\frac{\partial^{i+i'}}{\partial\beta^i\partial\beta'^{i'}}m_{\beta,\beta'}[\tilde{f}]
=m_{\beta,\beta',i,i'}[\tilde{f}]
=\eta_{\beta,\beta',i,i}[f],
\]
where $\eta_{\beta,\beta',i,0}$ is a signed measure on $\mcX\times\mcX$ with density,
\[
\eta_{\beta,\beta',i,i}(\dee x,\dee x')=\frac{\partial^{i+i'}}{\partial\beta^i\partial\beta'}m_{\beta,\beta'}(x')\dee x \dee x'.
\]
\eprf

\subsubsection{Invariant/MCMC kernels}\label{ex:kernel_invariant}

Suppose $M_{\beta,\beta'}(x,\dee x')=K_{\beta'}(x,\dee x')$, where for all $\beta\in[0,1]$, $K_{\beta'}$ is a $\pi_{\beta'}$-invariant Markov kernel corresponding to an MCMC move. The canonical choice of backward kernel $L_{\beta',\beta}=K^*_{\beta'}$ is the quasi-reversal of $K_{\beta'}$. The incremental weight $G_{\beta,\beta'}(x,x')$ reduces to the Radon--Nikodym derivative between $\pi_{\beta'}$ and $\pi_\beta$ and does not depend on $x'$:
\[ \label{def:incremental_weight_invariant}
G_{\beta,\beta'}(x,x')=\frac{\pi_{\beta'}(x)}{\pi_\beta(x)}.
\]
The incremental discrepancy equals the Rényi $2$-divergence between $\pi_\beta$ and $\pi_{\beta'}$, independent of the forward kernel:
\[
D(\beta,\beta')=D(\pi_{\beta},\pi_{\beta'}),
\]
where the Rényi $2$-divergence between probability distributions $p\ll q$ with Radon-Nikodym derivative $\dee p/\dee q$ is
\[\label{def:discrepancy_renyi}
D(q,p)=\log\left(1+\var_q\left[\frac{\dee p}{\dee q}\right]\right).
\]

Provided $\beta\mapsto \pi_\beta(x)$ is strictly positive and continuously differentiable in $\beta$, the function $\dot{G}_{\beta}$ equals the Fisher score function for $\pi_\beta$, independent of $x'$:
\[\label{eq:S-invariant}
\dot{G}_{\beta}(x,x')=\frac{\dee}{\dee\beta}\log \pi_\beta(x).
\]
Taking expectations of $\dot G_\beta$ with respect to $\pi_{\beta}\otimes M_{\beta,\beta}$, the local discrepancy $\delta(\beta)=\var_{\beta,\beta}[\dot G_{\beta}]$ equals the Fisher information for the annealing distributions:
\[
\delta(\beta)=\var_{\pi_\beta}\left[\frac{\dee}{\dee\beta}\log \pi_\beta\right].
\]

\cref{prop:regularity-MCMC} provides sufficient conditions for \cref{assump:regular-proposal,assump:regular-weights} to hold.

\bprop\label{prop:regularity-MCMC}
Suppose for all $x\in \mcX$, $\beta\mapsto \pi_\beta(x)$ is strictly positive and $C^3([0,1])$. Let $M_{\beta,\beta'}=K_{\beta'}$ and $L_{\beta',\beta}=K_{\beta'}^*$ be the MCMC kernels, and let $\mcF^\text{pre}$ be the prior class constructed in \cref{sec:prior_class}. If $(1+\bar{G})^2\in \mcF^\text{pre}_3$, then \cref{assump:regular-proposal,assump:regular-weights} are satisfied. 
\eprop

\bprf
\cref{assump:regular-proposal} follows from \cref{lem:prior_class}, so we only need to verify \cref{assump:regular-weights}. Let $U_{\beta,\beta'}(x,x')=\log G_{\beta,\beta'}(x,x')$, and for $\textbf{i}=(i,i')$ define
\[
U_{\beta,\beta',\textbf{i}}(x,x')=\frac{\partial^{i+i'}}{\partial\beta^i\partial\beta'^{i'}}\log G_{\beta,\beta'}(x,x'),\quad \bar{U}_{\textbf{i}}(x,x')=\sup_{\beta,\beta'\in[0,1]}\left|U_{\beta,\beta',\textbf{i}}(x,x')\right|.
\]
Since $G_{\beta,\beta'}(x,x')=\pi_{\beta'}(x)/\pi_\beta(x)$ and $\beta\mapsto\pi_\beta(x)$ is positive and $C^3([0,1])$, we have
\[
U_{\beta,\beta',\textbf{i}}(x,x')
=\frac{\partial^{i+i'}}{\partial\beta^i\partial\beta'^{i'}}\log \pi_{\beta'}(x)-\frac{\partial^{i+i'}}{\partial\beta^i\partial\beta'^{i'}}\log\pi_\beta(x).
\]
For $i=0,\dots,3$, we have $U_{\beta,\beta',i,0}(x,x')=-V_{\beta,i}(x)$, $U_{\beta,\beta',0,i}(x,x')=V_{\beta',i}(x)$, and when $i,i'\geq 1$, $U_{\beta,\beta',i,i'}(x,x')=0$. In all cases, for $|\textbf{i}|\leq 3$, $\bar{U}_\textbf{i}(x,x')$ is independent of $x'$ and $\bar{U}_\textbf{i}(x,x')\leq \bar{V}_{|\textbf{i}|}(x)$. 

Therefore, for all $k=0,\dots,3$ and $\textbf{i}_1=(i_1,i_1'),\dots, \textbf{i}_j=(i_j,i_j')$ with $|\textbf{i}_1|+\dots+|\textbf{i}_j|\leq k$, the product $(1+\bar{G}(x,x'))^2\bar{U}_{\textbf{i}_1}(x,x')\cdots \bar{U}_{\textbf{i}_j}(x,x')$ is independent of $x'$ and 
\[
(1+\bar{G}(x,x'))^2\bar{U}_{\textbf{i}_1}(x,x')\cdots \bar{U}_{\textbf{i}_j}(x,x')\leq (1+\bar{G}(x,x'))^2\bar{V}_{|\textbf{i}_1|}(x)\cdots \bar{V}_{|\textbf{i}_j|}(x).
\]
Since $(1+\bar{G})^2\in\mcF^\text{pre}_3$, by construction, $(1+\bar{G})^2\bar{V}_{|\textbf{i}_1|}\cdots \bar{V}_{|\textbf{i}_j|}\in \mcF^{\text{pre}}_{3-k}$. Since $\mcF^\text{pre}_{3-k}$ is closed under pointwise products, $(1+\bar{G})^2\bar{U}_{\textbf{i}_1}\cdots \bar{U}_{\textbf{i}_j}\in \mcF^\text{pre}_{3-k}$.
\eprf

\paragraph{MCMC kernels with Linear path.}\label{sec:linear-example}
Suppose $\pi_\beta= \eta^{1-\beta}\pi^\beta/Z_\beta$ is the linear path. Then the annealing distributions form an exponential family with natural parameter $\beta$, sufficient statistic $V(x)=\log \pi(x)-\log \eta(x)$, and log-partition function $A(\beta)=\log Z_\beta$:
\[\label{eq:linear-path-exponential-family}
\pi_\beta(x)=\eta(x)\exp(\beta V(x)-A(\beta)).
\]
The incremental weight equals
\[
G_{\beta,\beta'}(x,x')=\exp((\beta'-\beta)V(x)-A(\beta')+A(\beta).
\]
By \citet[Lemma 1]{nielsen2013chi}, $D(\beta,\beta')$ equals the discrete second derivative of $A(\beta)$ at $\beta'$ with step size $\Delta\beta=\beta'-\beta$:
\[
D(\beta,\beta')=A(\beta'+\Delta\beta)+A(\beta'-\Delta\beta)-2 A(\beta').
\]
The function $\dot G_{\beta}$ reduces to the centered log-density of $\pi_\beta$:
\[
\dot G_\beta(x,x')=V(x)-\dot{A}(\beta)
=V(x)-\E_{\beta}[V],
\]
and $\delta(\beta)$ equals the variance of $V$ with respect to $\pi_\beta$:
\[
\delta(\beta)=\ddot{A}(\beta)=\var_{\beta}[V].
\]

\bprop\label{prop:regularity-linear}
Suppose the forward/backward kernels are MCMC kernels and $\pi_\beta$ is the linear path between $\eta$ and $\pi$. If $(1+\exp(V))^2|V|^3$ is integrable with respect to both $\eta$ and $\pi$, then \cref{assump:regular-proposal,assump:regular-weights} hold.
\eprop

\bprf
By \cref{prop:regularity-MCMC}, we need to verify that $(1+\bar{G})^2\in \mcF^\text{pre}_3$, which holds if for all $i_1,\dots, i_j\geq 1$ with $i_1+\dots+i_j\leq 3$, 
\[
\int_\mcX\left(1+\tilde{G}(x)\right)^2\bar{V}_{i_1}(x)\cdots \bar{V}_{i_j}(x)\bar{\pi}(x)\dee x<\infty,
\]
where $\bar{G}(x,x')=\tilde{G}(x)=\sup_{\beta,\beta'\in[0,1]}\pi_{\beta'}(x)/\pi_\beta(x)$. 

\paragraph{Bounding $\bar{\pi}(x)$.}
By the AM-GM inequality, 
\[
\bar{\pi}(x)
&=\sup_{\beta\in[0,1]}\frac{\eta(x)^{1-\beta}\gamma(x)^\beta}{Z_\beta}
\leq\sup_{\beta\in[0,1]}\frac{(1-\beta)\eta(x) + \beta \gamma(x)}{Z_\beta}\\
&\leq\frac{1}{\inf_{\beta\in[0,1]}Z_\beta}\left(\eta(x)+\gamma(x)\right).
\]

\paragraph{Bounding $\tilde{G}(x)$.}
We have
\[
\tilde{G}(x)
&=\sup_{\beta,\beta'\in[0,1]}\frac{\pi_{\beta'}(x)}{\pi_\beta(x)}
=\sup_{\beta,\beta'\in[0,1]}\frac{Z_\beta}{Z_{\beta'}}\exp((\beta'-\beta) V(x))\\
&\leq\frac{\sup_{\beta\in[0,1]}Z_\beta}{\inf_{\beta\in[0,1]}Z_{\beta}}\exp(\max\{0,V(x)\})
\leq \frac{\sup_{\beta\in[0,1]}Z_\beta}{\inf_{\beta\in[0,1]}Z_{\beta}}(1+\exp(V(x))).
\]

\paragraph{Bounding $\bar{V}_{i}(x)$ for $i=1,2,3$.}
For $i=1$,
\[
\bar{V}_1(x)=\sup_{\beta\in[0,1]}\left|\frac{\dee}{\dee\beta}\log \pi_\beta(x)\right|=\sup_{\beta\in[0,1]}|V(x)-A'(\beta)|\leq |V(x)|+\sup_{\beta\in[0,1]}|A'(\beta)|.
\]
For $i=2,3$, we have $\bar{V}_i(x)=\sup_{\beta\in[0,1]}|\dee^i A(\beta)/\dee\beta^i|$, which is independent of $x$. Since $V$ has finite third moments with respect to $\pi$ and $\eta$, the log-partition function $A(\beta)$ is $C^3([0,1])$ and $\sup_{\beta\in[0,1]}|\dee^i A(\beta)/\dee\beta^i|<\infty$, provided $\sup_{\beta\in[0,1]}\pi_\beta[|V|^3]<\infty$. This holds since
\[
\pi_\beta[|V|^3] 
\leq \frac{1}{\inf_{\beta\in[0,1]}Z_\beta}\left(\eta[|V|^3]+\gamma[|V|^3]\right)<\infty.
\]

\paragraph{Verifying $(1+\bar{G})^2\in \mcF^\text{pre}_3$.}
Finally,
\[
&\int_\mcX (1+\exp(V(x)))^2 |V(x)|^3\bar{\pi}(x)\dee x\\
&\leq \frac{1}{\inf_{\beta\in[0,1]}Z_\beta}\left(\eta[(1+\exp(V))^2 |V|^3]+\gamma[(1+\exp(V))^2 |V|^3]\right)<\infty,
\]
which establishes $(1+\bar{G})^2\in \mcF^\text{pre}_3$.
\eprf

\subsubsection{Deterministic kernels}\label{ex:kernel_deterministic} 
Suppose for each $\beta,\beta'$ we have a bijection $F_{\beta,\beta'}:\statespace\mapsto\statespace$. This defines a forward kernel $M_{\beta,\beta'}(x,\dee x')$ which deterministically transports $x$ to $x'= F_{\beta,\beta'}(x)$ and a backward kernel $L_{\beta',\beta}(x',\dee x)$, which deterministically transports $x'$ to $x=F^{-1}_{\beta',\beta}(x')$,
\[\label{eq:kernels-determinisitic}
M_{\beta,\beta'}(x,\dee x') 
=\delta_{F_{\beta,\beta'}(x)}(\dee x'),\qquad
L_{\beta',\beta}(x',\dee x)
=\delta_{F_{\beta,\beta'}^{-1}(x')}(\dee x).
\]
In this case, the incremental weight equals,
\[\label{eq:incrmental_weight_deterministic}
G_{\beta,\beta'}(x,x')
=\frac{\pi_{\beta'}( x')}{F_{\beta,\beta'}\pi_\beta(x')},
\]
where $F_{\beta,\beta'}\pi_\beta$ is the probability measure on $\mcX$ defined by the pushforward of $\pi_\beta$ by $F_{\beta,\beta'}$. By taking expectations with respect to $\pi_\beta\otimes M_{\beta,\beta'}$, the incremental discrepancy $D(\beta,\beta')$ is the Rényi $2$-divergence between the pushforward $F_{\beta,\beta'}\pi_\beta$ and $\pi_{\beta'}$,
\[\label{eq:dicrepancy_deterministic}
D(\beta,\beta')=D(F_{\beta,\beta'}\pi_\beta,\pi_{\beta'}).
\]

Notably, $D(\beta,\beta')=0$ if and only if $F_{\beta,\beta'}$ is a transport map between $\pi_\beta$ and $\pi_{\beta'}$. If $F_{\beta,\beta'}$ is an approximate transport map then $D(\beta,\beta')$ can be small even when $D(\pi_\beta,\pi_{\beta'})$ is large, but the opposite can also be true if $F_{\beta,\beta'}$ is poorly chosen. 

In general, we want to choose $F_{\beta,\beta'}$ to approximately transport samples from $\pi_\beta$ to $\pi_{\beta'}$, and use SMC samplers to correct the error of the approximation. We can also augment the deterministic kernels with a $\pi_{\beta'}$-invariant kernel to obtain the annealed flow transport algorithm \citep{arbel2021annealed, matthews2022continual}.

\paragraph{Continuous normalising flows.}
An important example is when $\mcX$ is a manifold, and $F_{\beta,\beta'}$ is a normalising flow defined as a diffeomorphism with Jacobian determinant $\det(\nabla F_{\beta,\beta'})(x)\neq  0$. In this case, the push-forward measure has density,
\[\label{eq:density_flow}
F_{\beta,\beta'}\pi_\beta(x')=\pi_\beta(F^{-1}_{\beta,\beta'}(x'))\det(\nabla F_{\beta,\beta'}^{-1}(x')).
\]
By substituting \cref{eq:density_flow} into \cref{eq:incrmental_weight_deterministic}, and using $x'=F_{\beta,\beta'}(x)$, we can express $G_{\beta,\beta'}(x,x')$ as a function of $x$ 
\[\label{def:incremental_weight_deterministic} 
G_{\beta,\beta'}(x,x')
=\frac{\pi_{\beta'}(F_{\beta,\beta'}(x))}{\pi_\beta(x)}\det(\nabla F_{\beta,\beta'}(x)).
\]
See \cite{arbel2021annealed} for more details.

\cref{def:incremental_weight_deterministic} holds for any choice of $F_{\beta,\beta'}$. However, to obtain a tractable dense schedule limit, we must assume some structure on $F_{\beta,\beta'}$. Suppose $\beta\mapsto \dot F_\beta(x)$ is a vector field, and for all $\beta$, $\beta'\mapsto F_{\beta,\beta'}$ is a first-order integrator for the flow generated by $\dot F_\beta$ solving the ordinary differential equation,
\[
\frac{\dee x_\beta}{\dee \beta}=\dot F_\beta(x_\beta).
\]
This implies for all $x\in \mcX$, as $\Delta\beta\to 0$, 
\[\label{eq:vector_feild_flow}
F_{\beta,\beta'}(x)=x+\dot F_\beta(x)\Delta\beta+O(\Delta\beta^2).
\]
If $\pi_\beta(x)$ is a continuously differentiable in $\beta\in[0,1]$ and $x\in \mcX$, then we have $\dot{G}_\beta(x,x')$ exists, is independent of $x'$ and equals, 
\[\label{eq:S-deterministic}
\dot{G}_\beta(x,x')
&=\der{}{\beta}\log \pi_\beta(x) +\nabla\log \pi_\beta(x)\cdot \dot F_\beta(x) + \mathrm{div}(\dot F_{\beta})(x),
\]
where $\mathrm{div}(\dot F_\beta)=\tr(\nabla \dot F_\beta)$ is the divergence of the vector field $\dot F_\beta$. By taking the variance with respect to $\pi_\beta\otimes M_{\beta,\beta}$, $\delta(\beta)$ equals,
\[\label{eq:delta-deterministic}
\delta(\beta)=\var_{\pi_\beta}\left[\der{}{\beta}\log \pi_\beta+\nabla \log \pi_\beta \cdot\dot F_\beta + \mathrm{div}(\dot F_\beta)\right].
\]
Intuitively, $\delta(\beta)$ measures the instantaneous residual error accrued by integrating $X_\beta\sim \pi_\beta$ by the flow generated by the vector field $\dot F_{\beta}$. Notably, $\delta(\beta)=0$ if and only if $\dot F_\beta$ satisfies the Liouville PDE, also known as the continuity equation \citep{gardiner2002},
\[\label{eq:continuity}
\der{}{\beta}\pi_\beta+\nabla (\pi_\beta \dot F_\beta)=0.
\]

\cref{prop:regularity-deterministic} identifies sufficient conditions to ensure that \cref{assump:regular-proposal,assump:regular-weights} hold with respect to the prior regularity class constructed in \cref{sec:prior_class}.

\bprop\label{prop:regularity-deterministic}
Suppose $\mcX$ is a manifold. For all $x\in \mcX$ suppose $x,\beta\mapsto \pi_\beta(x)$ is positive and $3$-times continuously differentiable in $\beta$ and $x$. For all $\beta,\beta'$, let $F_{\beta,\beta'}(x)$ denote a normalising flow that is 3-times continuously differentiable in $\beta,\beta'$ and $x$. Suppose $M_{\beta,\beta'}$ and $L_{\beta,\beta'}$ are the deterministic kernels generated by the flow $F_{\beta,\beta'}$ defined by \cref{eq:kernels-determinisitic}. If for all $i_1,\dots,i_j\geq 1$ and $i_1+\dots+i_j=k\leq 3$, we have $(1+\barG)^2 \bar{U}_{i_1}\cdots\bar{U}_{i_j}\in \mcF^\pre_{3-k}$, then \cref{assump:regular-proposal,assump:regular-weights} hold.
\eprop

\bprf[Proof of \cref{prop:regularity-deterministic}]
We will show \cref{assump:regular-proposal,assump:regular-weights} hold with the prior regularity class $\mcF^\pre$. 
\cref{assump:regular-weights} is true since $G_{\beta,\beta'}(x,x')$ satisfies \cref{def:incremental_weight_deterministic}, and we have $G_{\beta,\beta'}(x,x')$ and its partial derivatives are independent of $x'$. We have $\beta,\beta'\mapsto G_{\beta,\beta'}$ is in $C^3([0,1]^2)$ since  $\pi_\beta(x)$ and $F_{\beta,\beta'}(x)$ are three-times continuously differentiable in $\beta,\beta',x$. \cref{assump:regular-proposal} holds by \cref{lem:prior_class}.
\eprf

\subsubsection{Optimal kernel}
Given $N$ particles, a fixed annealing schedule $\mcB=\beta_{0:T}$, and forward kernels $M_t=M_{\beta_{t-1},\beta_t}$, 
it follows from \citet[Proposition 1]{del_moral_sequential_2006} that the \emph{optimal backward kernel} $L_{t-1}$, in the sense of minimising the variance of the weights $\textbf{w}_t$, is equal to
\[\label{eq:optimal_kernel}
L_{t-1}(x_{t},\dee x_{t-1})
=\frac{\eta^N_{t-1}(\dee x_{t-1}) M_t(x_{t-1},\dee x_t)}{\eta^N_t(\dee x_{t})},
\]
where $\eta_{t-1}^N$ denotes the law of particle $X_{t-1}^n$ after iteration $t-1$ and $\eta_t^N=\eta^N_{t-1}M_t$ the law of particle $X_t^n$ after the forward propagation. Using \cref{eq:optimal_kernel}, SMC samplers reduce to IS with reference $\eta_t^N$ and target $\pi_t$, with weighted particles $(\textbf{X}^N_t,\textbf{w}^N_t)$ satisfying
\[
X^n_t\sim \eta^N_t,\quad w^n_t=\frac{\gamma_{\beta_t}(X^n_t)}{\eta^N_t(X^n_t)}.
\]
We cannot compute $\eta_t^N(x)$, making implementing the optimal backward kernel impractical. However, it provides insights into the theoretical optimal performance SMC samplers can achieve. 

Under \cref{assump:integrability,assump:independent_weights_time,assump:independent_weights_particles,assump:ELE}, the optimal backward kernel that minimises the variance of $\textbf{w}_t$ reduces to the ``suboptimal backward kernel'' from \citet[Section 3.3.2]{del_moral_sequential_2006}, and with incremental weight function equal to the Radon--Nikodym derivative between $\pi_{\beta'}$ and the forward approximation $\pi_\beta M_{\beta,\beta'}( x')=\int\pi_\beta(\dee x)M_{\beta,\beta'}(x,\dee x')$, 
\[\label{def:incremental_weight_optimal}
G_{\beta,\beta'}(x,x')=\frac{\pi_{\beta'}( x')}{\pi_\beta M_{\beta,\beta'}( x')}.
\]
By taking the expectation with respect to $\pi_\beta\otimes M_{\beta,\beta'}$, we get that the incremental discrepancy is given by:
\[
D(\beta,\beta')=D(\pi_{\beta}M_{\beta,\beta'},\pi_{\beta'}).
\]
Provided $\pi_{\beta'}(x')$ and $\pi_\beta M_{\beta,\beta'}(x')$ are differentiable in $\beta'$, and $M_{\beta,\beta}$ is $\pi_\beta$-invariant, then $\dot{G}_\beta(x,x')$ exists, and satisfies 
\[\label{eq:S-optimal}
\dot G_\beta(x,x')
=\der{}{\beta} \log \pi_\beta(x')-\left.\frac{\partial}{\partial\beta'} \log \pi_{\beta}M_{\beta,\beta'}(x')\right|_{\beta'=\beta}.
\]
By taking the expectation with respect to $\pi_\beta\otimes M_{\beta,\beta'}$ and using the assumption that $M_{\beta,\beta}$ is $\pi_\beta$-invariant, we have,
 \[
 \delta(\beta)=\var_{\pi_\beta}\left[\der{}{\beta} \log \pi_\beta(x')-\left.\frac{\partial}{\partial\beta'} \log \pi_{\beta}M_{\beta,\beta'}(x')\right|_{\beta'=\beta}\right].
 \]
This implies that $\delta(\beta)$ captures the instantaneous residual error between the annealing distribution and the forward approximation. Note that this deviates from the Fisher information, provided $M_{\beta,\beta'}(x)$ is not locally equal to the identity kernel.

\cref{prop:regularity-optimal} identifies sufficient conditions to ensure that \cref{assump:regular-proposal,assump:regular-weights} hold with respect to the prior regularity class $\mcF^\post$ constructed in \cref{sec:posterior_class}.

\bprop\label{prop:regularity-optimal}
For all $x\in \mcX$ suppose $\beta,\beta'\mapsto \pi_\beta, \pi_\beta M_{\beta,\beta'}(x)$ are positive and in $C^3([0,1]^2)$. Suppose  $L_{\beta,\beta'}$ is the optimal backward kernel defined in \cref{eq:optimal_kernel}.If for all $i_1,\dots,i_j\geq 1$ and $i_1+\dots+i_j=k\leq 3$, we have $(1+\barG)^2 \bar{U}_{i_1}\cdots\bar{U}_{i_j}\in \mcF^\post_{3-k}$, then \cref{assump:regular-proposal,assump:regular-weights} hold.
\eprop

\bprf[Proof of \cref{prop:regularity-optimal}]
We will show \cref{assump:regular-proposal,assump:regular-weights} hold with the posterior regularity class $\mcF^\post$. \cref{assump:regular-proposal} is true by \cref{lem:prior_class}. To verify \cref{assump:regular-weights}, we note that $G_{\beta,\beta'}(x,x')$ satisfies \cref{def:incremental_weight_deterministic}. Therefore for all  $x,x'\in\mcX$, $G_{\beta,\beta'}(x,x')$ is independent of $x$, and $x,x'\in\mcX$, $\beta,\beta'\mapsto G_{\beta,\beta'}(x,x')$ is in $C^3([0,1]^2)$. 
\eprf

\paragraph{Example: MCMC kernels}

When $M_{\beta,\beta'}$ is $\pi_{\beta'}$-invariant, then the data processing inequality for the Rényi $2$-divergence \citep[Theorem 1]{van2014renyi} implies,
\[
D(\beta,\beta')
=D(\pi_{\beta}M_{\beta,\beta'},\pi_{\beta'})
=D(\pi_{\beta}M_{\beta,\beta'},\pi_{\beta'}M_{\beta,\beta'})
\leq D(\pi_{\beta},\pi_{\beta'}).
\]
Therefore, the discrepancy of the optimal kernel is always less than the discrepancy of the MCMC kernels with equality if and only if $M_{\beta,\beta'}(x,\dee x')=\delta_x(\dee x')$ is the identity kernel. 

\paragraph{Example: Deterministic kernels}
If $M_{\beta,\beta'}=\delta_{F_{\beta,\beta'}(x)}(\dee x')$ is the deterministic kernel, then $\pi_\beta M_{\beta,\beta'}=F_{\beta,\beta'}\pi_\beta$, which implies 
\[
D(\beta,\beta')
=D(\pi_{\beta}M_{\beta,\beta'},\pi_{\beta'})
=D(F_{\beta,\beta'}\pi_\beta,\pi_{\beta'}).
\]
Therefore, the deterministic backward kernel is optimal.
\paragraph{Example: Mixture kernel}
Suppose $M_{\beta,\beta'}$ is a mixture between the independent kernel for $\pi_{\beta'}$ and the identity kernel for some $\theta\in[0,1]$,
\[
M_{\beta'}(x,\dee x')=\theta\pi_{\beta'}(\dee x')+(1-\theta) \delta_x(\dee x').
\]
We have $\pi_\beta M_{\beta,\beta'}$ is a mixture of $\pi_\beta$ and $\pi_{\beta'}$, with density,
\[
\pi_\beta M_{\beta,\beta'}(x')=\theta \pi_{\beta'}(x')+(1-\theta)\pi_\beta(x').
\]
By taking derivatives with respect to $\beta'$ we have,
\[
\left.\frac{\partial}{\partial\beta'} \log \pi_{\beta}M_{\beta,\beta'}(x')\right|_{\beta'=\beta}
=\theta  \der{}{\beta}\log \pi_\beta(x').
\]
Therefore $\dot G_\beta(x,x')$ equals,
\[
\dot G_\beta(x,x')=\der{}{\beta}\log \pi_\beta(x') - \left.\frac{\partial}{\partial\beta'} \log \pi_{\beta}M_{\beta,\beta'}(x')\right|_{\beta'=\beta}
=(1-\theta) \der{}{\beta}\log \pi_\beta(x').
\]
By taking expectations with respect to $\pi_\beta$ we obtain that $\delta(\beta)$ is equal to $(1-\theta)^2$ times the Fisher information of $\Pi$,
\[
\delta(\beta)=(1-\theta)^2\var_{\pi_\beta}\left[ \der{}{\beta}\log \pi_\beta\right].
\]
Therefore, the optimal global barrier $\Lambda$ is $(1-\theta)$ times the global barrier for the MCMC kernels.

\subsection{Strategies for overcoming the barrier}\label{sec:barrier}
\cref{thm:dense_limit_variance} implies that when $\Lambda$ is large relative to $N$ and $T$, estimating $Z$ can be prohibitively challenging. Since the computational resources required for stable performance scale with $\Lambda$ (\cref{sec:criticality}), improving ASMC efficiency for challenging problems requires decreasing $\Lambda$. We outline two complementary strategies.

\paragraph{Expanding the annealing parameter space.} One approach is to expand $\Pi$ to a $d$-dimensional statistical manifold $\bar{\Pi}\supset\Pi$, \citep{gelman1998simulating,grosse2013annealing,syed2021parallel,masrani2021q,barzegar2024optimal}. Parametrising $\bar{\Pi}=(\pi_\beta)_{\beta\in B}$ by a $d$-dimensional manifold $B$, the Riemannian metric becomes the $d\times d$ matrix $\bar{\delta}(\beta)=[\delta_{i,j}(\beta)]_{i,j\in[d]}$, where $\delta_{i,j}(\beta)$ are the entries of the covariance matrix of the vector-valued $\dot G_\beta(x,x')=\left.\nabla_{\beta'}\log  G_{\beta,\beta'}(x,x')\right|_{\beta'=\beta}$ with respect to $\pi_\beta\otimes M_{\beta,\beta}$. The geodesic $\varphi^*:[0,1]\to B$ from $\eta$ to $\pi$ minimising the kinetic energy $\bar{E}(\varphi)=\int_0^1\dot\varphi(u)^\intercal\bar{\delta}(\varphi(u))\dot\varphi(u)\dee u$ can be found by solving a non-linear differential equation \citep{gelman1998simulating,rotskoff2017statistical}. 

\paragraph{Modifying the kernels.} Alternatively, one can decrease $\Lambda$ without changing the annealing distributions by reducing the local barrier $\lambda(\beta)$ through optimising the forward and backward kernels, for example by incorporating learnable deterministic flows or diffusions \citep{vaikuntanathan_escorted_2011,arbel2021annealed,barzegar2024optimal,phillips2024particle,chen2024sequential}.

\subsubsection{Overcoming the global barrier for Gaussian paths}\label{ex:gaussian_det}

Suppose $\mcX=\reals$ and the reference and target are Gaussian distributions $\eta=N(\mu_0,\sigma^2)$ and $\pi=N(\mu_1,\sigma^2)$ for some $\mu_0,\mu_1\in\reals$ variance $\sigma^2>0$. Suppose the annealing distributions $\pi_\beta\propto \eta^{1-\beta}\pi^\beta$ are the linear path between $\eta$ and $\pi$. Then $\pi_\beta=N(\mu_\beta,\sigma^2)$ is also a Gaussian distribution with $\mu_\beta= (1-\beta)\mu_0+\beta \mu_1$ linearly interpolating the means between $\mu_0$ and $\mu_1$ and with a fixed variance $\sigma^2$. If we use the MCMC kernels then $\delta(\beta)$ equals,
\[
\delta(\beta)= \left(\frac{\mu_1-\mu_0}{\sigma}\right)^2.\]
The corresponding global barrier equals $\Lambda=|\mu_1-\mu_0|/\sigma$, and the optimal schedule generator is the identity $\varphi^*(u)=u$ that uniformly generates the annealing parameters between $[0,1]$. We can improve $\Lambda$ by (1) choosing a better annealing path $\pi_\beta$ between $\pi_0$ and $\pi_1$ or (2) augmenting the forward/backward kernels with a deterministic flow, as we will demonstrate in the remainder of this Section.

\subsubsection{Non-linear Gaussian paths}\label{sec:better_path}
Suppose $\pi_\beta = N(\mu_\beta,\sigma_\beta^2)$ is  a curve in the two parameter family, $\Pi=\{N(\mu,\sigma^2):(\mu,\sigma)\in \reals\times \reals_+\}$. Then $\delta(\beta)$, induced by the MCMC kernels, coincides with the Fisher information metric for $\Pi$,
\[
\delta(\beta)=\left(\frac{\dot\mu_\beta}{\sigma_\beta}\right)^2+2\left(\frac{\dot\sigma_\beta}{\sigma_\beta}\right)^2.
\]
Notably, $(\Pi,\delta)$ viewed as a Riemannian manifold is isometric to the Poincar\'e upper half plane, and the geodesic curve between $\pi_0$ and $\pi_1$ has $\Lambda=O(\log z)$ as $z=|\mu_1-\mu_0|/\sigma\to\infty$, compared to $\Lambda=O(z)$ for the linear path. This translates to an exponential reduction in $\Lambda$, and hence, an exponential improvement in the efficiency of \cref{alg:AMCS}. See \citet{costa2015fisher} for details.

\subsubsection{Example: Gaussian path with deterministic transport}\label{sec:betta_kernel}
Let us suppose now $F_{\beta,\beta'}(x)$ is an affine flow generated by the linear vector field $\dot F_\beta(x)$,
\[
F_{\beta,\beta'}(x)=x+\Delta\beta  \dot F_\beta(x),\quad 
\dot F_\beta(x)= a_\beta x+ b_\beta.
\]
By taking expectations with respect to $\pi_\beta$, we obtain
\[
\delta(\beta)
&=\left(\frac{\dot\mu_\beta-\dot F_\beta(\mu_\beta)}{\sigma_\beta}\right)^2+ 2\left(\frac{\dot\sigma_\beta}{\sigma_\beta}-a_\beta\right)^2.
\]
Note that even if the Fisher information is large using an MCMC kernel, we can achieve $\delta(\beta)=0$ provided $a_\beta$ and $b_\beta$ equal,
\[
a_\beta= \frac{\dot\sigma_\beta}{\sigma_\beta},\quad b_\beta=\dot\mu_\beta-\frac{\dot\sigma_\beta}{\sigma_\beta } \mu_\beta.
\]
This choice of $a_\beta,b_\beta$, ensures that $F_{\beta,\beta'}$ approximates a transport map along $\pi_\beta$ when $\beta'\approx\beta$.

\section{Supplement for 
\cref{sec:criticality}}

\subsection{Proof of \cref{thm:geodesics}}
\benum
\item[(a)]
By \cref{thm:incremental_discrepancy_estimate}, the incremental discrepancy satisfies
\[
\delta_{\min} \Delta\beta^2\leq D(\beta,\beta')\leq \delta_{\max} \Delta\beta^2
\]
for all $\beta,\beta'\in[0,1]$ with $\Delta\beta=\beta'-\beta$, where $\delta_{\min}=\inf_\beta \delta(\beta)$ and $\delta_{\max}= \sup_\beta \delta(\beta)$. 

For the optimal schedule, we have $\varphi^*(u)=\Lambda^{-1}(\Lambda u)$, which gives $\dot{\varphi}^*(u)=\Lambda/\lambda(\varphi^*(u))$. Since $\lambda(\beta)=\sqrt{\delta(\beta)}$ is positive and continuous by \cref{thm:incremental_discrepancy_estimate}, it follows that
\[
\inf_u \dot{\varphi}^*(u)=\frac{\Lambda}{\sqrt{\delta_{\max}}}>0, \quad
\sup_u \dot{\varphi}^*(u)= \frac{\Lambda}{\sqrt{\delta_{\min}}}<\infty.
\]
By the mean value theorem, there exists $\tilde{u}\in [u,u']$ such that $\Delta\beta = \dot{\varphi}(\tilde{u})\Delta u_t$, hence
\[
\Delta u_t \inf_u \dot{\varphi}(u)\leq |\Delta\beta| \leq \Delta u_t \sup_u \dot{\varphi}(u).
\]
Substituting $\beta^*_t=\varphi^*(u_t)$ with $u_t=t/T$ and $\Delta u_t=1/T$ gives the desired bounds for all $t\in[T]$:
\[
\frac{\Lambda^2}{\kappa T^2}\leq D(\beta^*_{t-1},\beta^*_t)\leq \frac{\kappa\Lambda^2}{T^2}.
\]

\item[(b)] 
By \cref{thm:variance_ERS}, the relative variance under the optimal schedule satisfies
\[
\var\left[\frac{\hZ}{Z}\right]=\left(1+\frac{\exp\left(D(\mcB^*_T)/\rho\right)-1}{N}\right)^{\rho}-1.
\]
We establish upper and lower bounds separately.

\benum
\item[(i)] Summing the upper bound from part (a) over $t\in[T]$ gives $D(\mcB_T^*)\leq \kappa\Lambda^2/T$. Therefore, when $N \geq\rho\log(1+\epsilon)^{-1}\left(\exp(\kappa\Lambda^2/(\rho T))-1\right)$,
\[
\var\left[\frac{\hZ}{Z}\right]
&\leq \left(1+\frac{\exp\left( \kappa\Lambda^2/(\rho T)\right)-1}{N}\right)^{\rho}-1\\
&\leq \left(1+\frac{\log(1+\epsilon)}{\rho}\right)^{\rho}-1\\
&\leq \exp(\log(1+\epsilon))-1\\
&=\epsilon,
\]
where we used the inequality $(1+x/r)^r\leq \exp(x)$ for all $r>0$.

\item[(ii)] Summing the lower bound from part (a) over $t\in[T]$ gives $D(\mcB_T^*)\geq \kappa^{-1}\Lambda^2/T$. Therefore, when $N \leq \rho\epsilon^{-1}\left(\exp(\Lambda^2/(\kappa\rho T))-1\right)$,
\[
\var\left[\frac{\hZ}{Z}\right]
&\geq \left(1+\frac{\exp\left(\Lambda^2/(\kappa \rho T)\right)-1}{N}\right)^{\rho}-1\\
&\geq \left(1+\frac{\epsilon}{\rho}\right)^{\rho}-1\\
&\geq 1+\epsilon-1\\
&=\epsilon,
\]
where we used the inequality $(1+x/r)^r\geq 1+x$ for all $r>0$.
\eenum
\eenum

\subsection{High-dimensional scaling}\label{ap:high-dim-details}
We will compute the high dimensional scaling when we have the $d$-dimensional annealing distributions, 
\[
\pi_{\beta}^{(d)}(x_{1:d})=\prod_{i\in[d]}\pi_\beta(x_{i})=\frac{1}{Z^{(d)}(\beta)}\prod_{i\in[d]}\gamma_\beta(x_{i}),
\]
with normalising constant $Z^{(d)}(\beta)$ for $\pi^{(d)}_\beta$ given by: 
\[
Z^{(d)}(\beta)=\int \prod_{i\in [d]}\gamma_\beta(x_{i})\dee x_{1:d}=\prod_{i\in[d]}\int\gamma_\beta(x_i)\dee x_i=Z_\beta^d.
\]
For $\beta,\beta'$ we also define the $d$-dimensional forward/backward kernels as the products:
\[
M^{(d)}_{\beta,\beta'}(x_{1:d},\dee x'_{1:d})
&=\prod_{i=1}^d
M_{\beta,\beta'}(x_i,\dee x_i'),\\ 
L^{(d)}_{\beta,\beta'}(x_{1:d},\dee x'_{1:d})
&=\prod_{i=1}^d
L_{\beta,\beta'}(x_i,\dee x_i').
\]
This results in incremental weights $G^{(d)}_{\beta,\beta'}$,
\[
G^{(d)}_{\beta,\beta'}(x_{1:d},x_{1:d}')
&=\frac{\pi^{(d)}_{\beta'}(\dee x'_{1:d})\otimes L^{(d)}_{\beta',\beta}(x'_{1:d},\dee x_{1:d}) }
{\pi^{(d)}_{\beta}(\dee x'_{1:d})\otimes M^{(d)}_{\beta,\beta'}(x_{1:d},\dee x'_{1:d})}\\
&=\prod_{i=1}^d\frac{\pi_{\beta'}(\dee x'_{i})\otimes L_{\beta',\beta}(x'_i,\dee x_i) }
{\pi_{\beta}(\dee x'_i)\otimes M_{\beta,\beta'}(x_i,\dee x'_i)}\\
&=\prod_{i=1}^d G_{\beta,\beta'}(x_i,x_i').
\]
If $(X_{1;d},X'_{1:d})\sim \pi^{(d)}_\beta\otimes M^{(d)}_{\beta,\beta'}$ then the product structure implies, $(X_i,X_i')\iidsim \pi_\beta\otimes M_{\beta,\beta'}$. Given $\beta,\beta'$ the incremental discrepancy $D^{(d)}(\beta,\beta')$ satisfies,
\[
D^{(d)}(\beta,\beta')
&=\log\left(\E\left[G_{\beta,\beta'}^{(d)}(X_{1:d},X_{1:d}')^2\right]\right)\\
&=\log\left(\prod_{i=1}^d\E\left[G_{\beta,\beta'}(X_i,X_i')^2\right]\right)\\
&=\sum_{i=1}^d \log\left(\E\left[G_{\beta,\beta'}(X_i,X_i')^2\right]\right)\\
&=d D(\beta,\beta').
\]
This implies the total discrepancy for the $d$-dimensional target satisfies $D^{(d)}(\mcB_T) = d  D(\mcB_T)$. Given $N$ particles, an annealing schedule $\mcB_T$ and resampling schedule with effective resample size $\rho$, we will denote $\hZ^{(d)}_\SMC$ as the unbiased estimator for the normalising constant $Z^d$ with variance, 
\[
\var\left[\frac{\hZ^{(d)}_\SMC}{Z^d}\right]=\left(1+\frac{\exp(dD(\mcB_T)/\rho)-1}{N}\right)^{\rho}-1.
\]
By differentiating $D^{(d)}(\beta,\beta')$, we have that $\delta^{(d)}(\beta)$ satisfies, 
\[
\delta^{(d)}(\beta)=\frac{1}{2}\frac{\partial^2}{\partial\beta'^2}D^{(d)}(\beta,\beta')=d\frac{1}{2}\frac{\partial^2}{\partial\beta'^2}D(\beta,\beta')=d \delta(\beta).
\]
Therefore, $E^{(d)}(\varphi)=dE(\varphi)$, and the global barrier $\Lambda^{(d)}=\sqrt{d}\Lambda$.
Since $\delta^{(d)}(\beta)\propto \delta(\beta)$ and $\Lambda^{(d)}\propto \Lambda$, we have the optimal $\varphi^*$ minimising the kinetic energy $E^{(d)}(\varphi)$ equals to $\varphi^*(u)=\Lambda^{-1}(\Lambda u)$ independent of the dimension. 

Suppose $T\sim d^{\alpha_T'}$ and $\rho \sim d^{\alpha_\rho'}$ as $d\to\infty$, where $0\leq\alpha_\rho'\leq \alpha_T'$. Since $\Lambda^{(d)}=\sqrt{d}\Lambda$, we can replicate our stability analysis as $d\to\infty$ by setting $\alpha_T'=\alpha_T/2$ and $\alpha_\rho'=\alpha_\rho/2$. This yields three regimes:

\paragraph{Unstable regime ($\alpha_T'+\alpha_\rho'<1$).} 
When the schedule is sparse relative to $d$ ($\alpha_T'<1/2$) or particles interact infrequently ($\alpha_\rho'<1-\alpha_T'$), obtaining variance less than $\epsilon>0$ requires at least an exponential number of particles in $d$. By \cref{thm:geodesics}, $N=\Omega\left(\epsilon^{-1}d^{\alpha_\rho'} \exp(\kappa^{-1} d^{1-\alpha_\rho'-\alpha_T'})\right)$ as $d\to\infty$.

\paragraph{Stable regime ($1/2\leq\alpha_T'\leq 1$ and $\alpha_\rho'\geq 1-\alpha_T'$).} 
With sufficient schedule density and adequate particle interaction, obtaining variance less than $\epsilon>0$ requires at most a linear number of particles in $d$. By \cref{thm:geodesics}, $N=O(\log(1+\epsilon)^{-1}d^{1-\alpha_T'})$ as $d\to\infty$.

\paragraph{Strongly stable regime ($\alpha_T'>1$).} 
When the schedule is sufficiently dense relative to $d$, obtaining variance less than $\epsilon>0$ requires only a constant number of particles: $N=O(1)$ as $d\to\infty$.

\section{Supplement for \cref{sec:methodology}}
\subsection{Asymptotic equivalence of round trip rate in parallel tempering}\label{sec:PT-details}
Let $\hZ_\PT$ be the normalising constant estimator obtained from NRPT run for $T'$ iterations with $N'$ annealing distributions generated by the schedule $\varphi$. Under stationarity, efficient local exploration, and integrability assumptions (A1--A3) from \citet{syed2019non}, $\hZ_\PT$ is equal in distribution to $\hZ$ from \cref{alg:AMCS} under \cref{assump:integrability,assump:independent_weights_time,assump:independent_weights_particles,assump:ELE} with $N=T'$, $T=N'$, MCMC forward kernels $M_{\beta,\beta'}$, and resampling at every iteration. 

Therefore, in the dense schedule limit as $N'\to\infty$, the relative variance of $\hZ_\PT$ is minimised by $\varphi^*$ given by
\[
\varphi^*(u)=\Lambda^{-1}(\Lambda u),\quad \Lambda(\beta)= \int_0^\beta\lambda(\beta')\dee\beta',
\] 
where $\lambda$ and $\Lambda$ denote the local and global barriers for the MCMC kernels:
\[
\lambda(\beta)= \sqrt{\var[V(X)]},\quad \Lambda = \int_0^1\lambda(\beta)\dee\beta,
\]
with $X\sim\pi_\beta$.

An alternative objective for measuring NRPT performance is the round trip rate, defined as the percentage of reference samples communicating with the target. \citet{syed2019non} showed that the optimal schedule generator maximising the round trip rate in the dense schedule limit is $\varphi_\PT^*$, where
\[
\varphi_\PT^*(u)=\Lambda_\PT^{-1}(\Lambda_\PT u),\quad \Lambda_\PT(\beta)= \int_0^\beta\lambda_\PT(\beta')\dee\beta'.
\] 
Here, $\lambda_\PT$ and $\Lambda_\PT$ are the PT local and global communication barriers respectively, defined as
\[
\lambda_\PT(\beta)=\frac{1}{2}\E|V(Y)-V(Y')|,\quad \Lambda_\PT=\int_0^1 \lambda_\PT(\beta)\dee\beta,
\] 
with $Y, Y'\overset{\text{iid}}{\sim} \pi_\beta$.

Notably, $\varphi^*=\varphi_\PT^*$ if and only if the two local barriers are proportional, i.e., $\lambda(\beta)\propto \lambda_\PT(\beta)$ for all $\beta\in[0,1]$. While this proportionality does not hold exactly in practice, empirically the two barriers are approximately proportional in the problems we considered (see \cref{fig:barriers}). Moreover, \cref{ap:numerical_schedules} provides numerical approximations of the proportionality constant, supporting the heuristic $\lambda(\beta)\approx 2\lambda_\PT(\beta)$. 

This relationship is more precise in the high-dimensional regime, where \cref{prop:NRPT} shows that $\lambda(\beta)\sim \sqrt{\pi}\lambda_\PT(\beta)$ with $\sqrt{\pi}\approx 1.77$. Therefore, both in theory and practice, the schedule generated to optimise the round trip rate is approximately equivalent to the schedule generated to optimise the variance of the normalising constant.

\cref{prop:NRPT} also implies that in the high-dimensional regime, $\Lambda_\PT = \Theta(\Lambda)$. Therefore, estimating the normalising constant with variance less than $\epsilon>0$ requires $N'=\Omega(\Lambda_\PT)$ annealing distributions and $T'=\Omega(\Lambda_\PT/\log(1+\epsilon))$ iterations. Fortunately, the choice $N'=\Theta(\Lambda_\PT)$ is identical to the recommended number of chains from round trip analysis in \citet[Section 5.3]{syed2019non}, making the recommendations from normalising constant estimation compatible with those from round trip analysis. This compatibility explains why NRPT can simultaneously explore difficult targets while accurately estimating their normalising constants (see \cref{fig:logz-estimates} and \citet[Figure 7(bottom)]{syed2019non}).

\bprop\label{prop:NRPT}
Suppose $\pi_\beta$ follows the linear path and $V$ has finite third moments with respect to $\eta$ and $\pi$. Then:
\begin{enumerate}
\item[(a)] There exists a problem-specific constant $C<\infty$ such that
\[
\sqrt{2}\lambda_\PT(\beta)\leq \lambda(\beta)\leq C \lambda_\PT(\beta),\quad \sqrt{2}\Lambda_\PT\leq \Lambda\leq C \Lambda_\PT.
\]

\item[(b)] Let $\lambda^{(d)}_\PT$ and $\Lambda^{(d)}_\PT$ denote the local and global communication barriers for the $d$-dimensional target from \cref{sec:high-d-scaling}. As $d\to\infty$,
\[
\lambda^{(d)}(\beta)\sim\sqrt{\pi}\lambda^{(d)}_\PT(\beta),\quad
\Lambda^{(d)}\sim \sqrt{\pi}\Lambda^{(d)}_\PT,
\]
where $\sqrt{\pi}=\sqrt{3.14\ldots}\approx 1.77$.
\end{enumerate}
\eprop

\bprf
\begin{enumerate}
\item[(a)] Since $|V|^3$ is integrable with respect to $\pi$ and $\eta$, both $\lambda(\beta)$ and $\lambda_\PT(\beta)$ are positive and continuous. Therefore,
\[
\lambda(\beta)\leq C\lambda_\PT(\beta),
\]
where $C=\sup_{\beta\in [0,1]}\lambda(\beta)/\lambda_\PT(\beta)<\infty$. For the lower bound, we use Jensen's inequality:
\begin{align*}
\lambda_\PT(\beta)
&=\frac{1}{2}\E[|V(X)-V(X')|]\\
&\leq \frac{1}{2}\sqrt{\E[(V(X)-V(X'))^2]}\\
&= \frac{1}{2}\sqrt{2\var[V(X)]}\\
&=\frac{1}{\sqrt{2}}\lambda(\beta).
\end{align*}

\item[(b)] This follows from combining \citet[Proposition 4, Appendix F]{syed2019non} with \cref{sec:high-d-scaling}:
\[
\lambda_\PT^{(d)}(\beta)\sim \sqrt{\frac{d}{\pi}\var[V(X)]}=\frac{1}{\sqrt{\pi}}\lambda^{(d)}(\beta).
\]
\end{enumerate}
\eprf

\section{Supplement for \cref{sec:numerical-experiments}}\label{ap:simulations}

Code and instructions for the Julia/GPU implementation are available at 
\url{https://github.com/alexandrebouchard/sais-gpu}, and for the Blang implementation, 
at \url{https://github.com/UBC-Stat-ML/blangSDK/blob/master/src/main/java/blang/engines/internals/factories/ISCM.xtend}.
All experiments were performed on the UBC ARC Sockeye cluster. 
Reproducible Nextflow scripts containing detailed versions, commits and 
parameters are available at \url{https://github.com/UBC-Stat-ML/iscm-nextflow} for the Blang experiments and 
\url{https://github.com/alexandrebouchard/OAIS-gpu/tree/main/nextflow} for the Julia experiments. 

\subsection{Models}\label{ap:models}

We provide a description of the models used for benchmarking; see also \citet[Appendix I]{syed2019non}. See also \cref{tab:models} for references to data sources as well as summary statistics including the PT and SMC global barrier estimates obtained from the experiments 
in \cref{sec:empirical-comparison}. 

\begin{figure}
	\begin{center}
	\scalebox{0.8}{
	\begin{tabular}{llllll}  
		\toprule
		Model (and dataset when applicable)    & $n$ & $d$ & $\hat \Lambda_\text{PT}$ & $\hat \Lambda$  \\
		\midrule
		Bayesian \emph{hierarchical} model (rocket launch data, \citep{McDowell_2019})  & $5\;667$ & $369$ & $8.2$ & $22.6$ \\
		Ising model  & N/A & $25$ & $3.0$ & $6.0$ \\
		Bayesian mixture model  & $300$ & $305$ & $6.6$ & $16.0$ \\
		Isotropic normal distribution & N/A & $5$ & $1.4$ & $2.5$ \\
		ODE parameters (mRNA data, \cite{leonhardt_single-cell_2014}) & $52$ & $5$ & $5.4$ & $16.3$ \\
		Phylogenetic species tree inference (mtDNA data, \cite{hayasaka_molecular_1988})  & $249$ & $10\;395$ & $6.1$ & $12.8$ \\ 
		Unidentifiable product parameterisation  & $100\;000$ & $2$ & $3.5$ & $8.1$ \\
		Rotor (XY) model, \cite{hsieh_finite-size_2013} & N/A & $25$ & $3.2$ & $5.8$ \\
		Spike-and-slab classification (RMS Titanic passengers data, \citep{Hind_2019}) & $200$ & $19$ & $4.3$ & $9.4$ \\
		\bottomrule
	\end{tabular}}
	\end{center}
	\caption{Summary of models used in the experiments, with the number of observations $n$ (when applicable), the number of latent random variables $d$, estimated NRPT global communication barrier $\hat \Lambda_\text{PT}$ and AIS/SMC communication barrier $\hat \Lambda$. For a more in-depth description, refer to \citet[Appendix I]{syed2019non}.}
	\label{tab:models}
\end{figure}

\paragraph{Bayesian hierarchical model}

A Bayesian hierarchical model employed to model rocket launch data \citep{McDowell_2019}. The dataset consists of $5,667$ rocket launch success indicators for $367$ types of rockets and is described as follows.

\begin{align}
   M &\sim \text{Uniform}(0, 1) \\
   S &\sim \text{Exponential}(0.1) \\
   P_k \mid M, S &\sim  \text{Beta}(MS, (1-M)S)\\
   Y_k \mid P_k, N_k &\sim  \text{Binomial}(N_k, P_k)
\end{align}
where $k \in \{1, 2, \dots, 367\}$ indexes the $k$-th rocket type, and $Y_k$ is the observed number of failed rocket launches out of $N_k$ launch attempts. 

\paragraph{Ising model}
Let $(V,\mathcal{E})$ be a graph with vertex set $V$ and edge set $\mathcal{E}$. The Ising model on $(V,\mathcal{E})$ at inverse temperature $\beta_\text{Ising}$ is defined by
\begin{align}
    \pi_\beta(x) &\propto \exp\left(-\beta_\text{Ising}\sum_{(i,j) \in \mathcal{E}} x_ix_j\right),
\end{align}
where each $x \in \{-1, 1\}^{|V|}$. In our experiments, $(V,\mathcal{E})$ constitute a $5 \times 5$ square lattice with $d=|V|=25$. Strictly speaking, this is a finite-lattice approximation to the Ising model which, in statistical physics, is defined on an infinite lattice. We used $\beta_\textrm{Ising} = 1$ in our experiments.

\paragraph{Rotor (XY) model}
A variant of the Rotor or XY model \cite{hsieh_finite-size_2013}, defined similarly to the Ising model in our experiments. 
We consider a model  given by the following distribution parameterised by $\beta_\text{xy}$:
\begin{align}
    \pi_\beta(x) \propto \exp\left(\beta_\text{xy}\sum_{(i, j) \in \mathcal{E}} \cos (x_i - x_j)\right),
\end{align}
where $x\in[-\pi,\pi]^d$ and $\mathcal{E}$ denotes the edge set of a $2$-dimensional $5\times5$ square lattice with $d=|V|=25$. We used $\beta_\textrm{xy} = 1$ in our experiments.

\paragraph{Simple mixture model}
The simple mixture model is defined as follows:
\begin{align}
    \pi &\sim \text{UniformSimplex}(1) \\
    \mu_1 &\sim \text{Normal}(150, 100^2)\\
    \mu_2 &\sim \text{Normal}(150, 100^2)\\
    \sigma_1 &\sim \text{Uniform}(0, 100)\\
    \sigma_2 &\sim \text{Uniform}(0, 100)\\
    K_i | \pi &\sim \text{Categorical}(\pi) \\
    Y_i | \mu, \sigma, K_i &\sim \text{Normal}(\mu_{K_i}, \sigma_{K_i}^2),
\end{align}
where $\text{UniformSimplex}(k)$ is a uniform distribution over the standard $k$-simplex, and $Y_i$ for $i \in \{1, 2, \dots, N=300\}$ is a set of real-valued observations.

\paragraph{Toy Mixture}
The toy mixture model is a uniform mixture of two Gaussian distributions with variance $0.01$ centred at $\pm 1$.

\paragraph{ODE Parameters}
In this application, we consider sampling parameters of an ordinary differential equation (ODE) for mRNA transfection data \citep{leonhardt_single-cell_2014}.
The model is defined as follows:
\begin{align}
    t_0 &\sim \text{LogUniform}(-2, 1, 10) \\
    k_0 &\sim \text{LogUniform}(-5, 5, 10) \\
    \beta_\text{ode} &\sim \text{LogUniform}(-5, 5, 10) \\
    \delta &\sim \text{LogUniform}(-5, 5, 10) \\
    \sigma &\sim \text{LogUniform}(-2, 2, 10) \\
    Y_t \mid k_0, \delta, \beta_\text{ode}, t_0, \sigma &\sim
 \text{Normal}(\mu(k_0, \delta, \beta, t, t_0),\sigma^2),
\end{align}
where $\mu(k_0, \delta, \beta_\text{ode}, t, t_0) = \frac{k_0}{(\delta - \beta_\text{ode})} (1 - \exp(-(\delta - \beta_\text{ode})  (t - t_0)))  \exp(-\beta_\text{ode}(t - t_0))$, $Y_t$ for $t\in \{1, \dots, 52\}$ are real-valued observations, and the parameters of the $\text{LogUniform}$ distribution are min, max, and the base of the log respectively.

\paragraph{Phylogenetic species tree inference}
We model a dataset of primate mitochondrial DNA (mtDNA)  \citep{hayasaka_molecular_1988} with a Bayesian phylogenetic model. The structure of the model is:
\begin{align}
    s &\sim \text{Exponential}(1) \\
    r &\sim \text{Exponential}(1) \\
    \tau \mid s, r &\sim \text{TreeDistribution}(s, r) \\
    Y \mid \tau &\sim \text{CTMC}(\tau),
\end{align}
where given $(s, r)$, the tree distribution is uniform over unrooted topologies, while branch lengths are distributed according to a gamma with shape $s$ and rate $r$. Conditionally on a tree $\tau$, the distribution over the observations is given by the leaf marginals of a tree shaped stationary, continuous-time Markov chain (CTMC) with rate matrix given by the Kimura model \citep{kimura_simple_1980}. Matrix exponentiation and the Felsenstein pruning algorithm are used to compute the likelihood (see \citet{chen2014bayesian} for background on Bayesian phylogenetic inference). 

\paragraph{Annealed Multivariate Normal}
The annealed multivariate normal distribution is defined as follows.
\begin{align}
   p(x) &\propto \prod_{i=1}^{N} \text{Normal}(x_i; 0, \tau_0^{-1})f(x_i) \\
   f(x_i) &= \exp\Big(\frac{\tau_0 x_i^2 - \tau(x_i -\mu)^2}{2}+\log\tau -\frac{\log \tau_0}{2}\Big)
\end{align}
where $p$ denotes the unnormalised density of $x \in \mathbb{R}^5$, $\tau = 5, \tau_0 = 1$ and $\mu=0$.
\paragraph{Discrete Multimodal}
The discrete multimodal distribution is a probability mass function defined over $\{0, 1, 2\}$ and proportional to $100$ for even integers and $1$ for odd integers.

\paragraph{Diffusion}
The diffusion model is based on a Euler-Maruyama discretisation of the Wright-Fisher stochastic differential equation (e.g., \cite{tataru2017statistical}):
\begin{align}
    X_0 &\sim \text{Normal}(0.1, \sigma^2) \\
    X_i \mid X_{i - 1} &\sim \text{Normal}(X_{i-1}, \sigma^2 X_{i-1} (1-X_{i-1})),
\end{align}
where $X_9$ is observed to be $0.9$.  

\paragraph{Unidentifiable product parameterisation}
The \emph{unidentifiable product parameterisation model} is a simple example of an unidentifiable model creating a difficult sampling problem due to high variability and poor conditioning of the posterior density's Hessian matrix. The model is defined as follows:
\begin{align}
    X &\sim \text{Uniform}(0, 10)\\
    Y &\sim \text{Uniform}(0, 10)\\
    N \mid X, Y &\sim \text{Binomial}(m, XY),\\
\end{align}
where $N=5,000$ is an observed random variable and $m=10,000$.

\paragraph{Spike-and-slab classification}
The RMS Titanic passengers dataset \citep{Hind_2019} consists of indicators for passenger survival status amongst other continuous, discrete, and categorical features, e.g., fare, age, and ticket class.
The dataset subset we use consists of $N=200$ observations and $8$ features where one feature, \emph{ticket class}, has been duplicated.
The motivation for the duplication is to illustrate the effectiveness of a spike-and-slab model \citep{mitchell1988bayesian} for classification tasks where data exhibits symmetries (e.g., collinearity).
The model is as follows.
\begin{align}
    Y_i &\sim \text{Bernoulli}(\text{Logistic}(\beta_0 + \langle x_i, \theta\rangle)) \\
    \sigma &\sim \text{Exponential}(1) \\
    \beta_0 \mid \sigma &\sim \text{StudentT}(1.0, 0.0, \sigma)\\
    \beta_j \mid \sigma &\sim \text{StudentT}(1.0, 0.0, \sigma) \\
    \rho &\sim \text{Uniform}(0, 1) \\
    B_j | \rho &\sim \text{Bernoulli}(\rho) \\
    \theta &= \begin{bmatrix}
        \beta_1 B_1 & \dots & \beta_8 B_8
    \end{bmatrix}^\intercal \\
\end{align}
for $i = 1, \dots, N$, $j = 1, \dots, 8$, where $Y_i$ are survival indicator random variables, $x_i$ are the corresponding feature vectors of $Y_i$, $\langle a,b \rangle$ denotes the dot product of $a, b$, and $\text{Logistic}(a) = (1+\exp(-a))^{-1}$ is the logistic function.
The $\rho$ variable represents the active probability of a feature in the model.
The $B_j$ variables indicate whether feature (indexed by) $j$ is active in the model.

\paragraph{Logistic regression}\label{ap:sonar}
The sonar dataset \citep{sonar1988} consists of measurements resulting from bouncing sonar signals off a metal cylinder and rocks.
Various patterns are captured under varying conditions.
The objective is to predict whether an object detected by sonar signals is a mine (metal cylinder) or a rock.
The dataset consists of $N=208$ observations of binary variables $b_i \in \{0, 1\}$ ($111$ mines and $97$ rocks) and real-valued features $x_i \in \mathbb{R}^{60}$ for $i \in \{1, \dots, N\}$.
We model the dataset with a logistic regression model as follows.
\begin{align}
\beta_0 &\sim \text{Normal}(\mu=0, \sigma^2=400) \\
\beta_1, \beta_2, \dots, \beta_{60} &\sim \text{Normal}(\mu=0, \sigma^2=25) \\
b_i \mid \beta_0, \dots, \beta_{60} &\sim \text{Bernoulli}(\text{Logistic}(\beta_0 + \langle \beta, x_i \rangle))
\end{align}
for all $i$ where $\beta = \begin{bmatrix} \beta_1 & \beta_2 & \dots & \beta_{60} \end{bmatrix}^\intercal$. 
Here, $x_i$ have been preprocessed and standardized to have zero mean and $0.5$ standard deviation as in \cite{dau2022waste}.

\subsection{Probabilistic programming implementation}\label{ap:experiment-details}

We implemented our OASMC and OAIS methods as plug-ins to the Blang open source modelling programming language \citep{Bouchard2022Blang}. 
As a result, the algorithms described in Section~\ref{sec:methodology}
 are available to practitioners using a modelling syntax similar to BUGS \citep{Lunn2000} and Stan \citep{stan2023stan}. 
Since both the algorithms described in this work as well as the Blang language do not make structural assumptions on the state space, our implementation can be used for state spaces other than $\mathbb{R}^d$, for example those containing combinatorial spaces. 
To demonstrate this capability we included an example where the posterior distribution is defined over the space of phylogenetic trees to complement the other examples defined over Euclidean spaces. 

The default exploration kernel $M_{\beta, \beta'}$ used in the Blang  experiments (all sections except for \cref{sec:gpus} and \cref{ap:wfsmc}) consists in a within-Gibbs slice sampler with doubling and shrinking \citep{neal2003slice}. 
For the phylogenetic model, an additional kernel is used to 
sample the space of trees, namely a subtree prune-regraft move \citep{lakner_efficiency_2008}. 
We used the time reversal of the forward kernel as the backward kernel, and systematic resampling. 
The building blocks of our implementation have been thoroughly tested for correctness using strategies such as enumeration of the program traces to ensure that unbiasedness of the normalising constant estimate numerically holds on small problems, and validation checks described in \citet[Sec. 10.5]{Bouchard2022Blang}.  
The correctness of the full algorithm is also tested by checking that the normalising constant estimates converge to the same value as a well tested PT implementation, see \cref{sec:empirical-comparison}.

\subsection{GPU implementation}\label{ap:GPU-details}

The GPU and CPU compared are respectively NVIDIA Tesla V100-SXM2-16GB and Intel Xeon Silver 4116.
We implemented OAIS and ZJA in the Julia programming language. 
Using the package \texttt{KernelAbstractions.jl}, we wrote code that 
compiles to both CPU and CUDA GPU. 
Parallelisation was achieved by assigning particles to threads. 
Instead of accumulating the moments $\hg_{t,i}$ in the inner loop as in \cref{alg:AMCS}, 
for the Julia implementation of OAIS, we stored the matrix $g_t^n$ to avoid 
communication in the inner loop (storing this matrix is not needed and is skipped for ZJA). 
We used efficient GPU functions provided by \texttt{CUDA.jl}, 
namely \texttt{cumsum} and \texttt{sum} to compute $\hg_{t,i}$ from $g_t^n$
(all adaptation costs are included in the experimental results' timings).
Details can be found in the file \texttt{barriers.jl} in the source code. 
For the GPU experiments' forward kernels, we used an alternation of three random walk MH algorithms, each with isotropic normal proposals, cycling between proposal standard deviations of $10^{-1}$, $10^0$, and $10^1$.
We used Strong Parallelism Invariance 
\citep{surjanovic2023pigeons} to assist software validation: for a given seed the GPU and CPU backends are 
carefully constructed via split random number generators \citep{jdk_splittable} so that 
they produce the same realisation. 
We also verified that the first two moments of the output of the Blang and Julia implementations match up to 
Monte Carlo error. 
The two models used in the GPU experiments, `SimpleMixture' and `Unid,' are based on the Blang models 
`SimpleMixture' and `UnidentifiableProduct', but we analytically marginalised the 
indicator latent variables in the Bayesian mixture model since our GPU 
implementation does not support discrete random variables at the time of writing.
The `Unid' model used $n = 10^{11}$ and $k = 10^{10}$. 
To avoid under-flows, before exponentiation 
we subtracted from the 
log-normalising estimates a constant, obtained for each target via a long run
(residual error on these constant offsets does not affect interpretation of the results since 
it translates all the curves in a given facet column of the figure).

\section{Additional numerical simulations}\label{ap:numerical}

\subsection{Comparison of round-based and online schedule tuning }\label{ap:numerical_schedules}

\begin{figure}[t]
\centering
\includegraphics[width=\textwidth]{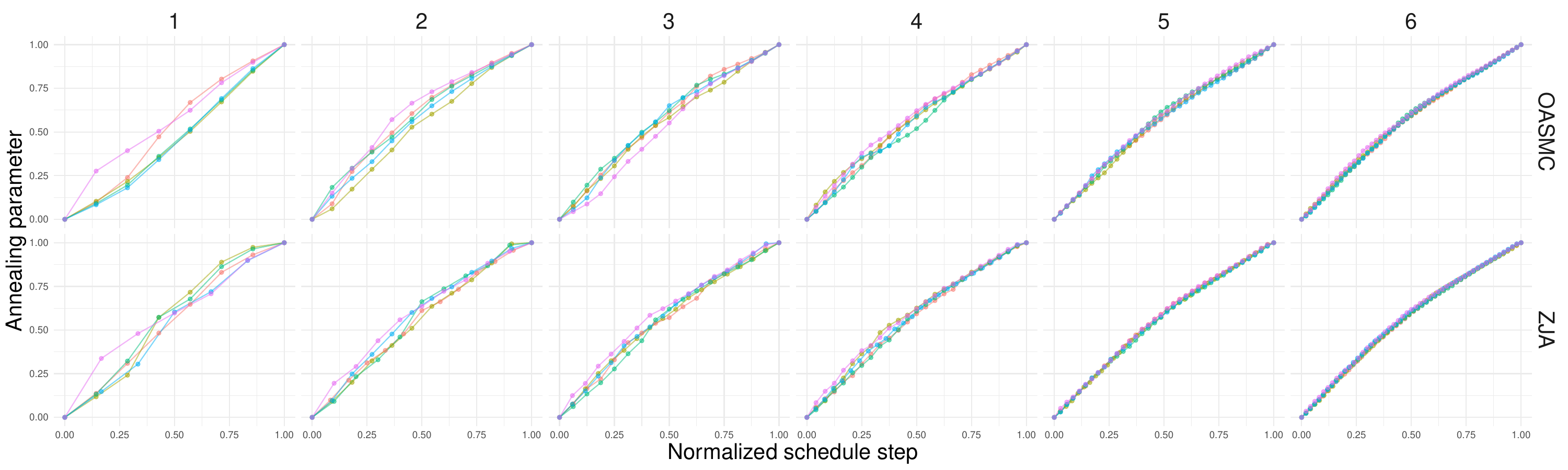}
\caption{Same as Figure~\ref{fig:rockets-schedules} but for the Ising model ($\Lambda \approx 4$).} 
\label{fig:ising-schedules}
\centering 
\end{figure}

\begin{figure}[t]
\centering
\includegraphics[width=\textwidth]{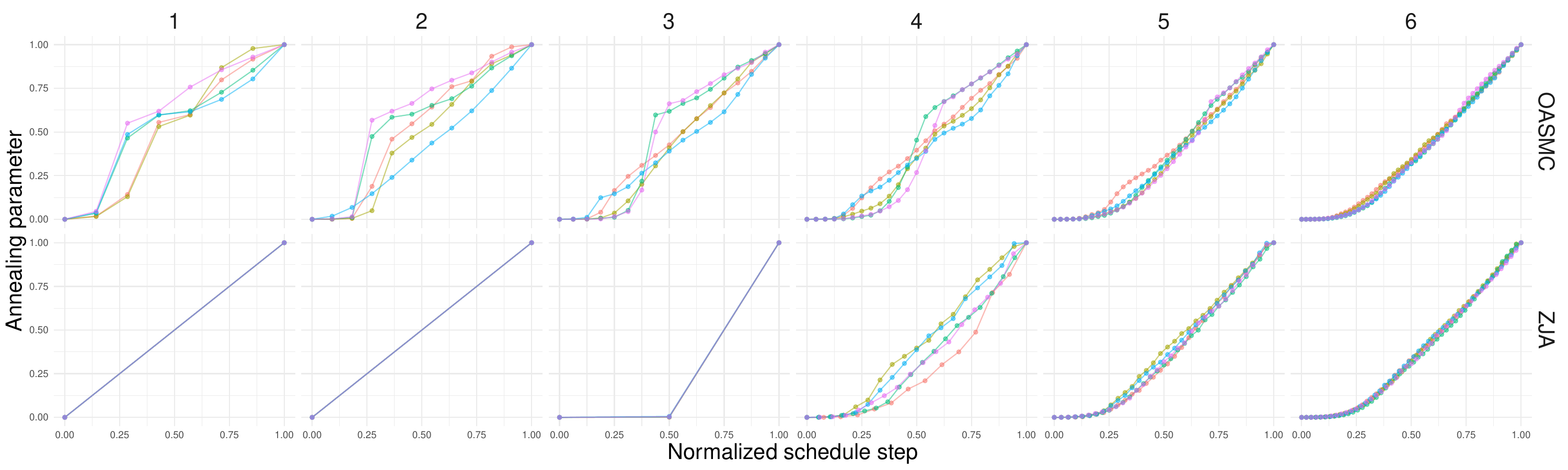}
\caption{Same as Figure~\ref{fig:rockets-schedules} but for a Bayesian hierarchical model with rocket launch data \citep{McDowell_2019} ($\Lambda \approx 23$).} 
\label{fig:ode-schedules}
\centering 
\end{figure}

We compare in Figures~\ref{fig:rockets-schedules}--\ref{fig:ode-schedules} the schedules obtained by OASMC to those from \citet{Zhou2016}. These results illustrate several useful points. First, they confirm that for sufficiently large budget, the schedules from \citet{Zhou2016} and OASMC coincide as predicted by the theory (\cref{sec:connection-online-cess}). Second, the results exemplify the challenge with setting line search thresholds in an online fashion. Even if $\Lambda$ were known (in practice, it is not), controlling the online threshold to get a coarse approximation with a prescribed number of time steps is not possible in the first few rounds, providing motivation for our doubling scheme. Finally, the flip side is that in OASMC, the information used to plan the schedule in a given round comes from the previous round which is more coarse, hence the slightly higher volatility after the first few rounds. But fortunately this penalty is a small constant (two for a doubling scheme).

\begin{figure}[h]
\centering
\includegraphics[width=0.95\textwidth]{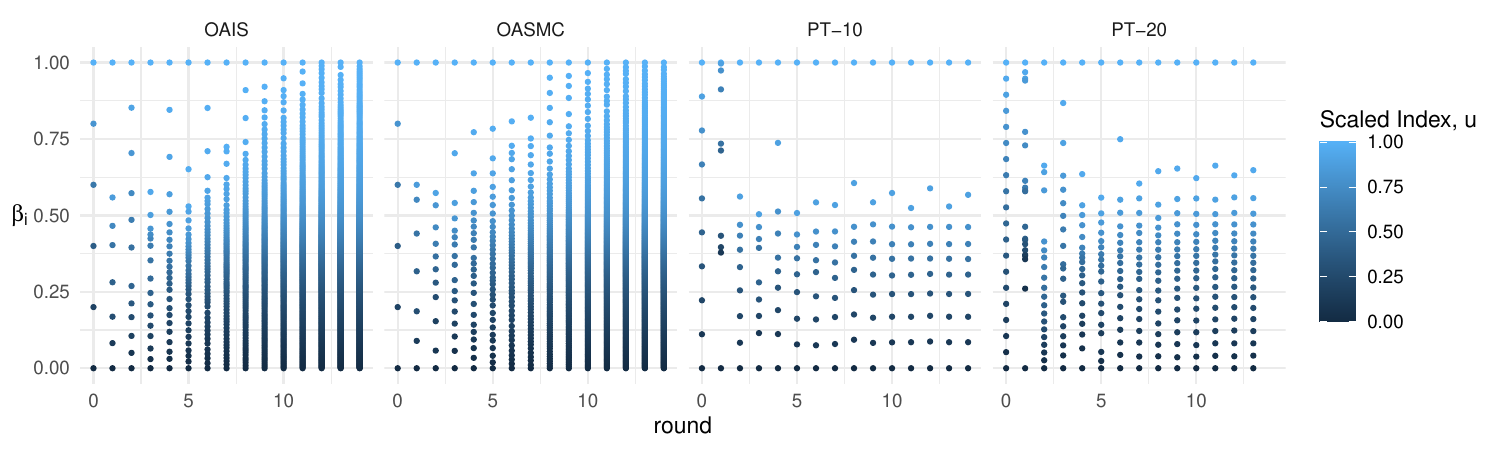}
\caption{Adaptation of annealing schedules for an Ising model (indices $t$ for $\beta_t$ represented as colours) as a function of the adaptation round for three algorithms (facets). In all algorithms, notice the concentration in a region bounded away from the endpoints in the neighbourhood of the Ising phase transition. PT algorithms typically use a fixed number of intermediate distributions $T$ since $T \gg 2\Lambda_\text{PT}$ leads to deteriorated round trip rates. In contrast, for the purpose of the normalising constant estimation, OAIS and OASMC are not inherently penalised by large $T$ values unless they lead, e.g., to prohibitive memory requirements (see Section~\ref{sec:budget}). In this example, we use $N_{k+1}, T_{k+1} \gets \sqrt{2} (N_k, T_k)$ until ESS is uniformly above a threshold (here 0.5), at which point we switch to $N_{k+1}, T_{k+1} \gets 2 N_k, T_k$ (the switch occurs at round 10 and 9 for OAIS and OASMC respectively).} 
\label{fig:annealing-param}
\centering 
\end{figure}

\begin{figure}[h]
\includegraphics[width=\textwidth]{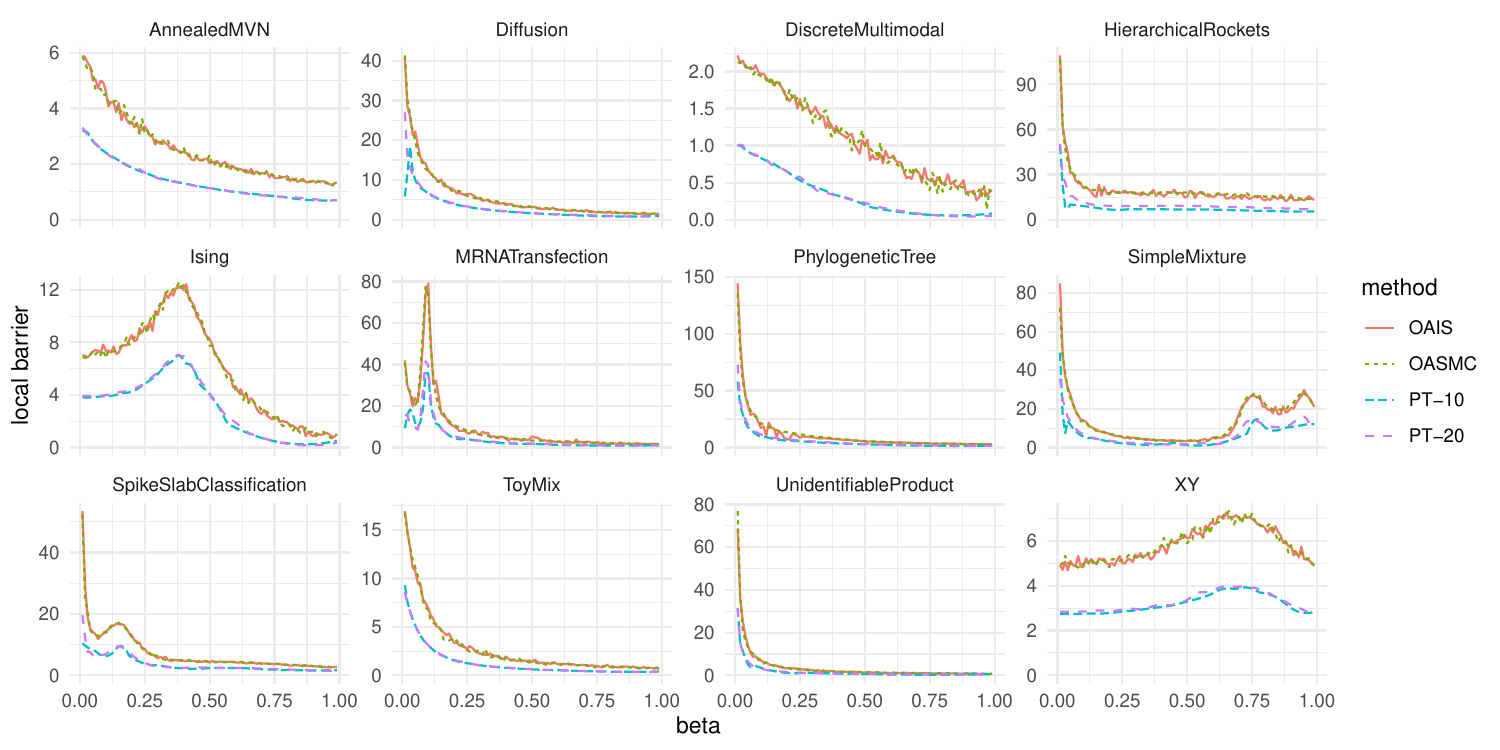}
\caption{Local barriers on 12 inference problems (facet). PT refers to the parallel tempering local communication barrier $\lambda_\text{PT}(\beta)$. OAIS and OASMC refer to the local barrier $\lambda(\beta)$ estimated by the two algorithms. The high overlap of OAIS and OASMC confirms that the local barrier is not sensitive to the resampling schedule.}
\label{fig:barriers}
\end{figure}

We then show an example in \cref{fig:annealing-param} of the evolution of schedules as a function of the round, comparing OAIS, OASMC, PT-10 and PT-20. Next, we show in \cref{fig:barriers} more examples of local barriers, again for OAIS, OASMC, PT-10 and PT-20.

\subsection{Computing expectation: numerical results}\label{ap:rejuvenation}

\begin{figure}[h]
\includegraphics[width=\textwidth]{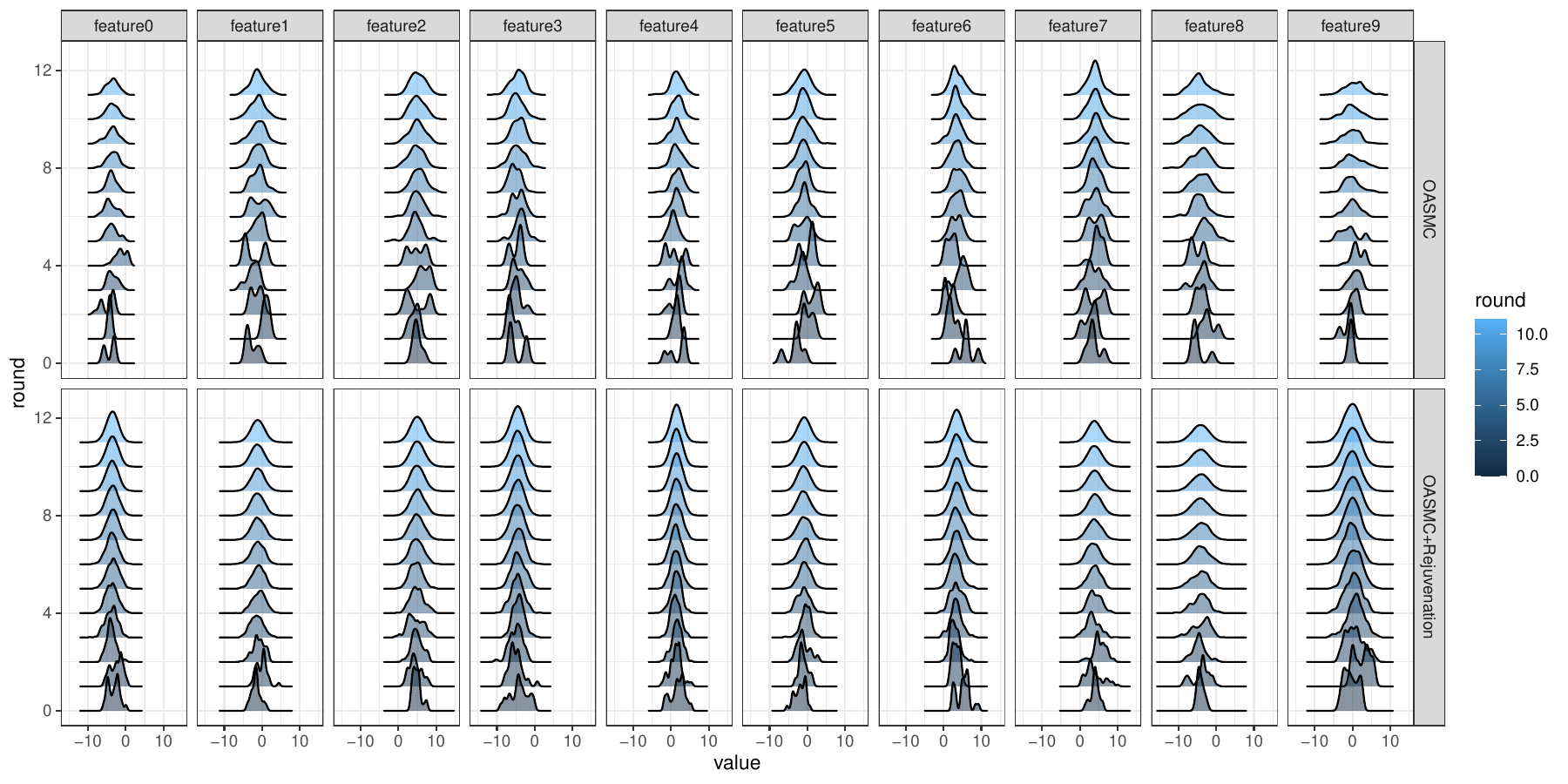}
\caption{Sequence of posterior distribution approximations for different model variables (facet columns, only first 10 variable marginals shown) from the sonar problem. Each element in the sequence of posterior approximations is obtained from successive rounds (y-axis) produced by our OASMC algorithm. Top facet row: OASMC with expectations estimated from the particles at the last ASMC iteration of each round. Bottom facet row: OASMC with expectations estimated using final iteration rejuvenation (round $k$ of OASMC+Rejuvenation has twice the cost of round $k$ of OASMC alone). Notice that with rejuvenation, after as few as 6 rounds, there is very little difference across subsequent rounds' posterior approximations. In contrast, OASMC requires at least 12 rounds to get approximations of similar quality. }
\label{fig:ridges}
\end{figure}

In this section we apply the final iteration rejuvenation discussed in \cref{sec:expectations}. Recall this is a way to compute expectations as a by-product of OASMC. We compare this final iteration rejuvenation method (labelled ``OASMC+Rejuvenation'') to the standard approach of taking only the particles at the last SMC iteration of each round (``OASMC'') in \cref{fig:ridges}. The results support the use of final iteration rejuvenation.

\subsection{Connections with Waste-Free SMC}\label{ap:wfsmc}

Traditionally, increasing the effort of SMC sampling has often been interpreted as increasing only the number of particles. Both our scheme and other recent work go beyond this and can accommodate a relatively large number of MCMC steps in proposals. An earlier proposal exploring this idea in a different fashion is the Waste-Free SMC (WFSMC) approach of \citet{dau2022waste}; see also \citet{andrieu_monte_2025} for a related, recently developed method. In the case of WFSMC, economical use of a large number of MCMC steps is achieved by having, between successive resampling events, homogeneous MCMC chains which are fully utilised in the next resampling step. In the case of OASMC, a similar goal is achieved using an increasingly fine discretisation of the annealing schedule while performing few MCMC exploration iterations per propagation step for a slowly changing annealing parameter.

This connection between WFSMC and OASMC motivates a numerical comparison of the two approaches. Moreover, since WFSMC is implemented in the popular and excellent \emph{particles} package \citep{particles2018}, comparison between the two methods will be of interest to practitioners.

\begin{figure}[t]
\centering
\includegraphics[width=0.8\textwidth]{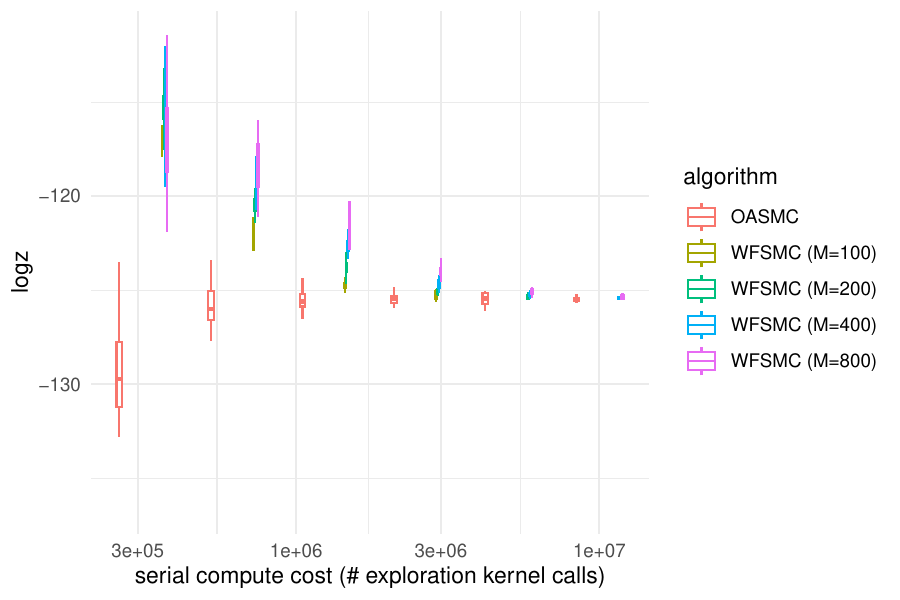}
\caption{Each box plot summarises $\log \hat Z$ estimates replicated over 20 seeds for WFSMC (Waste-Free SMC;  \citet{dau2022waste}) and our OASMC method. The x-axis indicates computational effort in terms of the number of MCMC exploration steps (for both methods, random walk MH with a dense covariance proposal estimated from the particle populations). For OASMC, the different computational efforts are obtained by recording the estimate at each round. For WFSMC, to modulate the computational effort, we follow the structure of Section 5.1 in \citet{dau2022waste}: the colours show values of $M$ in Dau and Chopin's notation (the number of resampled particles). Increased computational effort for each WFSMC series (colour) corresponds to increasing the number of recycled MCMC iterations per MCMC exploration steps.} 
\label{fig:wfsmc}
\centering 
\end{figure}

To this end, we have reproduced in the Blang PPL the first numerical example in \citet{dau2022waste}, and ensured that for large computational budgets, the approximation from the \emph{particles} package's WFSMC and our OASMC algorithm closely match. 
We used version 0.4 of the \emph{particles} package. Blang and the \emph{particles} package are implemented in different programming languages, but since they are based on the same MCMC exploration kernel, we can use the total number of MH iterations as a fair metric of serial computational cost. 

The results are shown in \cref{fig:wfsmc}. In most situations, the regime of interest will be on the right of \cref{fig:wfsmc}'s x-axis, i.e., the ``low error'' region of this plot, where all methods perform similarly. In that regime, the advantage provided by our method is its predictable runtime and anytime property, which makes it more convenient for practitioners. The results also suggest an advantage for OASMC on the left of the x-axis, i.e. the ``low cost'' region of \cref{fig:wfsmc}, which may be useful in a massively parallel context (several independent $\hat Z$ estimators can be averaged thanks to their unbiasedness). 
As discussed at the beginning of this section, there is a homogeneous/inhomogeneous dichotomy that could explain the empirical performance gap in the ``low cost'' regime.

\section{Choice and adaptation of propagation kernels}\label{ap:explorer}

ASMC methods rely on the propagation kernels $M_{\beta_{t-1}, \beta_t}$ being able to adjust the particles to the changes in the distributions $\pi_\beta$ as  $\beta$ varies from zero to one. Because $\pi_\beta$ can vary significantly across values of $\beta$, using, say, the same MH proposal for all $\beta$ in random walk MH would be highly inefficient.
We provide here a short review of the existing literature on strategies to tune invariant MCMC kernels in the context of ASMC, and guidance on how these strategies can be incorporated into OASMC algorithms (recall that we do not require the propagation kernels to be based on invariant MCMC moves, see \cref{sec:analysis-forward-backward} for alternatives, however invariant propagation kernels are an important special case hence the motivation for this section). 

In order to face the fact that the distributions $\pi_\beta$ may be quickly changing with $\beta$, two strategies have been used in the ASMC context: online kernel adaptation (see, e.g., \citet{chopin_sequential_2002,jasra_inference_2011,fearnhead_adaptive_2013}), and the use of kernels that can work robustly across a range of targets without tuning (e.g., \citet{neal2003slice,Biron2024autoMALA}). The latter is straightforward to integrate into OASMC since these automatic kernels are by construction robust to a wide range of target distributions. This simplicity is the main motivation for taking this route in most of the numerical simulations in the present work. 
As for the adaptation approaches, they can be directly applied to OASMC, as long as we let the number of particles at round $k$, $N_k$, grow sufficiently to have robust MCMC adaptation. See \cref{ap:wfsmc} for one set of experiments using adaptation of propagation kernels (specifically, full covariance adaptation of random walk MH kernels). 
Note also that the round-based nature of OASMC opens the door to novel kernel adaptation strategies. For example, one could use data collected in the previous round and/or pool samples from several annealing parameters (via importance sampling). This type of adaptation has been used for parallel tempering \citep{Bouchard2022Blang,surjanovic2023pigeons} and the convergence of closely related round-based adaptation algorithms has been analysed in \citet{hofstadler_almost_2025}.

This section and our numerical results have focused on propagation steps based on  invariant kernels; for the tuning of unadjusted counterparts of the algorithms discussed here, see \cite{kim_tuning_2025} for a recently developed approach to tuning these unadjusted propagation kernels, including an up-to-date comprehensive review of that literature.

\end{appendix}

\end{document}